\documentclass[usenatbib,useAMS,a4paper,fleqn]{mnras}
\usepackage{amssymb}
\usepackage{newtxtext,newtxmath}
\usepackage[T1]{fontenc}
\usepackage{ae,aecompl}
\usepackage{times}
\usepackage{microtype}
\usepackage[usenames]{xcolor}

\usepackage{graphicx}
\date{Accepted XXX. Received YYY; in original form ZZZ}
\pubyear{2016}

\begin{document}

\label{firstpage}
\pagerange{\pageref{firstpage}--\pageref{lastpage}}

\title[External effects on inner systems]
  {The effects of external planets on inner systems: multiplicities, inclinations, and pathways to eccentric warm Jupiters}

\author[Mustill, Davies, \& Johansen]
  {Alexander J. Mustill$^1$\thanks{E-mail: alex@astro.lu.se}, 
    Melvyn B. Davies$^1$, 
    Anders Johansen$^1$\\
    $^1$Lund Observatory, Department of Astronomy \& Theoretical Physics, 
    Lund University, Box 43, SE-221 00 Lund, Sweden}

\maketitle

\begin{abstract}
  We study how close-in systems such as those detected by \textit{Kepler} are affected by the 
  dynamics of bodies in the outer system. We consider 
  two scenarios: outer systems of giant planets potentially unstable to planet--planet 
  scattering, and wide binaries that may be capable of driving Kozai or 
  other secular variations of outer planets' eccentricities. 
  Dynamical excitation of planets in the outer system reduces the 
  multiplicity of Kepler-detectable planets in the inner system 
  in $\sim20-25\%$ of our systems. 
  Accounting for the occurrence rates of wide-orbit planets and 
  binary stars, $\approx18\%$ of close-in systems could 
  be destabilised by their outer companions in this way. This 
  provides some contribution to the apparent excess 
  of systems with a single transiting planet compared to multiple; however, 
  it only contributes at most 25\% of the excess.
  The effects of the outer dynamics can generate 
  systems similar to Kepler-56 (two coplanar planets significantly 
  misaligned with the host star) and Kepler-108 (two significantly 
  non-coplanar planets in a binary). We also identify three pathways to 
  the formation of eccentric warm Jupiters resulting from the interaction between 
  outer and inner systems: direct inelastic collision between an eccentric outer and an 
  inner planet; secular eccentricity oscillations that may ``freeze out'' when 
  scattering resolves in the outer system; and scattering in the inner system 
  followed by ``uplift'', where inner planets are removed  
  by interaction with the outer planets. In these scenarios, the formation of 
  eccentric warm Jupiters is a signature of a past history of
  violent dynamics among massive planets 
  beyond $\sim1$\,au.
\end{abstract}


\section{Introduction}


\noindent The population of planet candidates detected by \textit{Kepler} shows a surplus of 
systems showing only one transiting planet \citep{Johansen+12,BallardJohnson16}, 
a finding that has been dubbed the ``\textit{Kepler} Dichotomy''. 
This translates into 
an excess of systems with only one \textit{planet} in the region probed by 
\textit{Kepler,} as altering the distribution of mutual 
inclinations amongst triple-planet systems cannot 
simultaneously account for the numbers of single-, double- and 
triple-transit systems \citep{Johansen+12}: 
a large fraction of the double-transit systems could be produced by 
intrinsically triple-planet systems, but this still requires an additional 
population of intrinsically single-planet systems to match the large 
observed number of single-transit systems. This suggests that Nature 
produces two distinct populations of inner planetary systems: one population 
of intrinsically single planets, and an additional population of multiple 
systems whose multiplicity peaks at three planets or higher.
There are three possible explanations for this excess:
\begin{itemize}
\item There is a high false positive rate amongst single-transit systems. 
  While most \textit{Kepler} multiple systems appear to be genuine 
  \citep{Rowe+14}, \cite{Santerne+16} find a $\sim50\%$ false-positive rate 
  for single giant \textit{Kepler} candidates. They speculate that the absolute number 
  of false positives 
  may be higher for the smaller candidates, although the rate could fall due to 
  the increased frequency of smaller planets.
\item Many systems form only one planet within $\lesssim1$\,au, while a smaller number 
  form multiple. \cite{ColemanNelson16} find that, when starting from a large number 
  of embryos, systems resembling the \textit{Kepler} singles only arise if one 
  planet grows to be massive enough to clear out its neighbours, and speculate 
  that many systems must form small numbers of embryos. Unfortunately, predicting 
  the formation times, locations and numbers of these embryos is challenging, 
  despite the significant effects that these initial conditions have on the 
  embryos' subsequent growth and migration \citep[e.g.,][]{Bitsch+15}. 
\item Many systems form multiple planets within $\lesssim1$\,au, but many are later 
  reduced in multiplicity by subsequent dynamical evolution as planets collide. 
  This route may be supported by an additional ``dichotomy'' in the distributions 
  of orbital eccentricities \citep[see][who argue for a two-component model for the eccentricity 
    distribution, with a low-$e$ ($\sim0.01$) and a high-$e$ ($\sim0.2$) 
    component; \citealt{Xie+16} argue for a similar ``eccentricity dichotomy'']{Shabram+16} and stellar obliquities 
  (\citealt{MortonWinn14} find that stars with a single transiting planet have higher 
  obliquity than those with multiple planets, while \citealt{Campante+16} favour a 
  mixture model for the obliquities of single-planet host stars but a single model 
  for hosts of multiple planets). 
  This evolution may be driven by the internal dynamics of the \textit{Kepler} multiples 
  \citep[e.g.,][]{Johansen+12,PuWu15,VolkGladman15} or by the effects of outer 
  bodies such as binary stars or outer giant planets on the inner system 
  \citep[e.g.,][]{Mustill+15}.
\end{itemize}
In this paper, we further explore the effects
a dynamically active outer system can have on systems of multiple inner planets. We build
on our previous work \citep{Mustill+15}, in which we considered the
effects of a planet with an arbitrarily-imposed eccentricity on an
inner system, by consistently modelling the dynamics in the outer
system leading to such eccentricity excitation, through Kozai cycles
or planet--planet scattering. We
gauge the contribution of disruptive outer bodies to the \textit{Kepler} multiplicity
function by destabilising and inclining inner systems, show that it is
possible to occasionally generate large mutual inclinations or obliquities as in
the tilted two-planet system Kepler-56 \citep{Huber+13} and the mutually 
inclined Kepler-108 \citep{MillsFabrycky16}, and identify several routes to forming
eccentric warm Jupiters at a few tenths of an au.

Can the \textit{Kepler} Dichotomy be resolved by appealing to instabilities 
driven by the internal dynamics of inner systems (henceforth, anything 
with an orbital period $P<240$\,d)? Probably not entirely: 
\textit{Kepler} triple-planet systems, for example, 
are robust to internal dynamical evolution. \cite{Johansen+12} 
showed that these triple-planet systems are too widely separated to undergo instability 
unless their masses are increased unrealistically, by a factor of around 100. Furthermore, when forced 
into instability in this way the outcome is typically only 
a reduction to a two-planet system. However, \cite{PuWu15} show that the higher-multiplicity 
systems are less stable, consistent with being the survivors from a continuous primordial 
population where the more closely-spaced systems were unstable. Meanwhile, 
\cite{BeckerAdams16} find that \emph{Kepler} multi-planet systems are 
inefficient at self-exciting their mutual inclinations: flat systems remain flat, 
and retain their high probablility of multiple transits.

A high occurrence rate of inner planetary systems ($\sim50\%$) has been revealed
by both RV surveys \citep{Mayor+11} and \textit{Kepler} \citep{Fressin+13}. 
But many of these inner systems do not exist 
in isolation. They may have wide-orbit companion planets, as in the case of 
Kepler-167, which possesses three super-Earths within 0.15\,au together with 
a transiting giant planet at 1.9\,au \citep{Kipping+16}. A number of studies 
have found wide-orbit candidates in the \textit{Kepler} light curves which transit 
only a small number of times and therefore are excluded from the KOI listings 
\citep{Wang+15,Osborn+16}; \cite{Uehara+16} 
estimate that at least 20\% of compact multi-planet systems also host 
giant planets beyond 3\,au, based on single-transit events in the KOIs; 
and \cite{Foreman-Mackey+16} 
estimate an average of 2 planets per star with periods between 2 and 25 
years and radii between $0.1$ and $1\mathrm{\,R_J}$, $0.4$ planets per star 
in the same period range with radii between $0.4$ and $1\mathrm{\,R_J}$,
and that these wide-orbit planets occur disproportionately often around 
stars already hosting inner planet candidates. 
\cite{Knutson+14} find that $50\%$ of hot Jupiter hosts also have a giant planet
companion between 1 and 20\,au, while \cite{Bryan+16} similarly find an
occurrence rate of $50\%$ for outer planetary companions to RV-detected
inner planets of a range of masses, although their sample is more metal-rich
than the \textit{Kepler} targets.
\cite{Wang+15} found that half of their long-period
\textit{Kepler} candidates exhibited transit timing variations, suggesting multiplicity.

Regarding the presence of planetary systems in wide binaries, 
some \textit{Kepler} systems, 
such as Kepler-444 \citep{Campante+15} and Kepler-108 
\citep{MillsFabrycky16}, reside in wide binaries. \cite{Ngo+15,Ngo+16} estimate that $50\%$
of hot Jupiters have a stellar companion between 50 and 2\,000\,au, around twice the rate 
for the average field star.
There is currently debate about the extent to which the presence of an outer binary 
companion affects the existence of inner \textit{Kepler} planets 
\citep[e.g.,][]{Wang+14,Deacon+16,Kraus+16}.

Statistics from systems without detected inner planets also reveal the 
prevalence of outer bodies. RV surveys 
reveal a population of ``Jupiter analogues'' (variously defined as low-eccentricity
$\sim$Jupiter-mass planets at several au) of a few
percent \citep{Rowan+16,Wittenmyer+16}. Direct imaging surveys are sensitive
to super-Jovian planets at tens of au, finding an occurrence rate of around
$10\%$ \citep{Vigan+12} for stars more massive than the Sun, falling to $1-2\%$ 
for Solar-type stars \citep{Galicher+16}.
Microlensing reveals an occurrence rate of $\sim50\%$ for ice-line planets
more massive than Neptune, where the host stars were typically sub-Solar in mass
\citep{Shvartzvald+16}. Around half of Sun-like stars are in members of
multiple stellar systems, with a period distribution peaking at $\sim10^5$ days
\citep{Raghavan+10,DucheneKraus13}. Compared to the statistics in 
the previous paragraph, it may be that stars with known inner 
planets are more likely than other stars to host wide-orbit giant 
planets, although one should be wary of biases such as for example 
in the stellar metallicities.

The configuration and evolution of bodies in the outer system can have 
significant dynamical effects on these inner systems. In \cite{Mustill+15}, 
we showed that a high-eccentricity giant planet \textit{en route} to becoming a 
hot Jupiter will destroy any existing close-in planets, thus explaining why 
hot Jupiters are typically not seen with close, low-mass companions. \cite{Mustill+15} 
also showed that, as the orbital binding energy of the eccentric giant can be 
comparable to that of the inner planets, the giant can in fact be ejected as a result 
of the interactions with the inner system, which may itself be reduced in multiplicity. 
Although hot Jupiters are relatively rare, being found around only $\sim1\%$ of stars 
\citep{Mayor+11,Howard+12,Fressin+13,Santerne+16}, 
models of high-eccentricity migration of hot Jupiters typically find that 
many more migrating giants are tidally disrupted than go on to become hot Jupiters 
\citep[e.g.,][]{Petrovich15b,Anderson+16,Munoz+16,PetrovichTremaine16}. Lower-mass planets may well 
be injected into the inner systems by the same dynamical mechanisms---scattering and 
Kozai perturbations---that give rise to hot Jupiters, and many outer planets 
thus sent inwards will attain pericentres insufficiently small for tidal 
circularisation, yet small enough to interact with inner planets at a few 
tenths of an au. All this motivates a general investigation into the influence 
of outer systems on inner \textit{Kepler}-detectable planets.

While the bulk of \textit{Kepler}-detected planets lie at a few tenths 
of an au, work has shown that instabilities in outer systems 
can be devastating for material in the habitable zone at 
$\sim1$\,au 
\citep{VerasArmitage05,VerasArmitage06,Raymond+11,Raymond+12,Matsumura+13,KaibChambers16}. 
\cite{Carrera+16} find that the survivability of bodies increases 
closer to the star, and \citep{Huang+16b} study the effects 
on the excitation of \textit{Kepler}-like super Earths. 
Direct scattering is the most obvious effect of eccentricity 
enhancement in the outer system, but secular resonances can also 
play a role in destabilising inner systems \citep{Matsumura+13,Carrera+16}. 
Secular interactions can also have more subtle effects on inner systems, 
resulting in gentle tilts \citep{GratiaFabrycky17}, or 
excitation of mutual inclination \citep{Hansen16,LaiPu16}, which 
in turn can contribute to the observed multiplicities seen by 
\textit{Kepler}.

Dynamical interaction between inner and outer systems may also account 
for the existence of eccentric warm Jupiters: giant planets with semi-major 
axes of a few tenths of an au and eccentricities of order $0.5$. While 
planet--planet scattering has long been recognised as a source of 
eccentricity excitation of giant planets 
\citep[e.g.][]{RasioFord96,WeidenschillingMarzari96,Chatterjee+08,JuricTremaine08,Raymond+11,Kaib+13}, 
\cite{Petrovich+14} showed that this process is ineffective at exciting 
eccentricities close to the star: on tight orbits, planets have a higher Keplerian 
velocity and so in order to impart a given change in velocity, 
a close encounter must occur at a smaller separation due to the reduced gravitational focusing, 
and such close encounters result instead in physical collision. 
Nor can eccentric warm Jupiters be explained by ``fast'' tidal migration of giant planets \emph{en route} 
to forming hot Jupiters, as the eccentricities of the observed planets lie below the tidal 
circularisation tracks along which such planets would migrate. Possible explanations 
are ``slow'' tidal migration, in which tidal dissipation only occurs 
briefly at the tip of a secular eccentricity cycle 
\citep{DawsonChiang14,Dong+14,Petrovich15b,PetrovichTremaine16}, 
and the physical collision of eccentric migrating giant planets with other planets 
on close-in orbits \citep{Mustill+15}. In this paper, we describe several other 
routes to the formation of eccentric warm Jupiters.

In summary, at least a few 10s of per cent of inner systems can be expected 
to host outer planets and/or stars. In this paper we study the effects of 
these outer bodies on inner systems with $N$-body integrations. 
We set up two scenarios: outer planets in binary systems that may be subject 
to Lidov--Kozai oscillations \citep{Lidov62,Kozai62,Naoz16}, and tightly-packed 
systems of outer planets that are unstable to scattering; 
\cite{JuricTremaine08} and \cite{Raymond+11} show that the eccentricity 
distribution of giant planets is consistent with around $75-83\%$ of them 
having originally come from unstable multiple systems. We investigate the effects that 
the dynamics of the outer system have on the multiplicities of the inner 
planets and on their mutual inclinations. In Section~\ref{sec:nbody} 
we describe the set-up of our $N$-body integrations. We 
describe the outcomes of a set of control integrations in 
Section~\ref{sec:control}. We give the results for 
planets in binary systems in Section~\ref{sec:binaries} and for 
unstable scattering systems in Section~\ref{sec:scattering}, 
describing the effects on the multiplicities and mutual inclinations 
of inner planetary systems. In Section~\ref{sec:warmj} we describe 
three mechanisms leading to the formation of eccentric warm Jupiters: 
collision between an inner and an eccentric outer planet (Section~\ref{sec:collide-warmj}), 
secular forcing aided by ``freeze-out'' (Section~\ref{sec:secular}), and 
\emph{in-situ} scattering aided by ``uplift'' from the outer system 
(Section~\ref{sec:uplift}). We discuss our results in 
Section~\ref{sec:discuss}, notably the effects on \textit{Kepler} systems' 
multiplicities (Section~\ref{sec:kepler-multi}) and mutual 
inclinations (Section~\ref{sec:kepler-inc}), the generation 
of large obliquities or mutual inclinations (Section~\ref{sec:inc}), 
and summarise in Section~\ref{sec:conclude}.

\section{Numerical methods and setup}

\label{sec:nbody}

\begin{figure*}
  \includegraphics[width=0.48\textwidth]{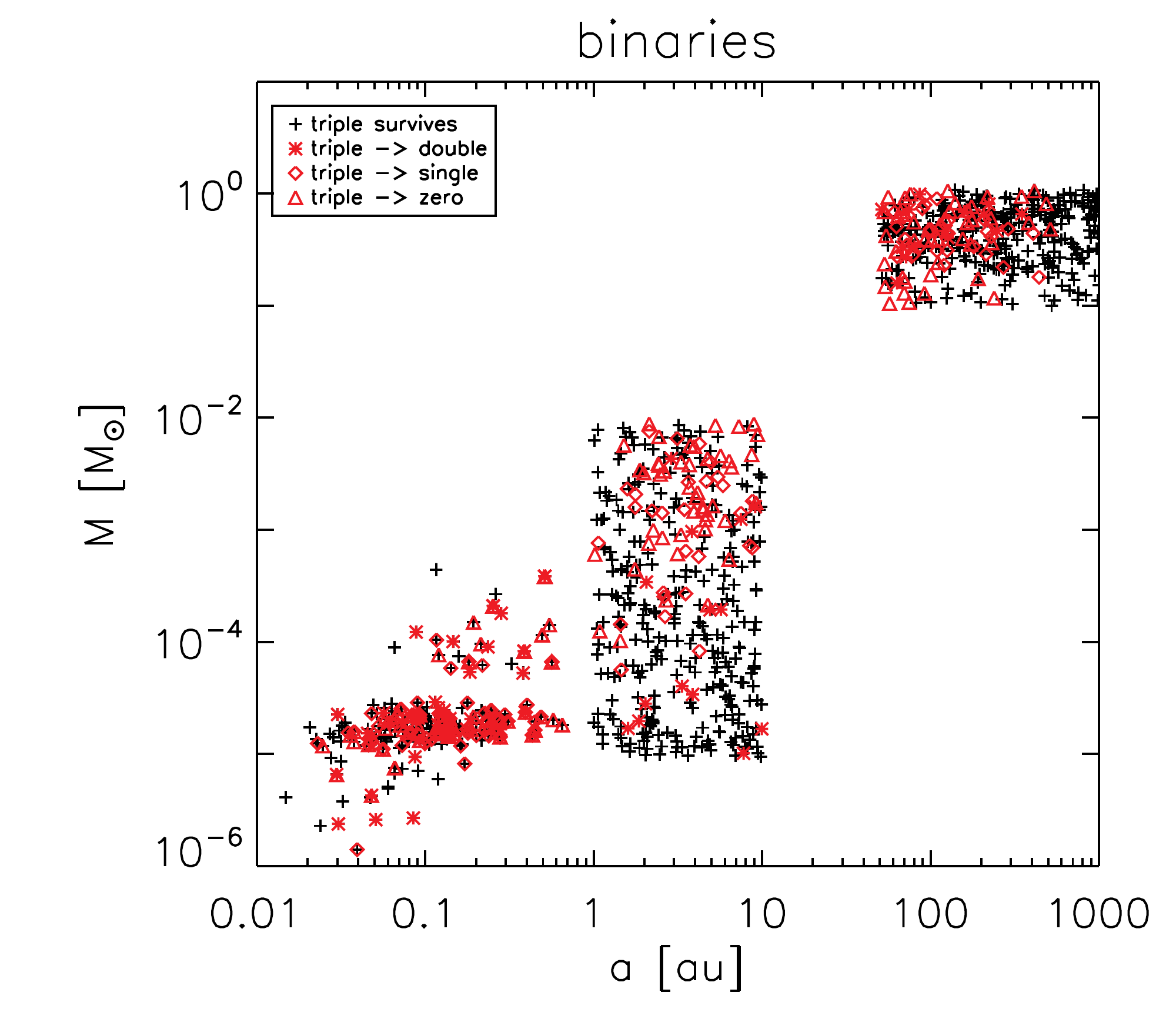}
  \includegraphics[width=0.48\textwidth]{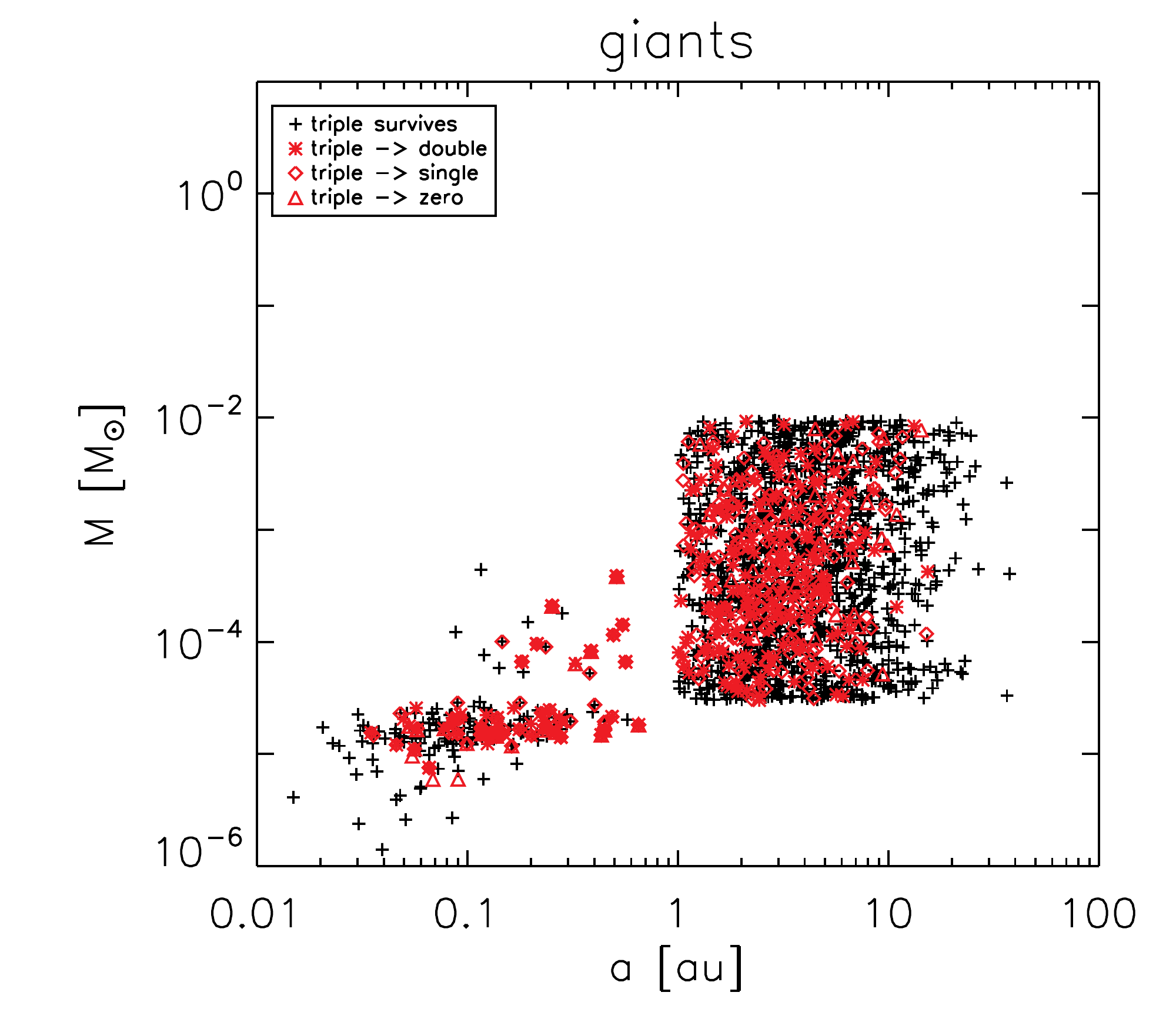}
  \caption{Initial semi-major axes and masses of all bodies in
    our \textsc{Binaries} and \textsc{Giants} simulations. There
    are three inner planets per system; inner planets are drawn from triple-planet
    KOI systems. In \textsc{Binaries} there is one additional outer
    planet and one stellar companion per system, while in \textsc{Giants}
    there are four outer planets. Bodies are colour-coded according to
    whether or not their system was unstable (defined as the loss of
    one or more of the inner triple). 88/400 of the \textsc{Binaries} 
    simulations, and 99/400 of the \textsc{Giants} simulations, 
    resulted in destabilisation of the inner system.}
  \label{fig:a-m}
\end{figure*}

We conduct our $N$-body integrations with the Bulirsch--Stoer integrator 
of the \textsc{Mercury} integrator package \citep{Chambers99}. 
The tolerance parameter is set to $10^{-13}$. 
Simulations are run for 10\,Myr. We 
incorporate leading-order post-Newtonian terms into the integrator: 
these are essential particularly for the binary systems to ensure 
that single-planet survivors correctly have their Kozai cycles suppressed, 
as a single planet 
in an inclined binary system will be protected from Kozai cycles by 
the relativistic precession \citep[e.g.][]{Ford+00,FabryckyTremaine07}. 
In this paper we treat 
collisions as perfect mergers; we intend to explore the effects of 
different collision prescriptions in a future paper. Bodies that 
pass within one stellar radius of the centre of the star are assumed to 
undergo an inelastic collision with this object; in reality, some may 
undergo tidal circularisation to become planets on tight circular orbits such as 
hot Jupiters.

Our systems are constructed from a combination of 
actual \textit{Kepler} Objects of Interest (KOIs) 
for the inner system and artificial planets and 
stellar companions for the outer system. 
For most integration sets we choose triple-planet KOI systems from the Q1--Q17 Kepler DR24 
for the inner planets \citep{Coughlin+16}. KOIs are 
subjected to the same cuts as in \cite{Lissauer+11} and 
\cite{Johansen+12}, and in addition KOIs labelled 
as false positives in the NASA exoplanet archive are removed. KOIs in candidate triple-planet systems 
were then assigned masses acording to the deterministic 
mass--radius relation of \cite{WeissMarcy14}. 
The probability of 
seeing each system as a triple-planet KOI system was calculated assuming an inclination 
distribution with $\beta=5^\circ$ 
\citep[see Fig~3 of][for this distribution]{Johansen+12}. 
When building a population of systems for the 
simulations described below, KOI systems were drawn weighted by the inverse of the probability 
of seeing them as a triple-transiting system, thus generating a model 
population closer to the actual debiased population of multiple systems.

Our simulations are divided into three main classes. First we integrate a 
\textsc{Control} set of the triple KOI systems in isolation. Systems which are 
stable are then used as templates for two sets of simulations with 
outer perturbers: \textsc{Binaries}, with binary stellar companions 
and outer planetary companions, and \textsc{Giants}, with outer 
planetary companions. The \textsc{Binaries} and \textsc{Giants} classes 
are themselves subdivided into several sets.

We refer the reader to Section~\ref{sec:initial} for a detailed 
discussion of our initial conditions. Here we note that our outer systems 
are motivated by current observational constraints on the occurrence 
rates and distributions of masses and semimajor axes of wide-orbit planets 
and binary stars, but that we assume that there is no correlation between 
the formation of the inner and the outer planets.

\subsection{The \textsc{Control} class}

\label{sec:control-setup}

We first evolve the triple-planet candidate KOI systems in isolation. They
were then cloned 8 times with zero eccentricities, inclinations assigned with $\beta=5^\circ$,
and randomised orbital phases, and integrated for 1\,Myr. Systems where one or more clones
experienced close encounters between the planets were removed from further consideration 
(see Section~\ref{sec:control} below). 
We identified one such unstable system (KOI00284). 
We also removed one system that lies very close to a 4:2:1 mean motion 
commensurability (KOI01426). 
The remaining 86 systems were used as templates for the 
inner triple-planet systems for the main integration sets described below.

\subsection{The \textsc{Binaries} class}

The \textsc{Binaries} class is divided into the \textsc{Binaries} simulation set proper, 
\textsc{Binaries-flat}, and \textsc{Binaries-0pl}. 

For our simulation 
set \textsc{Binaries} we add one extra planet and one wide binary companion. The planet 
has a mass ranging from $3-3000\mathrm{\,M}_\oplus$ drawn from a uniform distribution in log space, 
a semi-major axis ranging from $1-10$\,au drawn uniformly in log space, 
zero eccentricity, and an inclination 
assigned the same way as the inner planets'. The binary has a mass drawn uniformly from 
$0.1\mathrm{\,M}_\odot$ to the primary's mass, an eccentricity drawn uniformly between 0 and 1, 
an inclination drawn from an isotropic distribution and a period drawn form a lognormal 
distribution with a peak at $10^5$ days and a standard deviation of 2.3 dex. 
Binaries with semi-major axes smaller than 50 or greater than 1000 au were then resampled. 
See \cite{DucheneKraus13} for the justification for our binary population. 
A misalignment for the orbital inclinations of these binaries 
is consistent with observations of discs in young binary systems 
\citep{JensenAkeson14,Brinch+16} and expected from simulations of star 
formation \citep{Bate12}. The planet 
properties are harder to justify as the region beyond 1\,au is subject to selection 
biases. However, the mass and semi-major axis distributions of giant planets are 
approximately flat in log space \citep[e.g.,][]{Cumming+08}. The initial semi-major axes 
and masses of all bodies in our \textsc{Binaries} simulations are shown in 
Figure~\ref{fig:a-m}.

The \textsc{Binaries-flat} set is set up similarly to the \textsc{Binaries} set 
itself, save for the inclinations of the four planets, which are all set to $0^\circ$. 
The wide binary companion retains its isotropic distribution. The \textsc{Binaries-0pl} 
set is set up similarly to the \textsc{Binaries} set, except that there is no 
extra planet added: the systems comprise the triple KOIs and a wide binary. This 
is to verify that the effects of the wide binary star acting alone on the KOI systems 
are negligible. Finally, \textsc{Binaries2} uses the same outer systems (binary star 
plus extra planet) as the \textsc{Binaries} set, but the inner planetary systems are taken from 
\emph{double-planet,} not triple-planet, observed KOI systems.

\subsection{The \textsc{Giants} class}

As with the \textsc{Binaries} class, we divide \textsc{Giants} into 
\textsc{Giants} proper, \textsc{Giants-flat} and \textsc{Giants-1pl}.

For our simulation set \textsc{Giants} we add four extra planets. Masses are drawn 
uniformly in log space from $10-3000\mathrm{M}_\oplus$, eccentricities are zero, and 
inclinations assigned with $\beta=5^\circ$. The semi-major axis of the inner planet 
is drawn randomly in log space from $1-3$\,au, while subsequent planets are placed 
$4-6$ mutual Hill radii beyond this. This places the systems on the edge of stability 
and ensures that many systems will experience instability during the 10Myr integration time. 
\cite{JuricTremaine08} and \cite{Raymond+11} argue that the eccentricity distribution 
of giant planets is best reproduced if the majority of such planets come from 
unstable multiple systems that undergo scattering to excite planetary eccentricities. The 
initial semi-major axes and masses of all bodies in the \textsc{Giants} set are shown in 
the right-hand panel of Figure~\ref{fig:a-m}.
We further discuss the initial conditions for our simulations in Section~\ref{sec:initial}.

To complement \textsc{Giants}, we also run a \textsc{Giants-flat} 
set, with similar initial conditions but with inclinations of all planets reduced to 
$\beta=0.1^\circ$ (unlike in \textsc{Binaries-flat} we avoid exact coplanarity 
of the inner planets, as these have no intrinsic way to break the symmetry of 
exact coplanarity, a role filled by the inclined binary in \textsc{Binaries-flat}). 
\textsc{Giants2} has its outer planets initialised in the same way as \textsc{Giants}, 
but the inner triple-planet KOI systems are replaced by double-planet KOI systems 
as in \textsc{Binaries2}. We also run a \textsc{Giants-1pl} set, where only the first 
of the four outer giants is placed. This is to verify that a single unexcited 
outer planet does not significantly afect the inner system.

\section{The control sample}

\label{sec:control}

As described in Section~\ref{sec:control-setup}, we first integrated 
the \emph{Kepler} triple-planet systems in isolation. Each system 
was cloned 8 times, assigned different orbital inclinations and phases, 
and integrated for 1\,Myr. We identified one unstable system: 
KOI00284 (alias Kepler-132), which has planets at 6.18 and 6.41 days' period.
This system was also identified as possessing a binary companion by
\cite{Lissauer+14}, who inferred that not all of the planets
orbit the same star.

The majority of these isolated systems experienced very little eccentricity 
excitation, with the median of the maximum eccentricity attained over 
1\,Myr being $3.6\times10^{-4}$, and only $1.6\%$ of planets ever attaining 
$e>0.01$. The most excited system was KOI01426 (alias 
Kepler-297), in which one planet always attained a maximum eccentricity of $\approx0.07$. 
This system lies close to a 4:2:1 period commensurability, and we removed 
this system in case the resonant dynamics would be important for its stabilisation.

The stability of the vast majority of these \emph{Kepler} triple-planet 
systems agrees with previous analyses 
\citep{Johansen+12,Fabrycky+14}. 
The two systems identified above (KOI00284 and KOI01426) were discarded when 
setting up the inner systems for the main integration sets.

\section{Population synthesis I: Binaries}

\label{sec:binaries}

\begin{figure*}
  \includegraphics[width=0.48\textwidth]{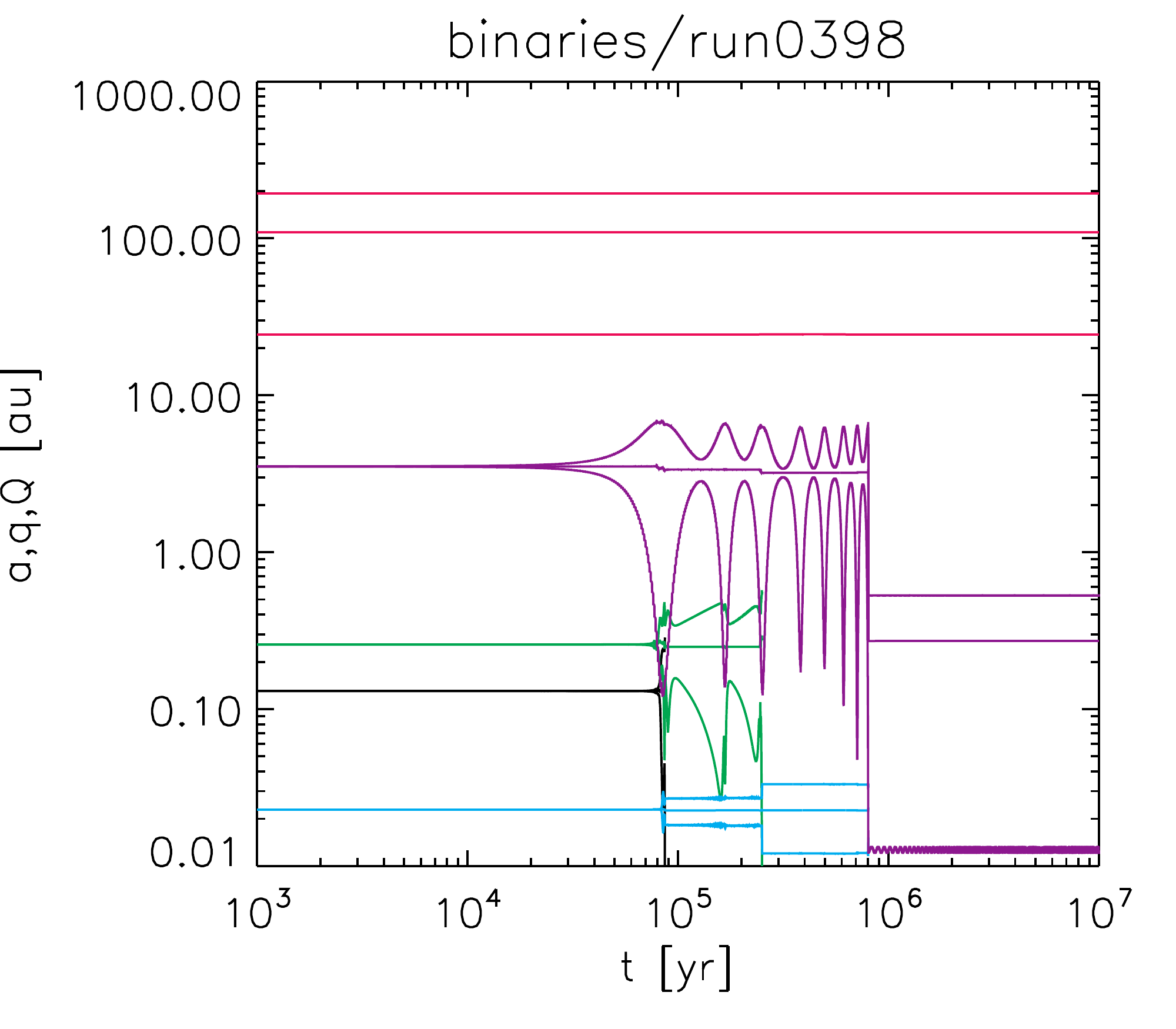}
  \includegraphics[width=0.48\textwidth]{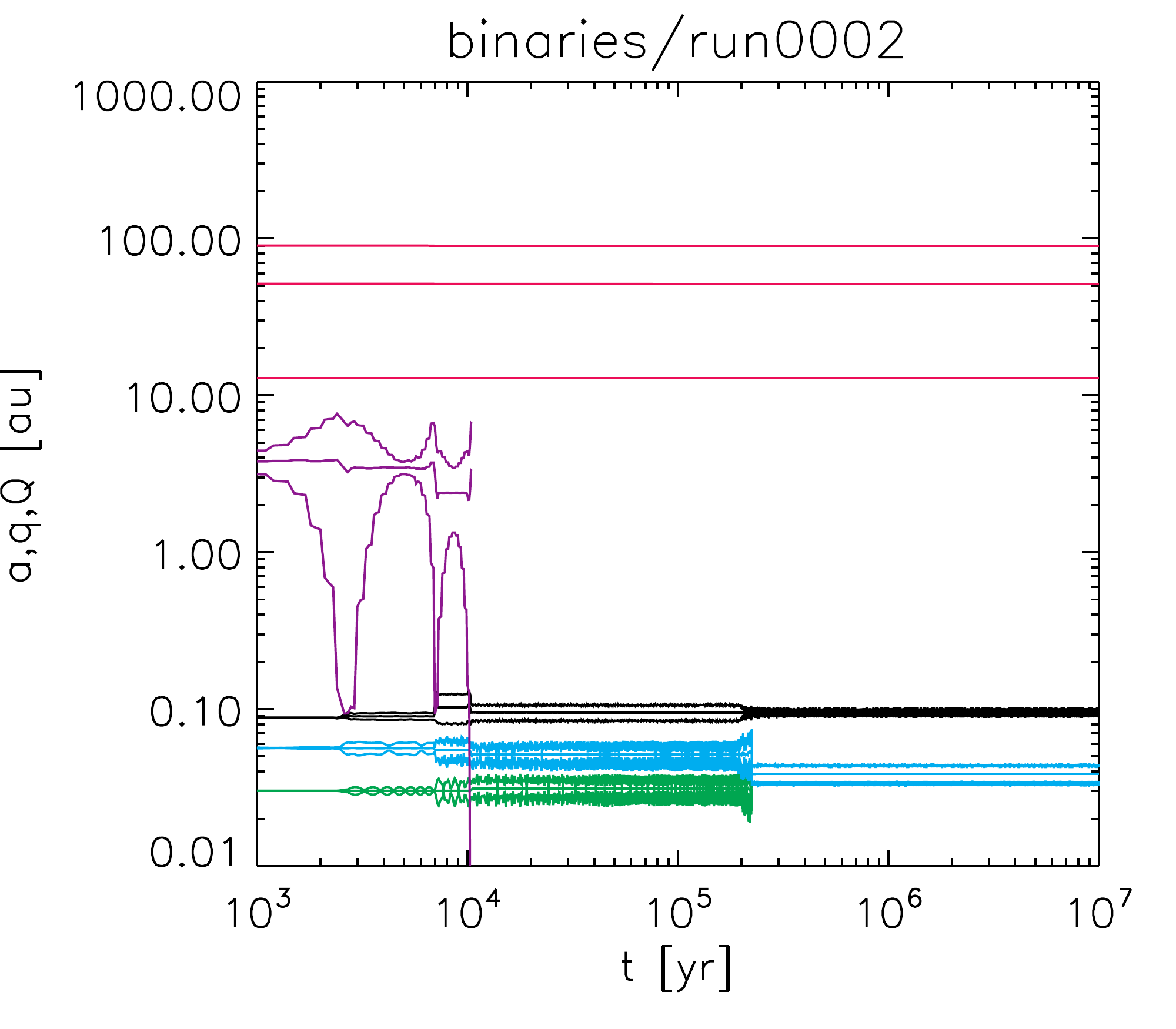}
  \caption{Example evolution of \textsc{Binaries} integrations, 
    showing semi-major axes, pericentres 
    and apocentres of all bodies in a system. The binary companion is in magenta, 
    the outer planet in purple, and the \emph{Kepler} triple-planet system in 
    blue, green and black. \textbf{Left: }A 
    giant planet is forced by Kozai cycles into the inner system, 
    clearing out the inner planets. This planet ends with a pericentre 
    $\sim0.01$\,au, and would in time circularise to form a hot Jupiter. 
    \textbf{Right: }Here the outer planet is swiftly forced into 
    the star, but triggers a delayed instability causing a reduction 
    in inner system multiplicity 200\,kyr later.}
  \label{fig:example}
\end{figure*}

In our \textsc{Binaries} simulation set we integrate 400 systems. 
Example evolution is shown in the Figure~\ref{fig:example}. 
In the left panel, a $0.6$ Jupiter-mass planet is sent into the inner 
system where it forces two super-Earths into the star, before colliding with 
the third. This inelastic collision drains specific orbital energy from the giant, 
leaving it as a highly-eccentric warm Jupiter with a pericentre of 
$\sim0.01$\,au whose Kozai oscillations 
have been shut off by relativistic precession. This planet 
would, in time, tidally circularise to form a 
hot Jupiter with no companions in the inner system. In the right panel, 
the outer giant is forced into the star by Kozai perturbations, but 
during each high-eccentricity phase it excites the eccentricities of the inner 
planets. This triggers a delayed instability about $0.2$\,Myr later, 
causing a merger of the two inner planets. The 
mutual inclination of the survivors is moderately excited, 
around $11^\circ$.

\subsection{Effects on outer system}

\begin{figure*}
  \includegraphics[width=.48\textwidth]{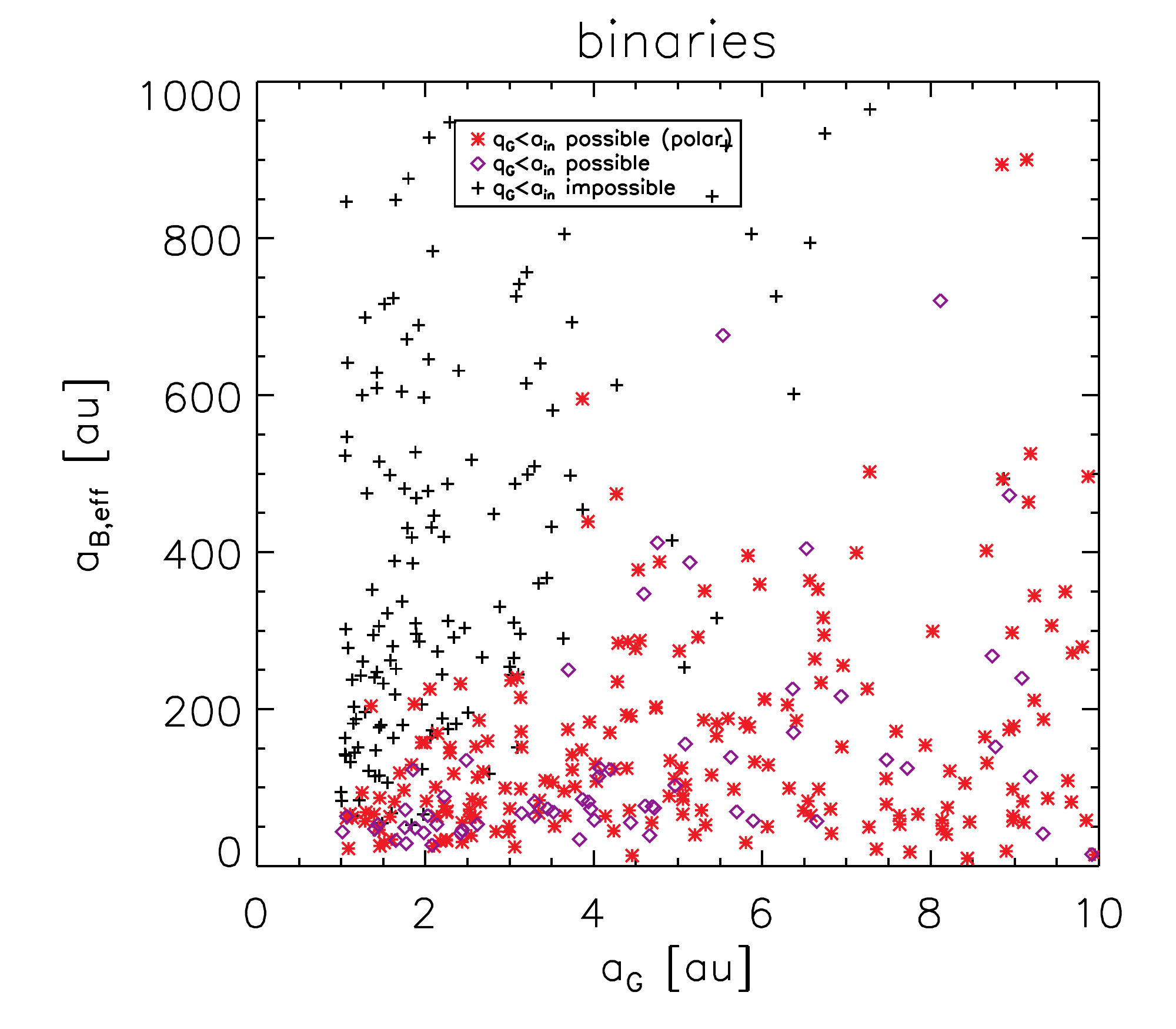}
  \includegraphics[width=.48\textwidth]{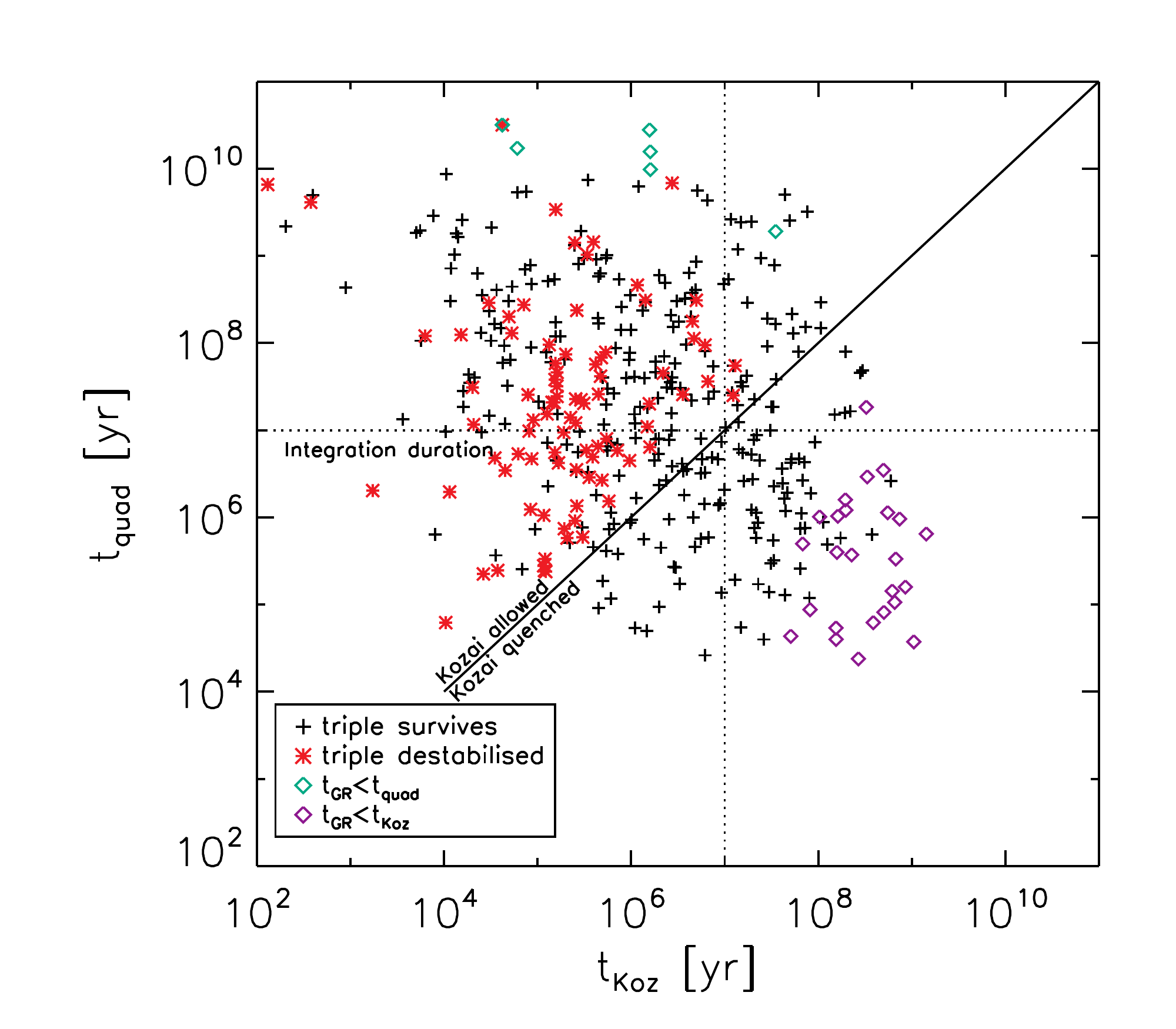}
  \includegraphics[width=.48\textwidth]{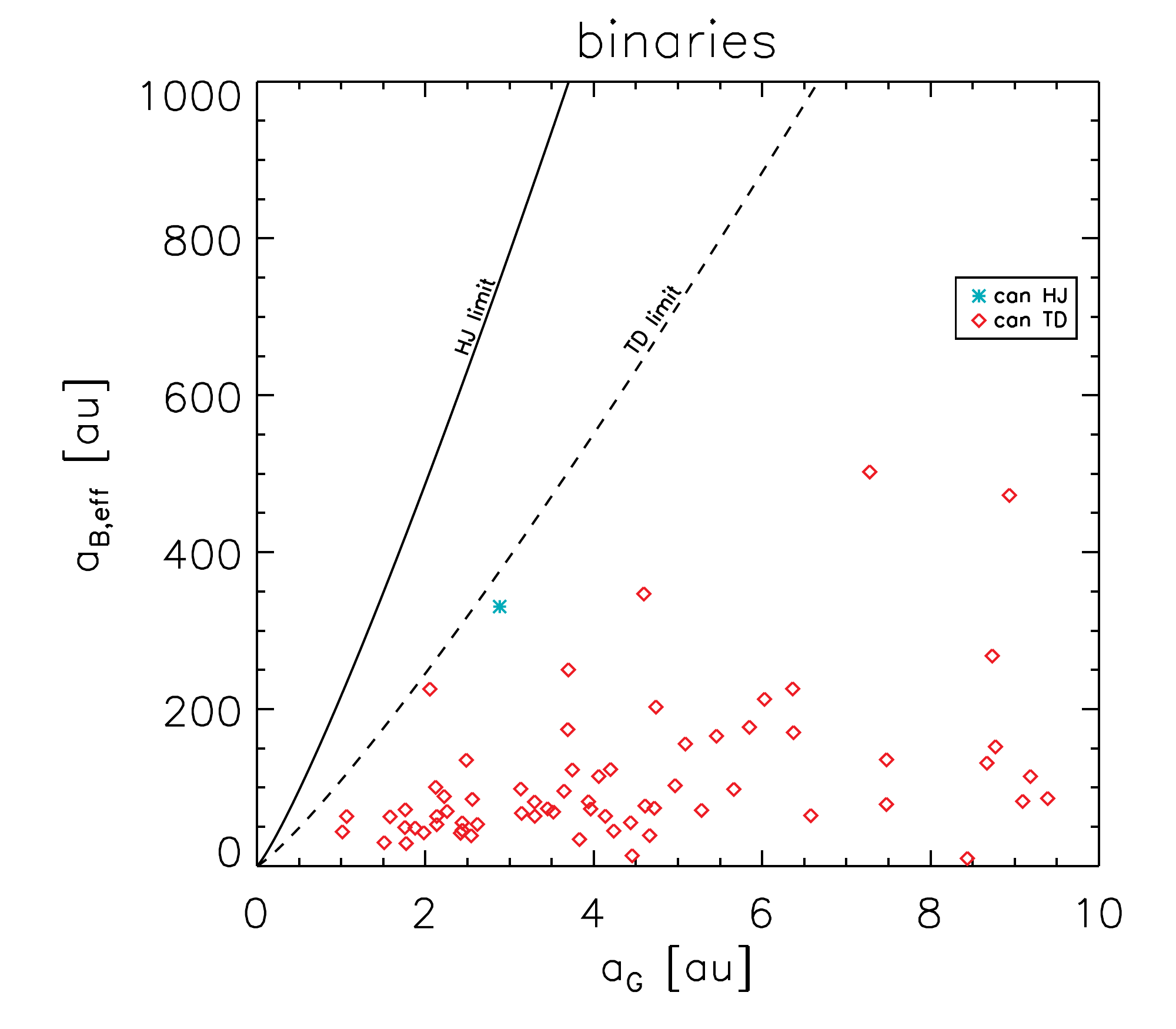}
  \includegraphics[width=.48\textwidth]{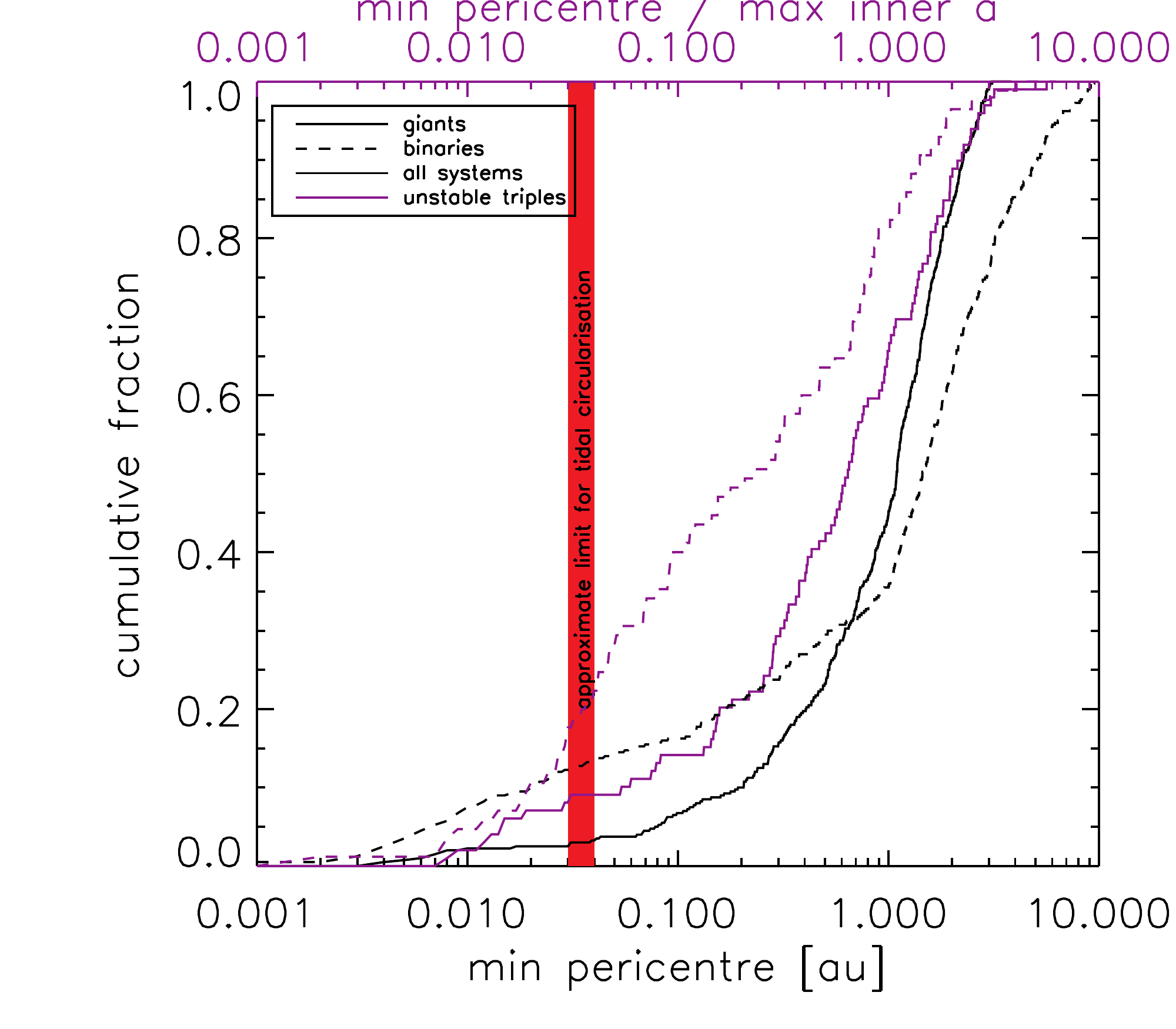}
  \caption{Prospects for the formation of Hot Jupiters in our integrations.
    \textbf{Top left: }Competition between Kozai forcing from the binary and secular perturbations 
    from the inner three planets for all outer planets in the \textsc{Binaries} simulations. 
    The ``effective''
    semimajor axis of the binary ($a_\mathrm{B,eff}=a_\mathrm{B}\sqrt{1-e_\mathrm{B}^2}$)
    is plotted against the outer giant planet's semimajor axis $a_\mathrm{G}$. The inner system 
    imposes an upper limit to the outer planet's eccentricity (see Eq~\ref{eq:j}); systems where this 
    eccentricity is sufficient to allow the outer planet's pericentre $q_\mathrm{G}$ to become 
    smaller than the outermost inner planet's semimajor axis $a_\mathrm{in}$ are shown in purple 
    (actual initial inclinations in the simulations) and red 
    (an initial inclination of $90^\circ$, the most effective case). This gives 
    an estimate of 62/400 systems possessing outer planets whose pericentres can be 
    driven to overlap the inner planets' orbits.
    \textbf{Top right: }Timescales for precession of the outer planet in the \textsc{Binaries} systems
    induced by
    the outer companion ($t_\mathrm{Koz}$) and by the inner planets
    ($t_\mathrm{quad}$; quadrupole approximation). Kozai cycles are
    quenched if $t_\mathrm{quad}\lesssim t_\mathrm{Koz}$. Red and black 
    points show whether the inner planets in a system were destabilised: this
    can only happen if $t_\mathrm{quad}> t_\mathrm{Koz}$ so that the Kozai
    mechanism can excite the giant planet's eccentricity. We also show as
    diamonds systems where the precession timescale of the outer planet due to general relativity
    is less than the Kozai timescale (purple points towards the right) or the quadrupole
    precession from the inner system (green points towards the top). For the former systems,
    GR precession would quench the Kozai effect even in the absence of the inner system, while
    for the latter systems, the GR precession dominates over that from the planetary quadrupole.
    \textbf{Bottom left: }Allowed parameter space for hot Jupiter formation for outer
    planets in the \textsc{Binaries} systems
    with $m>M_\mathrm{Saturn}$ based on our initial conditions. 
    We show only systems where the inner planets were destabilised.
    Turquoise stars represent
    planets that can be tidally circularised to form hot Jupiters; red diamonds represent
    planets that can be tidally disrupted at the Roche limit. The solid and dashed lines
    represent the limits for fiducial $m_\mathrm{pl}=M_\mathrm{Jupiter}$ and $m_\mathrm{B}=M_\odot$.
    \textbf{Bottom right: }Minimum pericentres attained over the course of the integration by the outer
    planet in the \textsc{Binaries} and \textsc{Giants} simulations. Around 10--15\% in
    \textsc{Binaries} attain either a sufficiently
    small pericentre to begin tidal circularistion of the orbit, and/or collide with the star,
    a figure comparable to studies of hot Jupiter migration by Kozai cycles and tidal friction.
    The \textsc{Giants} simulations are much less efficient at generating small pericentres.
    In purple the minimum pericentre is shown as a fraction of the 
    semimajor axis of the outermost inner planet (upper $x$-axis), 
    for systems where the inner triple was destabilised. In most of these systems the orbit 
    of an outer planet overlaps with those of the inner planets, but there is a minority of 
    systems where destabilisation occurs at a distance. }
  \label{fig:hj-formation-binaries}
\end{figure*}

Eccentricity forcing by Kozai cycles followed by dissipation by tidal friction 
has been proposed as a migration mechanism for hot Jupiters 
\citep{WuMurray03,FabryckyTremaine07}. As our integrations neglect tidal 
forces, it is worth verifying that our set-up provides a credible means of 
delivering hot Jupiters. We explore this in Figure~\ref{fig:hj-formation-binaries}. 
Kozai cycles can be suppressed by the introduction of additional precessional 
forces as arise from the mass of the inner planets 
\citep{Innanen+97,Malmberg+07,Kaib+11,BoueFabrycky14}, 
relativistic corrections and distortion of the planet or 
star. These forces are relatively stronger closer to the star and hence a 
system where the binary is on too wide an orbit will not be able to induce 
Kozai cycles. 
We first consider the effects of the mass quadrupole of the 
inner planets, which for most of these systems is the most important barrier 
to Kozai cycles. Write the total averaged potential for the outer planet as 
\begin{equation}
\Phi_\mathrm{tot}=\Phi_\mathrm{Koz}+\Phi_\mathrm{quad}+\Phi_\mathrm{GR},
\end{equation}
where $\Phi_\mathrm{Koz}$ is the Kozai potential from the binary companion
\begin{equation}
\Phi_\mathrm{Koz}=\frac{\Phi_0}{8}
\left(1-6e_\mathrm{G}^2-3j_\mathrm{G}^2\cos^2i_\mathrm{B}+15e_\mathrm{G}^2\sin^2\omega_\mathrm{G}\sin^2i_\mathrm{G}\right),
\end{equation}
$\Phi_\mathrm{quad}$ is the quadrupole of the inner planets
\begin{equation}
\Phi_\mathrm{quad}=-\frac{\Phi_0\epsilon_\mathrm{quad}}{4j_\mathrm{G}^3},
\end{equation}
and $\Phi_\mathrm{GR}$ is the post-Newtonian precession term. 
Here subscript $_\mathrm{B}$ refers to the binary and $_\mathrm{G}$ 
to the ``giant'' outer planet, $j=\sqrt{1-e^2}$, and 
\begin{equation}
\Phi_0=\frac{\mathcal{G}M_\mathrm{B}a_\mathrm{G}^2}{M_\star a_\mathrm{B}^3 \left(1-e_\mathrm{B}^2\right)^{3/2}}.
\end{equation}
The strength of the potential from the inner 
planets is parametrized by 
\begin{equation}
\epsilon_\mathrm{quad}=\frac{M_\star}{M_\mathrm{B}}
\left(\frac{a_\mathrm{B}}{a_\mathrm{G}}\right)^3
\left(1-e_\mathrm{B}\right)^{3/2}
\sum_{\mathrm{inner}}\frac{M_\mathrm{inner}}{M_\star}\left(\frac{a_\mathrm{inner}}{a_\mathrm{G}}\right)^2,
\end{equation}
which is $\propto a_\mathrm{G}^{-5}$, a very strong function of the outer planet's semimajor axis. 
Neglecting temporarily $\Phi_\mathrm{GR}$, conservation of energy and the $z$-component of angular 
momentum leads to the equivalent of Eq~50 of \cite{Liu+15}:
\begin{equation}
\frac{\epsilon_\mathrm{quad}}{4}\left(\frac{1}{j_\mathrm{min}^3}-1\right)
=\frac{9e_\mathrm{max}^2}{8j_\mathrm{min}^2}\left(j_\mathrm{min}^2-\frac{5}{3}\cos^2i_0\right)\label{eq:j}
\end{equation}
for the maximum eccentricity (and minimum $j$) attained by an initially 
circular orbit at an inclination of $i_0$. In the limit as 
$\epsilon_\mathrm{quad}\to0$, for an initially polar orbit, 
$j_\mathrm{min}\sim\sqrt[3]{2\epsilon_\mathrm{quad}/9}$.

The precession induced by the inner planetary system can be removed if this 
system is itself disrupted by the outer planet, as in \cite{Mustill+15}. We may 
estimate that this should happen if the pericentre distance of the outer planet, 
$q_\mathrm{G}$, should ever be less than the outermost planet's semimajor axis. 
We use Equation~\ref{eq:j} to calculate the minimum achievable $j_\mathrm{G}$ for each 
system in our simulations, and show whether the outer planet could have attained 
a pericentre lying within the inner system in the top right panel 
of Figure~\ref{fig:hj-formation-binaries}. Here, planets are shown in the space of 
initial planetary semimajor axis $a_\mathrm{G}$
and binary ``effective'' semimajor axis
$a_\mathrm{B,eff}=a_\mathrm{B}\sqrt{1-e_\mathrm{B}^2}$. Planets represented by 
black crosses cannot attain a pericentre lying within the inner system, while 
those presented by purple diamonds can. Those represented by red stars could if 
the initial binary inclination were $90^\circ$. Understanding the transition from 
black-dominated at small $a_\mathrm{G}$ and large $a_\mathrm{B,eff}$ to red-dominated 
at large $a_\mathrm{G}$ and small $a_\mathrm{B,eff}$ is straightforward: 
if the binary is wide or the outer planet close to the inner system, the Kozai 
forcing is weak and cannot excite large eccentricities. The distribution of the 
purple diamonds (including the constraint of the initial binary inclination) 
is more complicated. At low $a_\mathrm{B,eff}$, fewer planets on wider orbits can 
penetrate the inner system (the inclination becomes the main constraint), but 
at high $a_\mathrm{B,eff}$ \emph{only} planets on wider orbits can 
penetrate the inner system, since regardless of inclination the Kozai 
forcing cannot overcome the inner planets' quadrupole if the outer 
planet is on too tight an orbit. The outcomes of the integrations are 
shown in the top right panel of Figure~\ref{fig:hj-formation-binaries}, as 
a function of the timescales for Kozai cycles
\begin{equation}
  t_\mathrm{Koz}=\left(\frac{a_\mathrm{B}}{a_\mathrm{G}}\right)^3
  \frac{1}{\left(1-e^2\right)^{3/2}}
  \frac{M_\star}{M_\mathrm{B}}
  \left(\frac{a_\mathrm{G}}{\mathrm{au}}\right)^{3/2}
  \left(\frac{M_\star}{\mathrm{M}_\odot}\right)^{-1/2}
\end{equation}
and for precession induced by the inner planets
\begin{equation}
  t_\mathrm{quad}=\frac{4}{3}\left(\frac{M_\star}{\mathrm{M}_\odot}\right)^{-1/2}
  \left(\frac{a_\mathrm{G}}{\mathrm{au}}\right)^{3/2}
  \left(\sum_{\mathrm{inner}}\frac{m_\mathrm{inner}a_\mathrm{inner}^2}{M_\star a_\mathrm{G}^2}\right)^{-1}
\end{equation}
where for the latter we consider the quadrupole contributions of each
inner planet. Points are coloured according to whether or not the inner
planets were destabilised: destabilisation occurs only when
$t_\mathrm{Koz}\lesssim t_\mathrm{quad}$, else the Kozai cycles are quenched by
the relatively stronger planet--planet interactions. This plot also justifies our
10\,Myr integration duration, as few systems lie in the second octant:
systems whose Kozai timescales exceed the integration duration
(and hence would not yet have been driven to a small pericentre in the integrations)
would typically have their Kozai cycles quenched by the inner planets anyway.

Outer planets which succeed in destroying the inner planets may 
then proceed to yet smaller pericentres and possible tidal circularisation 
or disruption. We show in the bottom left panel of 
Figure~\ref{fig:hj-formation-binaries} the prospects for this in systems containing an 
outer planet more massive than Saturn and which lost one or more of the inner planets. 
The lines show, for each $a_\mathrm{G}$, the maximum $a_\mathrm{B,eff}$ that permits 
tidal circularistaion or disruption, based on
the competition between Kozai forcing and remaining short-range forces
\citep[equations 47 and 49 of][taking an optimistic $a_\mathrm{p,crit}=0.04$\,au
for tidal circularisation, and assuming a polar orbit]{Anderson+16}. These lines 
lie comfortably above all of the systems shown in this plot: systems which would 
be prevented by GR precession from forming a hot Jupiter in the absence of the 
inner system cannot destroy this inner system in the first place, as the planet-induced 
precession is too strong.

Actually predicting whether a given planet will tidally circularise or 
disrupt is not so simple, but studies of Kozai cycles plus tidal friction find 
a fraction of $\sim10-15\%$ of Jupiters forced by Kozai cycles being either 
circularised or disrupted, the balance between these two outcomes 
depending on planet mass, radius and the 
poorly-constrained tidal dissipation parameters \citep{Petrovich15b,Anderson+16,Munoz+16}. 
We find a similar fraction of our outer planets being forced onto sufficiently small 
pericentres that would allow one or the other of these outcomes 
(Fig~\ref{fig:hj-formation-binaries}, bottom right panel). We also see that $\sim30\%$ 
of the outer planets attain a minimum pericentre $<1$\,au, allowing them to 
interact with the inner systems through strong, direct scattering. 
The systems shown in Figure~\ref{fig:example} contain planets that would tidally 
circularise or disrupt, in each case reducing the number of inner 
planets in the system.

\begin{figure}
  \includegraphics[width=.5\textwidth]{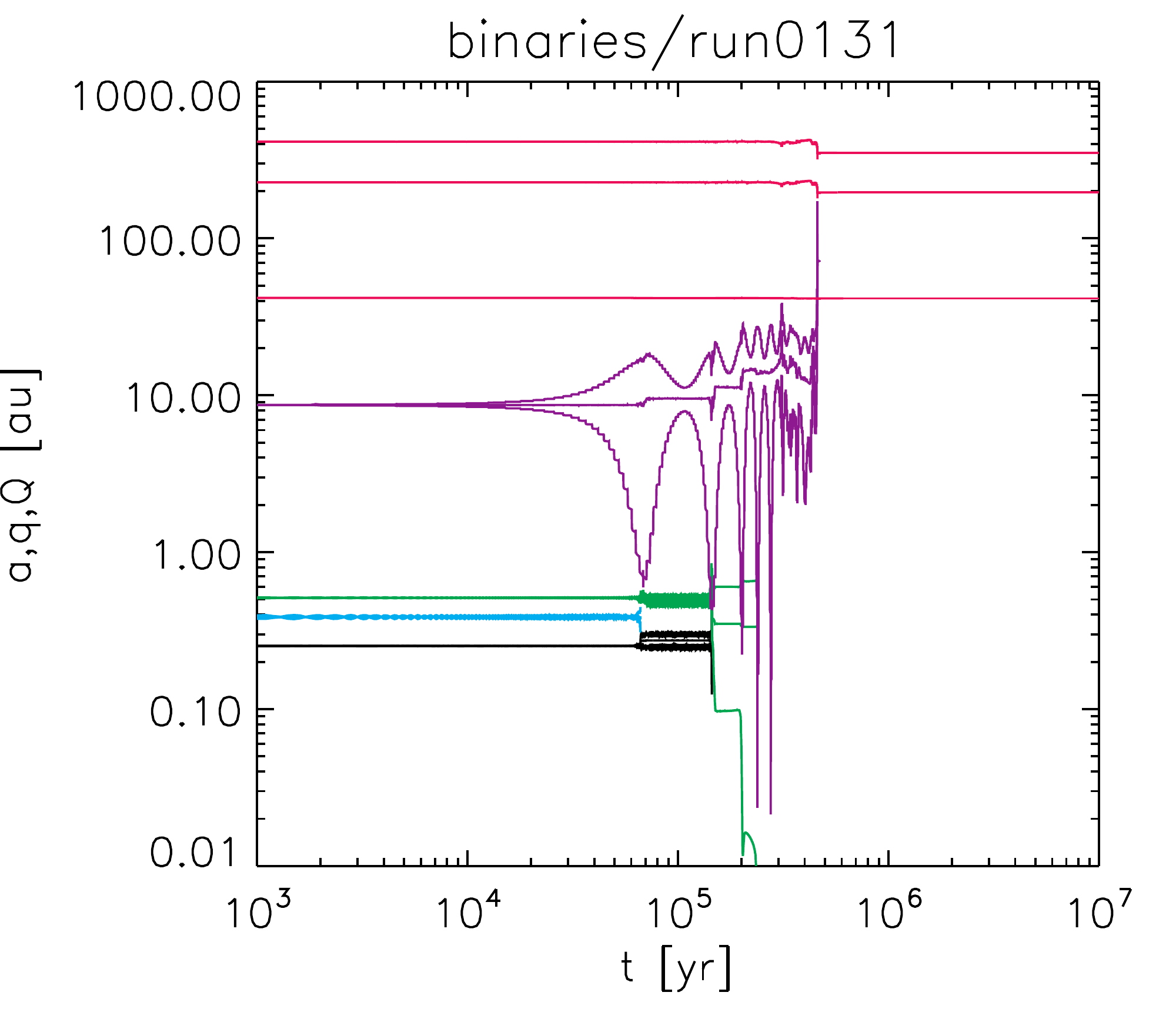}
  \caption{Ejection of the outer planet by the combined effects of the 
  inner system and the stellar binary. After the outer planet's 
  pericentre reaches the inner system, scattering off the inner 
  planets begins to raise its semi-major axis, ultimately leading 
  to strong interaction with the binary and subsequent ejection.}
  \label{fig:eject}
\end{figure}

Overall in our \textsc{Binaries} simulations, we lost 80 out of 400 outer planets: 
43 were ejected, 35 hit the star, and 2 hit a more massive inner planet. In addition, 
2 hit a less massive inner planet and survived. Considering only planets more massive 
than Saturn (194/400), we lost 32: 18 ejected and 14 hit the star, while 2 were hit by 
less massive inner planets and survived. Ejection of the outer planet can follow a 
similar route to that described in \cite{Mustill+15}: despite having a larger mass 
than the inner planets, a highly-eccentric planet with a large semi-major axis can have 
less orbital binding energy than the lower-mass inners, and comparatively small changes 
to the semi-major axes of the latter can cause a large change to semi-major axis of the 
former. An example is shown in Figure~\ref{fig:eject}, where the ejection of the outer 
planet is finally secured by the binary star after the inner planets have raised its 
semi-major axis.

\subsection{Effects on inner system}

\subsubsection{Intrinsic multiplicities}

\begin{figure*}
  \includegraphics[width=\textwidth]{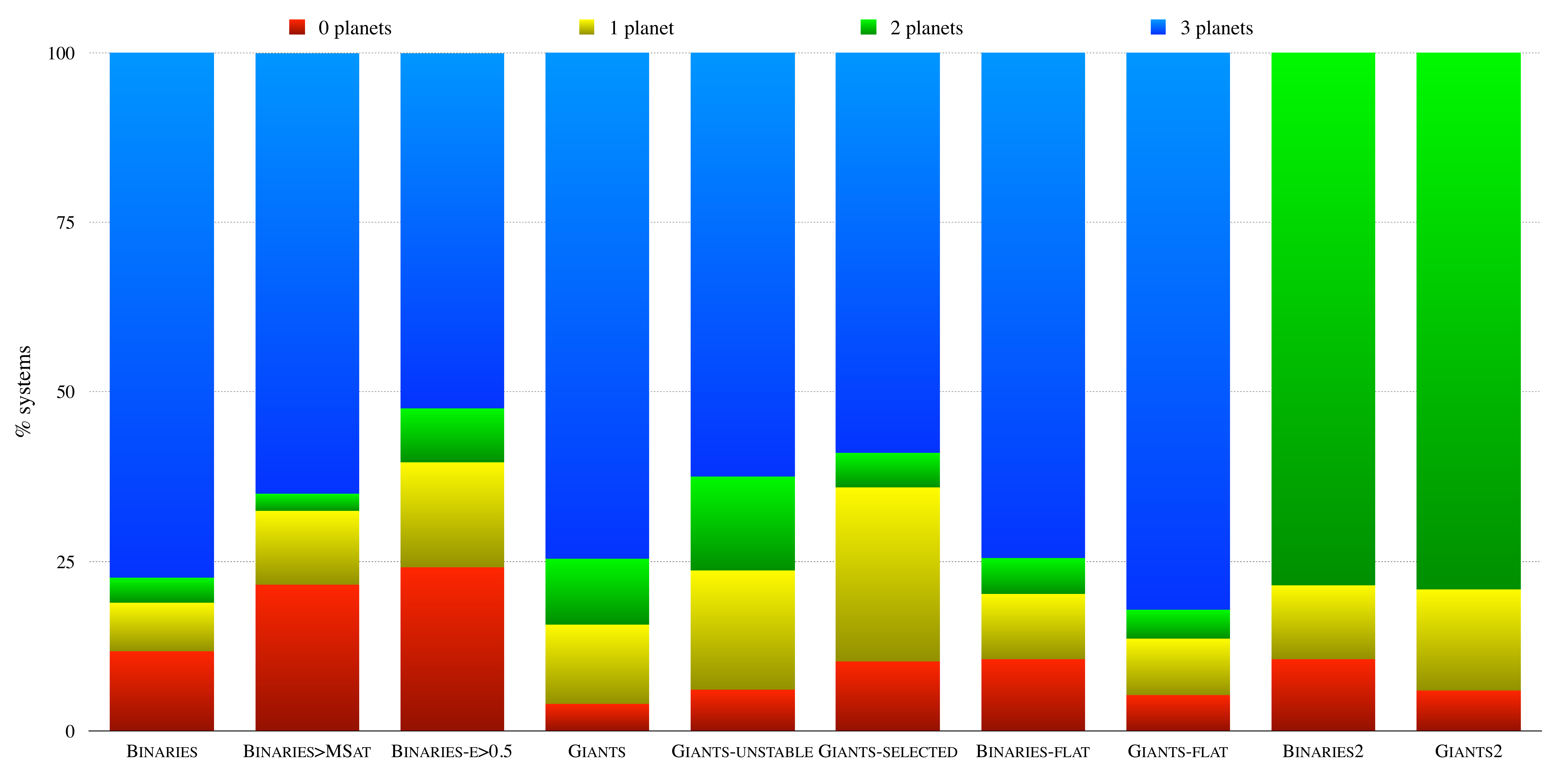}
  \caption{Multiplicities of inner systems arising from our simulation sets. See text 
  and caption to Table~\ref{tab:nsurv} for a description of the simulation sets.}
  \label{fig:bars}
\end{figure*}

\begin{table*}
  \begin{tabular}{lccccc}
    Integration set          & $N_\mathrm{sys}$ & $N_{0\mathrm{p}}$ & $N_{1\mathrm{p}}$ & $N_{2\mathrm{p}}$ & $N_{3\mathrm{p}}$ \\
    \hline
    \textsc{Binaries}        & 400 & 46 ($11.7\pm1.6\%$) & 28 ($7.2\pm1.3\%$) & 14 ($3.7\pm0.9\%$) & 312 ($77.9\pm2.1\%$)\\
    \textsc{\ \ Binaries-E$>0.5$} & 185 & 44 ($24.1\pm3.1\%$) & 28 ($15.5\pm2.6\%$) & 14 ($8.0\pm2.0\%$) & 99 ($53.5\pm3.6\%$)\\
    \textsc{\ \ Binaries$>$Msat} & 194 & 42 ($21.9\pm2.9\%$) & 21 ($11.2\pm2.2\%$) & 5 ($3.1\pm1.2\%$) & 126 ($64.8\pm3.4\%$)\\
    \textsc{Giants}          & 400 & 15 ($4.0\pm1.0\%$) & 46 ($11.7\pm1.6\%$) & 38 ($9.7\pm1.5\%$) & 301 ($75.1\pm2.2\%$)\\
    \textsc{\ \ Giants-unstable} & 259 & 15 ($6.1\pm1.5\%$) & 45 ($17.6\pm2.4\%$) & 35 ($13.8\pm2.1\%$) & 164 ($63.2\pm3.0\%$)\\
    \textsc{\ \ Giants-selected} &  39 &  4 ($12.1\pm5.0\%$) & 10 ($26.7\pm6.8\%$)&  2 ($7.3\pm4.0\%$) & 23 ($58.5\pm7.6\%$)\\
    \textsc{Binaries-flat}   & 300 & 31 ($10.6\pm1.8\%$) & 28 ($9.6\pm1.7\%$) & 15 ($5.3\pm1.3\%$) & 226 ($75.2\pm2.5\%$)\\ 
    \textsc{Giants-flat}     & 300 & 15 ($5.3\pm1.3\%$) & 24 ($8.3\pm1.6\%$) & 12 ($4.3\pm1.2\%$) & 249 ($82.8\pm2.2\%$)\\
    \textsc{Binaries2}       & 300 & 31 ($10.6\pm1.8\%$) & 32 ($10.9\pm1.8\%$) & 237 ($78.8\pm2.3\%$) & -   \\
    \textsc{Giants2}         & 300 & 17 ($6.0\pm1.4\%$) & 44 ($14.9\pm2.0\%$) & 239 ($79.5\pm2.3\%$)& -   \\
    \textsc{Binaries-0pl}    & 200 & 0  ($<1.5\%$) & 4  ($2.5\pm1.1\%)$    & 1  ($1.0\pm0.7\%$)    & 195 ($97.0\pm1.2\%$)   \\
    \textsc{Giants-1pl}      & 200 & 0  ($<1.5\%$) & 0 ($<1.5\%$)  & 0 ($<1.5\%$)  & 200 ($>98.5\%$)       
  \end{tabular}
  \caption{Number of simulations per set ($N_\mathrm{sys}$), and numbers with a given number 
    of inner planets after 10 Myr integrations ($N_{n\mathrm{p}}$). \textsc{Binaries} have 
    3 inner planets, one outer from 1 to 10\,au, and a binary companion; 
    \textsc{Binaries$>$tkoz} is the subset of these where the integration 
    time exceeded the Kozai time-scale; \textsc{Binaries$>$Msat} 
    the subset where the outer planet's mass is greater than Saturn's. 
    \textsc{Giants} have three inner planets and four outer planets; 
    \textsc{Giants-unstable} is the subset of these that lost at least one giant planet; 
    \textsc{Giants-selected} is a subset of \textsc{Giants} chosen to have 
    an eccentricity distribution consistent with the observed population.
    \textsc{flat} systems have initially zero mutual inclination between 
    the inner planets (binary companions remain isotropically distributed).
    \textsc{Binaries2} and \textsc{Giants2} have 
    initially only two inner planets. \textsc{Binaries-0pl} has a \textit{Kepler} 
    triple-planet system, a binary companion, but no extra planet, while \textsc{Giants-1pl} 
    has a \textit{Kepler} triple-planet system and a single outer giant planet. Percentages in brackets 
    give the mean and standard deviation for the occurrance rate of each outcome 
    from inverting the binomial distribution. $3\sigma$ upper/lower limits are given 
    where appropriate.}
  \label{tab:nsurv}
\end{table*}

The majority of our inner planetary systems remain stable in the \textsc{Binaries} 
simulations: 312/400 retain all three inner planets, while 15 are reduced to double-planet 
systems, 28 to singles, and 45 are cleared of all planets with $P<240$\,d. These statistics 
are tabulated in Table~\ref{tab:nsurv} and displayed graphically in Figure~\ref{fig:bars}; 
Figure~\ref{fig:a-m} shows the outcomes as a function of the masses and semimajor axes of 
bodies in the system. Tighter binaries are more destabilising, as are more massive outer 
planets. The more distant outer planets are also more destabilising for the inner system; 
this is attributable to their Kozai cycles not being quenched by the inner planets, as 
explained above. Of the inner planets lost, $52\%$ collide with the star, 
$5\%$ are ejected, while $43\%$ are lost to inelastic planet--planet collisions. 
If we restrict attention to only those systems where the outer planet 
attains a maximum eccentricity of at least 0.5 (\textsc{Binaries-e$>$0.5} in 
Figure~\ref{fig:bars} and Table~\ref{tab:nsurv}) 
then the fraction of destabilised inner systems rises to almost 50\%. If we consider systems 
where the outer planet is more massive than $3\times10^{-4}\mathrm{\,M}_\odot$, we find 
around 1 in 3 inner systems destabilised (\textsc{Binaries$>$Msat} in Table~\ref{tab:nsurv}, 
with 126 of 194 triple-planet systems remaining triples, 5 being reduced to doubles, 21 to singles and 42 
with no surviving inner planet). 

\begin{figure}
  \includegraphics[width=.5\textwidth]{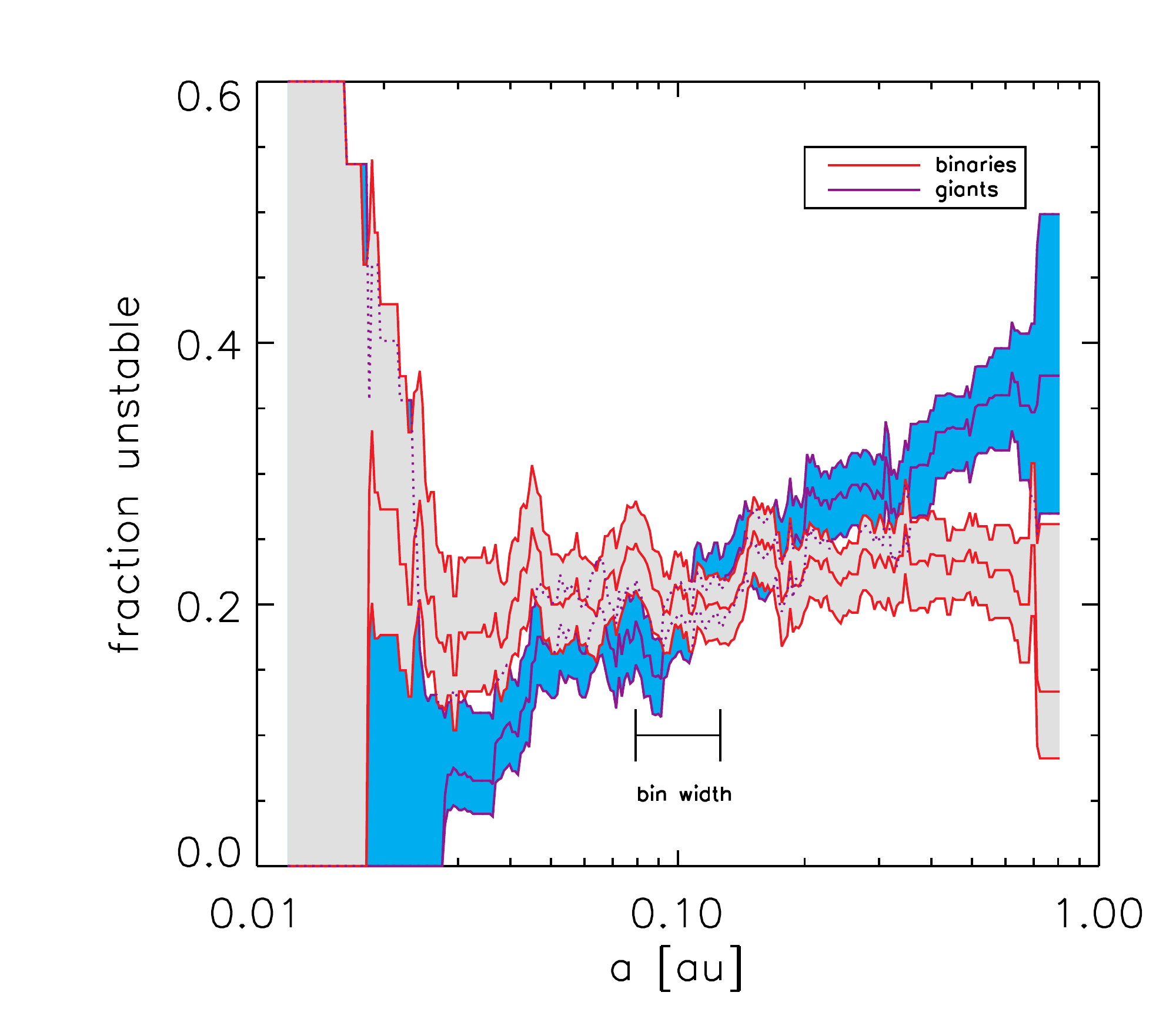}
  \caption{Rates of destabilisation as a function of 
    the inner planets' semi-major axes. Stability is counted as a 
    property of the whole inner system, so for example a system with 
    planets at $0.1$, $0.3$ and $0.5$\,au, which loses the outermost 
    planet, contributes to the destabilisation fraction at all three radii. 
    We show the fraction for running bins of the indicated width,
    together with $1\sigma$ confidence intervals. The \textsc{Binaries}
    simulations are equally destructive to all inner systems, whereas
    the \textsc{Giants} simulations are less destructive the smaller
    the inner system's semimajor axis.}
  \label{fig:funstable}
\end{figure}

The destabilisation fraction of $20-25\%$ is approximately constant across the range of 
semimajor axes of the inner planets: Figure~\ref{fig:funstable} shows the fraction of 
inner planets with given initial semimajor axes that reside in systems that lose at least 
one inner planet, which is relatively flat, in contrast to the \textsc{Giants} simulations, 
which we discuss in the next section. This flatness can be understood in terms of the 
minimum pericentre of the outer planet (Figure~\ref{fig:hj-formation-binaries}): if 
Kozai cycles are excited in the outer planet, it is easy for the pericentre to 
attain a very low value (roughly as many outer planets attain $q_\mathrm{min}<0.1$\,au as 
$q_\mathrm{min}\in[0.1,1]$\,au). Destabilisation of the inner system can sometimes 
occur when the outer planet's pericentre does not come inside the initial semimajor 
axis of the inner planet, this occurring in 19\% of unstable inner systems 
(Figure~\ref{fig:hj-formation-binaries}, bottom right panel).

Our simulations permit us to verify the main result of
\cite{Mustill+15}, where we showed that a highly-eccentric
proto-hot Jupiter would quickly destroy any other
planets in the inner system, before tidal circularisation could 
change its orbit. In 11 simulations in \textsc{Binaries}, an outer 
planet attained a pericentre less than $0.05$\,au at some point during 
the integration, while not being destroyed by ejection or collision. 
In all of these systems, the three inner planets were lost, mostly by 
collision with the star, and in 9 cases this occurred within 1\,Myr, long before 
tidal circularisation would cause the outer planet to become a 
hot Jupiter.

\begin{figure}
  \includegraphics[width=.5\textwidth]{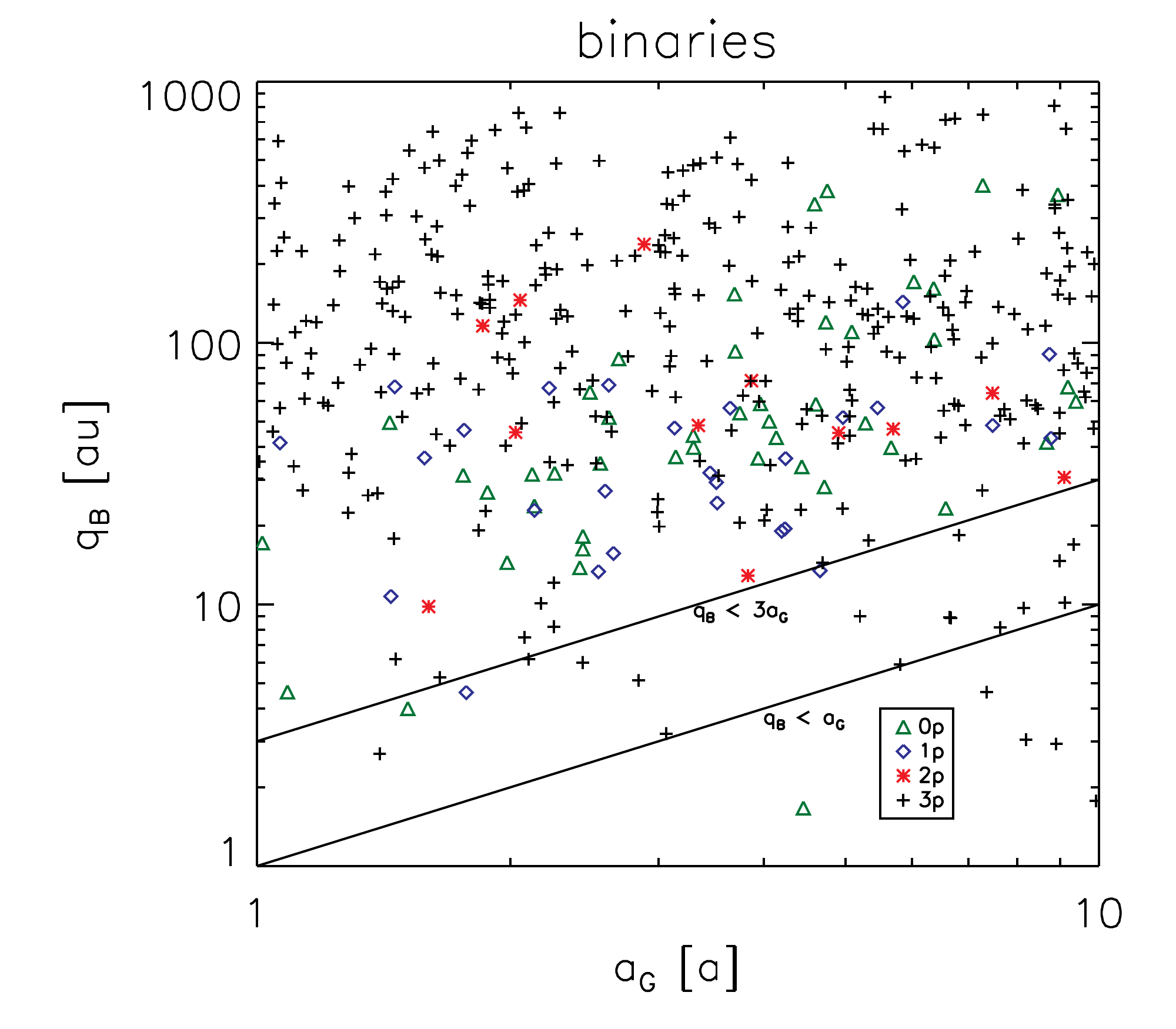}
  \caption{For outer planets with $M>10^{-4}\mathrm{\,M}_\odot$, 
    we show the initial semi-major axis of outer planet $a_\mathrm{G}$, 
    pericentre of binary $q_\mathrm{B}$, and resulting number of 
    inner planets. We also show the line where the binary's 
    pericentre reaches one or three times the outer planet's 
    orbit; this latter is an approximate stability limit for hyperbolic
    3-body encounters \citep[e.g.,][]{Pfalzner+05}.}
  \label{fig:ag-qb}
\end{figure}

We note that a small number of our systems are set up with 
the binary on an initial orbit taking it very close to 
the planetary region, due to the high eccentricities that can 
be randomly assigned. The binary pericentres, and the initial 
semi-major axes of the giant outer planets, are shown in Fig~\ref{fig:ag-qb}. 
Some binary companions overlap with or come within a factor 
of a few of the orbit of the outer planet, but this does not 
lead to a higher rate of destabilisation of the inner systems; 
giant planets in these systems are however removed considerably 
more quickly than in those with higher binary pericentres: the 
median time to ejection of the outer planet in systems with 
$q_\mathrm{B}<3a_\mathrm{G}$ was only $24$\,kyr compared to 
$1.06$\,Myr in systems with $q_\mathrm{B}>3a_\mathrm{G}$
While seemingly 
implausible in the context of planet formation in these systems, 
systems such as this might arise if binary orbits are 
changed or stellar companions exchanged by encounters with other stars 
in a cluster \citep[e.g.,][]{Malmberg+07b}.

\subsubsection{Effects on mutual inclinations}

\begin{figure}
  \includegraphics[width=0.5\textwidth]{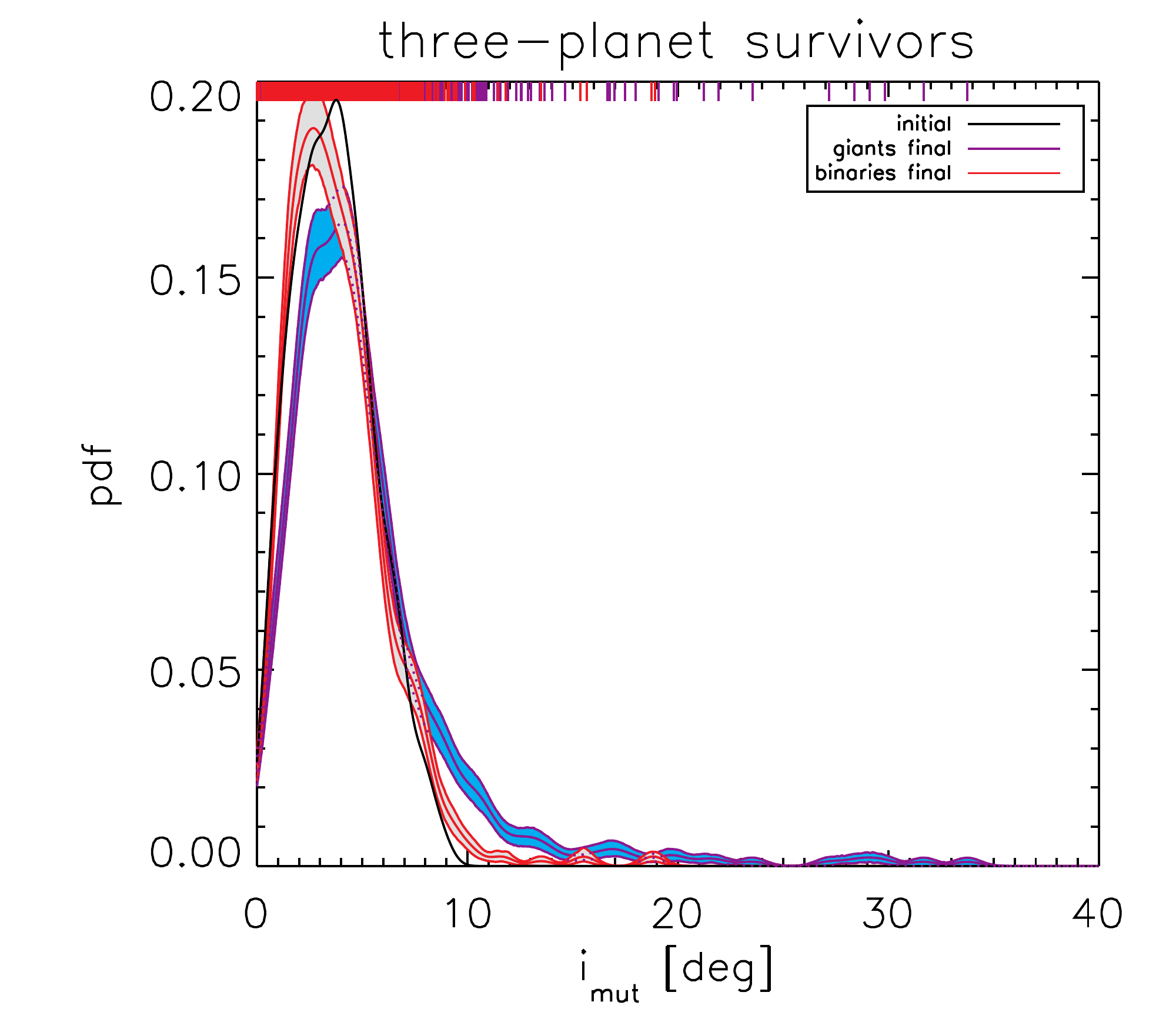}
  \includegraphics[width=0.5\textwidth]{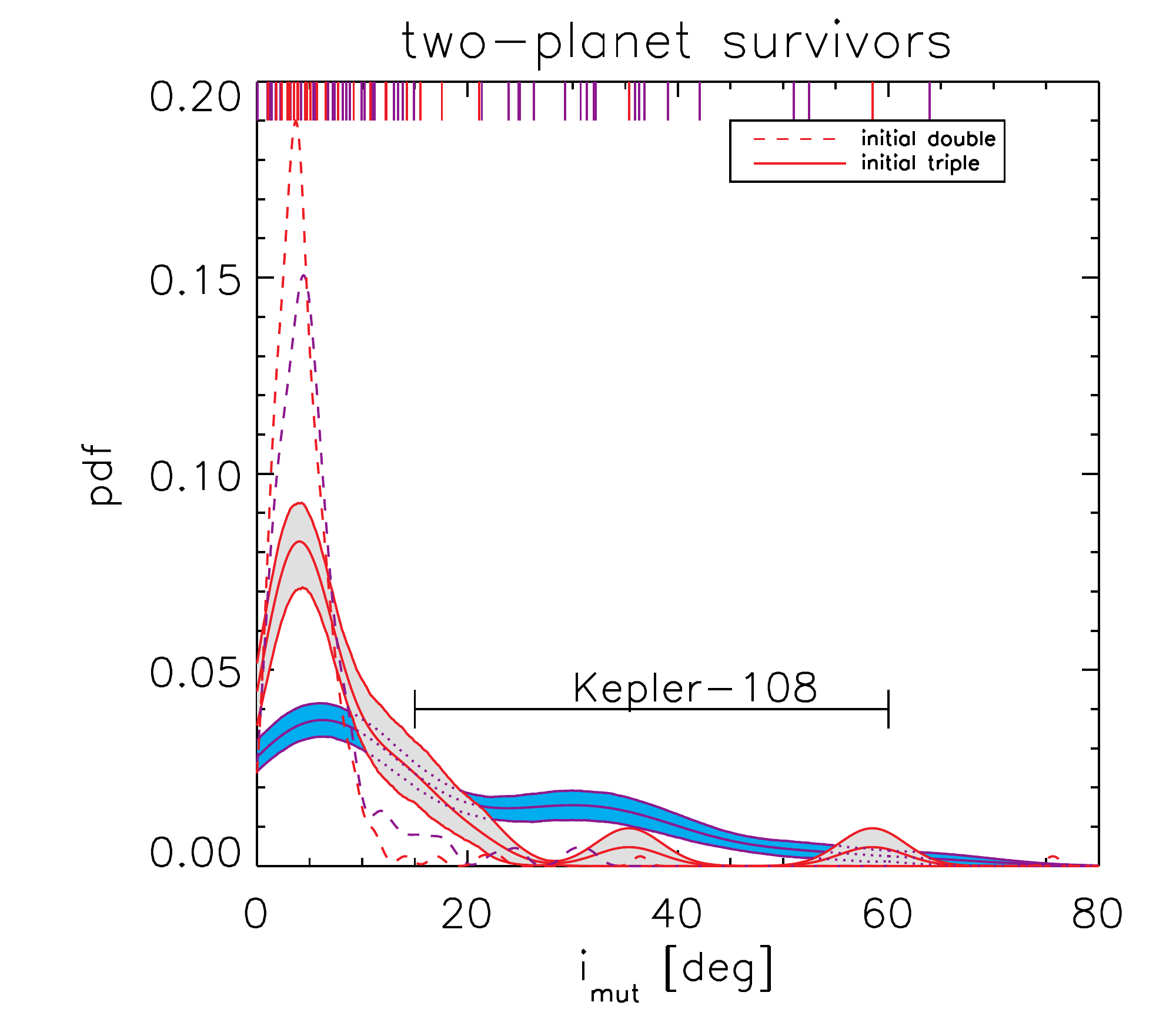}
  \caption{\textbf{Top: }Final mutual inclinations of adjacent inner planet pairs
    in the \textsc{Binaries} and \textsc{Giants} simulations, for systems that end with
    three inner planets. The initial distribution is also shown. \textbf{Bottom: }
    Final mutual inclinations in the systems that end up with two planets.
    Here we combine \textsc{Giants} with \textsc{Giants-flat}, and \textsc{Binaries} with
    \textsc{Binaries-flat}, for better statistics. We also
    show the range of mutual inclinations inferred for Kepler-108 \citep{MillsFabrycky16}.
    We also show with dashed lines the initial 2-planet systems 
    (\textsc{Binaries2} and \textsc{Giants2}).}
  \label{fig:imut}
\end{figure}

Direct loss of inner planets is the most violent but not the only effect the outer 
system can have on the inner. The mutual inclinations of the inner planets can also be 
affected. In Figure~\ref{fig:imut} we show the instantaneous mutual 
inclinations of surviving 3-planet systems from \textsc{Binaries} 
at 10\,Myr. While the bulk of the distribution 
is close to the initial distribution (between $0^\circ$ and $10^\circ$), a small number 
of systems are excited to a higher mutual inclination of up to $20^\circ$. The outer 
planet is incapable of exciting high mutual inclinations amongst the inner planets 
through secular means, as the inner planets typically are coupled together too strongly. 
We use Equation~29 of \cite{LaiPu16} to parametrise the strength of the inclination 
forcing from the outer planet compared to the coupling between the inner planets 
(their $\bar\epsilon$ we call $\epsilon$). $\epsilon\ll1$ implies strong coupling 
between inner planets with little excitation of mutual inclinations, while 
$\epsilon\gg1$ implies that the outer planet dominates, allowing mutual inclinations 
amongst the inners to be excited to up to twice the initial value of that between the 
inners and the outer planet. Secular resonance can exist in the region $\epsilon\sim1$ 
that can excite still higher values of mutual inclination \citep{LaiPu16}. 
All of our surviving triple-planet systems have $\epsilon<1$, with many around $10^{-3}$, and hence 
the outer planet cannot efficiently drive up mutual inclinations 
in the inner system. Similar results hold for the other integration sets.

More interesting 
is the case of two-planet survivors, which shows a larger tail of systems of high 
mutual inclination of up to $60^\circ$ (Figure~\ref{fig:imut}, bottom panel). This 
provides a means of generating misaligned systems such as Kepler-108 
\citep{MillsFabrycky16}, as we discuss below. The systems initially with two 
inner planets (\textsc{Binaries2} and \textsc{Giants2}) 
that retain both are less excited, as is shown by the dashed lines.

In summary, Kozai perturbations to outer planets disrupt inner systems in 
around $20-25\%$ of cases. Mutual inclinations of surviving triple-planet systems 
remain unexcited, but destabilised systems reduced to two planets 
can become mutually inclined up to several tens of degrees.

\section{Population synthesis II: Scattering}

\label{sec:scattering}

\begin{figure*}
  \includegraphics[width=0.48\textwidth]{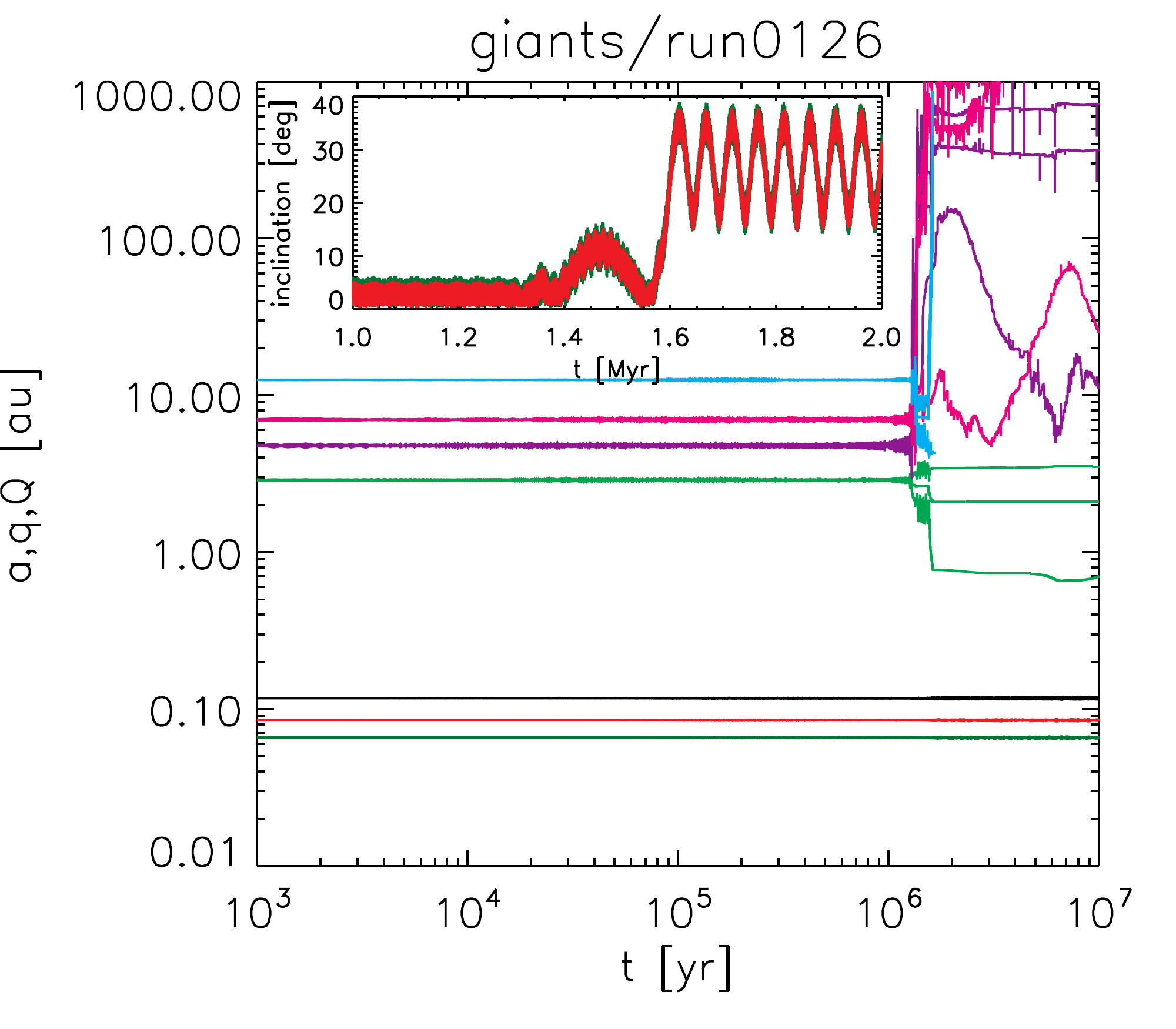}
  \includegraphics[width=0.48\textwidth]{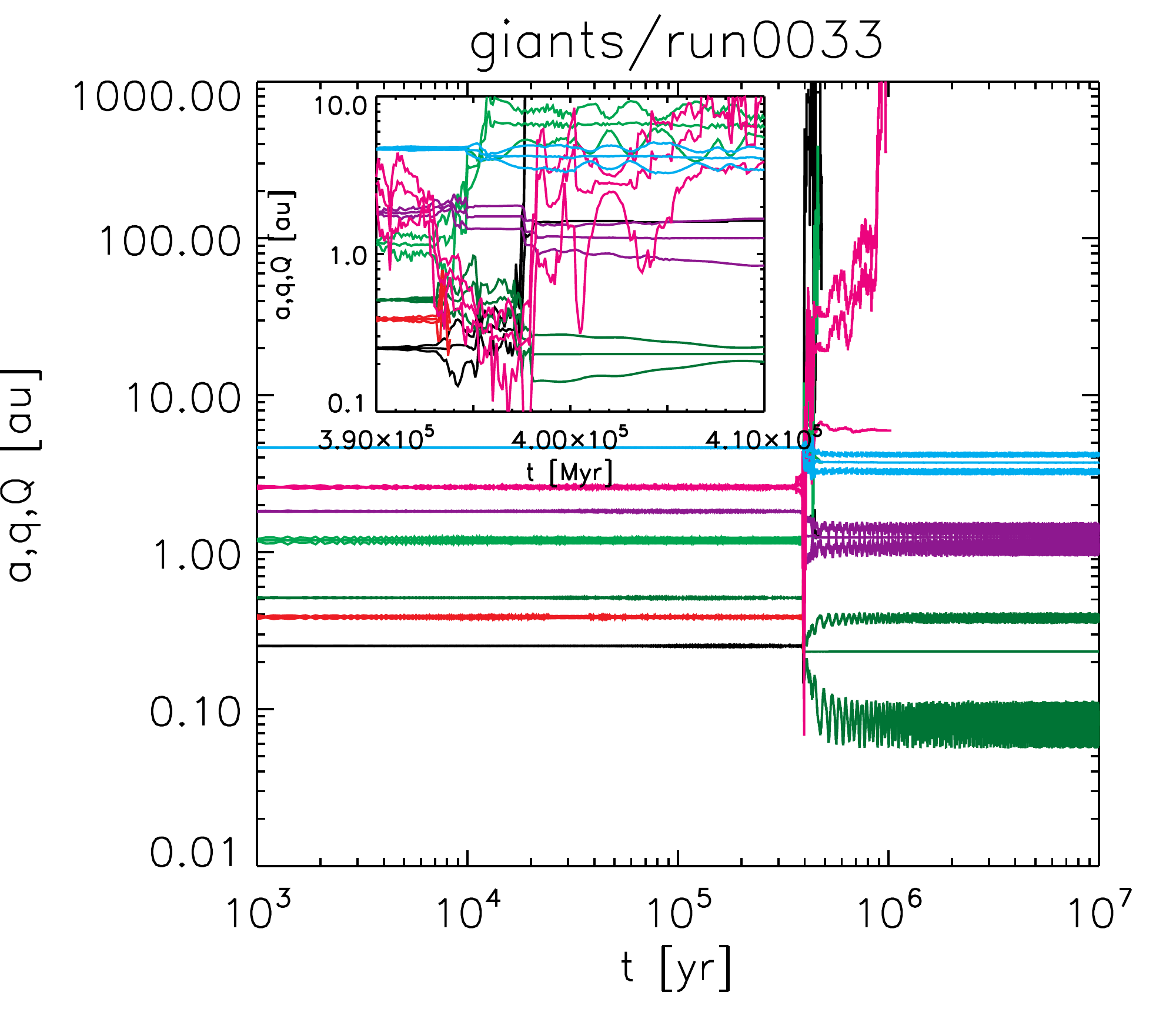}
  \caption{Example evolution of \textsc{Giants} systems, showing 
      the semi-major axis, pericentre and apocentre of all bodies. The 
      inner \emph{Kepler} triple-planet system is in dark green, red and 
      black, while the four outer planets are in light green, purple, 
      magenta and cyan. \textbf{Left: }
    Scattering in the outer system leaves the inner planets
    dynamically cold but induces a significant obliquity on
    the whole inner system; inclinations of the inner planets are shown in
    the inset. \textbf{Right: }A contribution
    to the \textit{Kepler} dichotomy: a single inner planet survives
    after scattering in the outer system induces scattering in the inner
    system. The inset shows a zoom-in around the strong scattering at 
      $\sim0.4$\,Myr.}
  \label{fig:example-giants}
\end{figure*}

Examples of the effects of scattering in the outer system in the 
\textsc{Giants} simulations are shown in Figure~\ref{fig:example-giants}. 
In the left panel, strong scattering leaves the inner system dynamically 
unexcited but induces a large obliquity on the set of three planets. 
In the right panel, we see a contribution to the 
\textit{Kepler} Dichotomy: the inner system is destabilised 
once scattering begins in the outer system and eventually 
only a single planet is left in the inner system. We now 
describe this integration set in more detail.

\subsection{Effects on outer system}

\label{sec:giants-outer}

Of our 400 systems, 141 were stable and retained all their giant planets. 
120 lost one giant, 130 two, and 9 lost three. One of the systems 
that retained its four giants was undergoing scattering at the 
end of the integration, with one planet having been ejected onto a 
wide ($a=120$\,au, $e=0.9$) orbit.

In addition to Kozai cycles, planet--planet scattering followed by tidal circularisation 
has also been proposed as a migration channel for hot Jupiters 
\citep{RasioFord96,WeidenschillingMarzari96,Nagasawa+08,BeaugeNesvorny12}. 
Combining N-body integrations with tidal forces, 
\cite{Nagasawa+08} found that 30\% of unstable 3-planet equal-mass systems form 
hot Jupiters, while with a small spread in masses (a factor 4 at most) 
\cite{BeaugeNesvorny12} found that 10\% of unstable three-planet and 23\% of four-planet 
systems form hot Jupiters. Our potential hot 
Jupiter formation rate---planets hitting the 
star, as well as planets attaining small pericentres---is much smaller, only a few 
per cent.\ (Figure~\ref{fig:hj-formation-binaries}, bottom right). We attribute this to the 
broader range of masses we use for the outer planets: in more hierarchical systems, 
the lower-mass planets can be ejected without the remaining large planets 
acquiring significant eccentricities, and equal-mass systems are far more disruptive 
to other bodies in the system \citep{Carrera+16}.

We also found that the eccentricities 
of our surviving outer planets were lower than the \emph{observed} population 
of giant exoplanets. We therefore construct a \textsc{Giants-selected} sample from our simulations 
in the following manner. We construct an empirical eccentricity distribution 
from \url{www.exoplanets.org} \citep{Han+14} for RV-discovered planets with mass greater than 
Saturn's and period greater than 50 days, and also construct the distribution 
for our surviving planets in the same mass range (Figure~\ref{fig:edist}, upper panel). We divide 
this up into 10 eccentricity bins of width $0.1$, assign each bin a weight 
of $N_\mathrm{empirical}/N_\mathrm{modelled}$, and normalise the weights so that 
the maximum is unity. For each model planet we then add it to our sample with a 
probability equal to its bin weighting. Thus, systems in a bin which is 
over-represented in the simulations will be assigned a low probability 
of being selected. In doing this, we reduce the 
size of our sample by a factor of roughly 10, but we avoid resampling the same 
system multiple times. 
This results in the \textsc{Giants-selected} 
distribution shown in the upper panel of Figure~\ref{fig:edist}, with 39 selected systems. The 
final states of these systems are 
displayed in the lower panel of Figure~\ref{fig:edist}.

\subsection{Effects on inner system}

\subsubsection{Intrinsic multiplicities}

Unsurprisingly, in the systems that retained all their giants, the inner 
system was nearly always unperturbed: only 4 of these 141 systems lost 
one of their KOIs, suggesting that long-range dynamical excitation 
(through secular resonances for example) is inefficient at 
destabilising inner systems, unless at least a moderate degree 
of excitation is reached in the outer system (the mean final eccentricity 
among the outer planets in these systems was $0.017$, and the median $0.006$), 
although our 10\,Myr integrations may miss instabilities that could 
occur on timescales of several Gyr.  
More significantly, most of the unstable giant 
systems also retained their three KOIs in  the inner system: only 
$95/259=37\%$ of the unstable systems lost one or more of their KOIs. Thus, 
even in dynamically active outer systems, destabilisation of the inner 
system occurs in roughly only 1 in 3 cases. 
Our \textsc{Giants-selected} runs are less hierarchical than their \textsc{Giants}
superset, with a median mass ratio of $2.1$ vs $2.7$. Unsurprisingly, they are
also more destructive of the inner systems than \textsc{Giants}, keeping only
$58.5\pm7.6\%$ of inner triple-planet systems intact, compared to $75.1\pm2.2\%$ for
\textsc{Giants} and $63.2\pm3.0\%$ for \textsc{Giants-unstable}. 
Of the inner planets lost, $51\%$ hit another planet, $38\%$ collided 
with the star, and $12\%$ were ejected. Though the number of events is small 
(21 ejections here, compared to 11 in \textsc{Binaries}), the larger 
fraction of ejections in the \textsc{Giants} simulations may be a signature 
of the ``uplift'' mechanism we discuss in the context of eccentric 
warm Jupiters in Section~\ref{sec:uplift}.

In contrast to \textsc{Binaries}, the fraction of destabilised 
inner systems rises with the semimajor axis of the inner planet 
(Figure~\ref{fig:funstable}). 
The outer planets in \textsc{Giants} rarely achieve such small 
pericentre as in our Kozai simulations (Figure~\ref{fig:hj-formation-binaries}, 
bottom-right panel), and only 11 outer planets managed to collide 
with the star here, compared to 35 in \textsc{Binaries}. 
In most cases of destabilisation of the inner 
system one or more of the outer planets' 
pericentres comes within the semimajor axis of the outermost inner planet 
(Figure~\ref{fig:hj-formation-binaries}, bottom-right panel, 
purple lines). 
However, in a minority of cases destabilisation occurs at a distance, 
without any outer planet's orbit penetrating the initial semimajor 
axes of the inner planets. This is more common in \textsc{Giants} 
than in 
\textsc{Binaries}, occurring in 34\% of cases of instability compared to 19\%.
This suggests 
that secular effects are more effective at exciting the inner system: 
as discussed in \cite{Matsumura+13} and \cite{Carrera+16}, 
as outer giant planets undergo scattering, secular resonances can jump around 
the inner system, exciting eccentricities of inner planets without 
the outer giants actually approaching the inner planets closely.

The stable inner systems experience a small degree of 
dynamical excitation. 
In Figure~\ref{fig:giants-einn} we show the eccentricity 
distribution of surviving triple KOIs, broken down by the number of 
surviving giants. While there is a trend towards higher eccentricities 
for more violent instabilities (as measured by the number of surviving 
giants), eccentricities remain low: 
in the unstable giant planet systems, 90\% of the 
KOIs in systems that retain all three of the inner planets 
have $e<0.09$ at the end of the integration. This comports 
with the majority of observed multi-planet \textit{Kepler} systems which 
appear to have similarly low eccentricity \citep{VanEylenAlbrecht15}.

\subsubsection{Mutual inclinations}

Effects on mutual inclinations are broadly similar to the 
\textsc{Binaries} runs, with little excitation among the 
triple-planet survivors (Fig~\ref{fig:imut}, top). The 
2-planet survivors are more excited than are the 2-planet 
survivors in \textsc{Binaries} (Fig~\ref{fig:imut}, 
bottom), although the sample size here is smaller. Interestingly, 
we see no correlation of the mutual inclination of the 
two-planet systems with the minimum pericentre attained by 
any of the giant planets, suggesting that in some systems at 
least the destabilisation and eccentricity excitation is 
a result of secular effects and not direct scattering by 
the outer planets. Secular effects could be amplified if 
secular resonances jump around the inner system during 
scattering among the outer planets \citep{Matsumura+13,Carrera+16}.

The diversity of final inclinations is shown in Figure~\ref{fig:i3-imut3}.
Here we show, for surviving double- and triple-planet systems, the mutual inclinations
between adjacent planet pairs against the inclination of each planet with
respect to the initial reference plane (each system is thus represented
by two or four points). Systems start in the dark shaded lower left
quadrant. A small number move rightwards, increasing the system's
inclination while keeping mutual inclinations low, to form systems similar
to Kepler-56. Other systems, often those destabilised and reduced to
double systems, move upwards and rightwards, gaining a mutual
inclination instead of remaining coplanar.

In contrast to the \textsc{Binaries} case, flattening the planetary system
does have an effect on the inner system. This is because by flattening the
outer system as well, a larger fraction of systems experience collisions
between the outer planets (64 cases in 300 systems, compared to 39 cases
in 400 systems), while only 3 outer planets collide with inner planets,
compared to 10 in the non-coplanar \textsc{Giants} runs, despite the
flatness of the systems.

\begin{figure}
  \includegraphics[width=.48\textwidth]{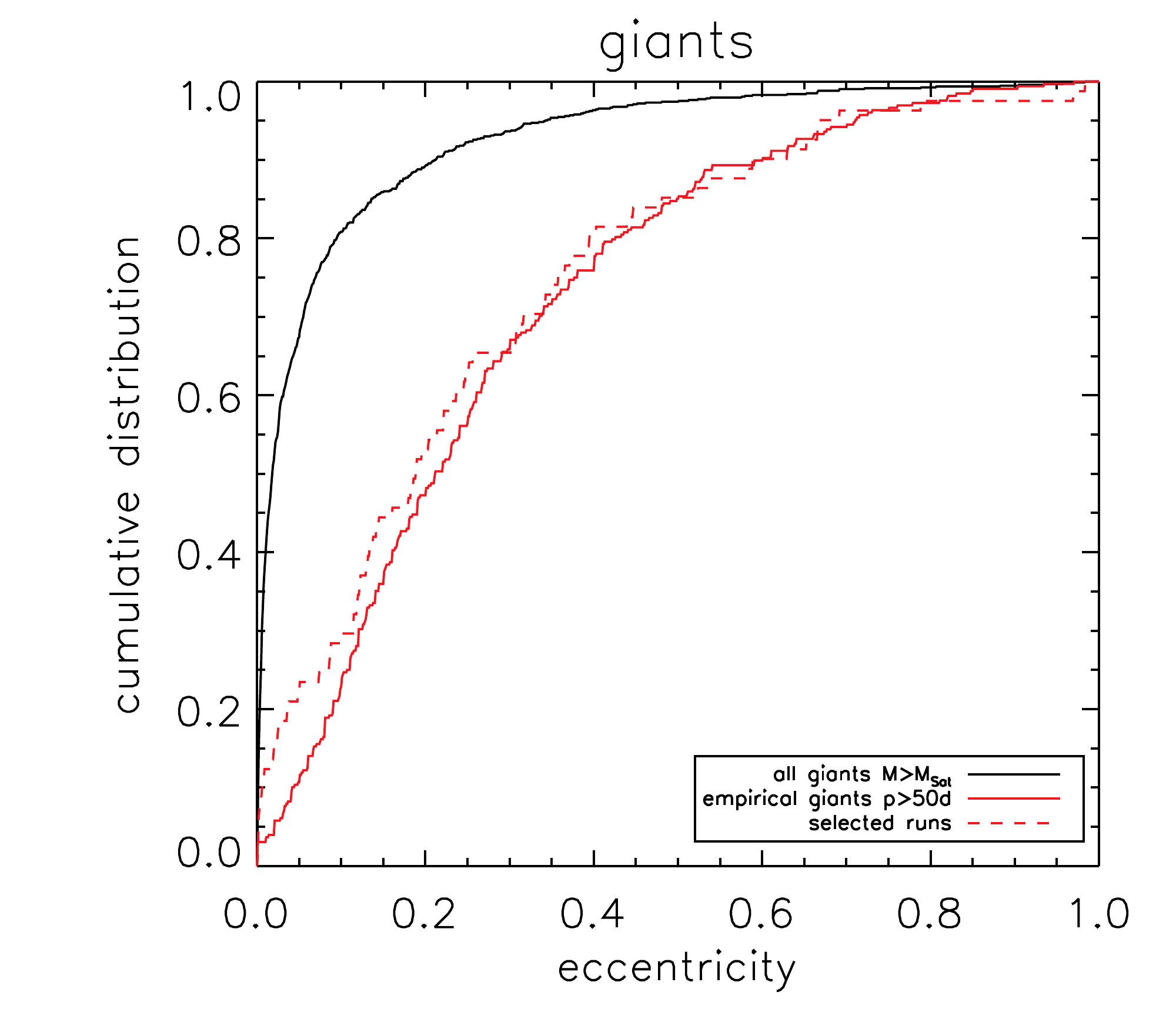}
  \includegraphics[width=.48\textwidth]{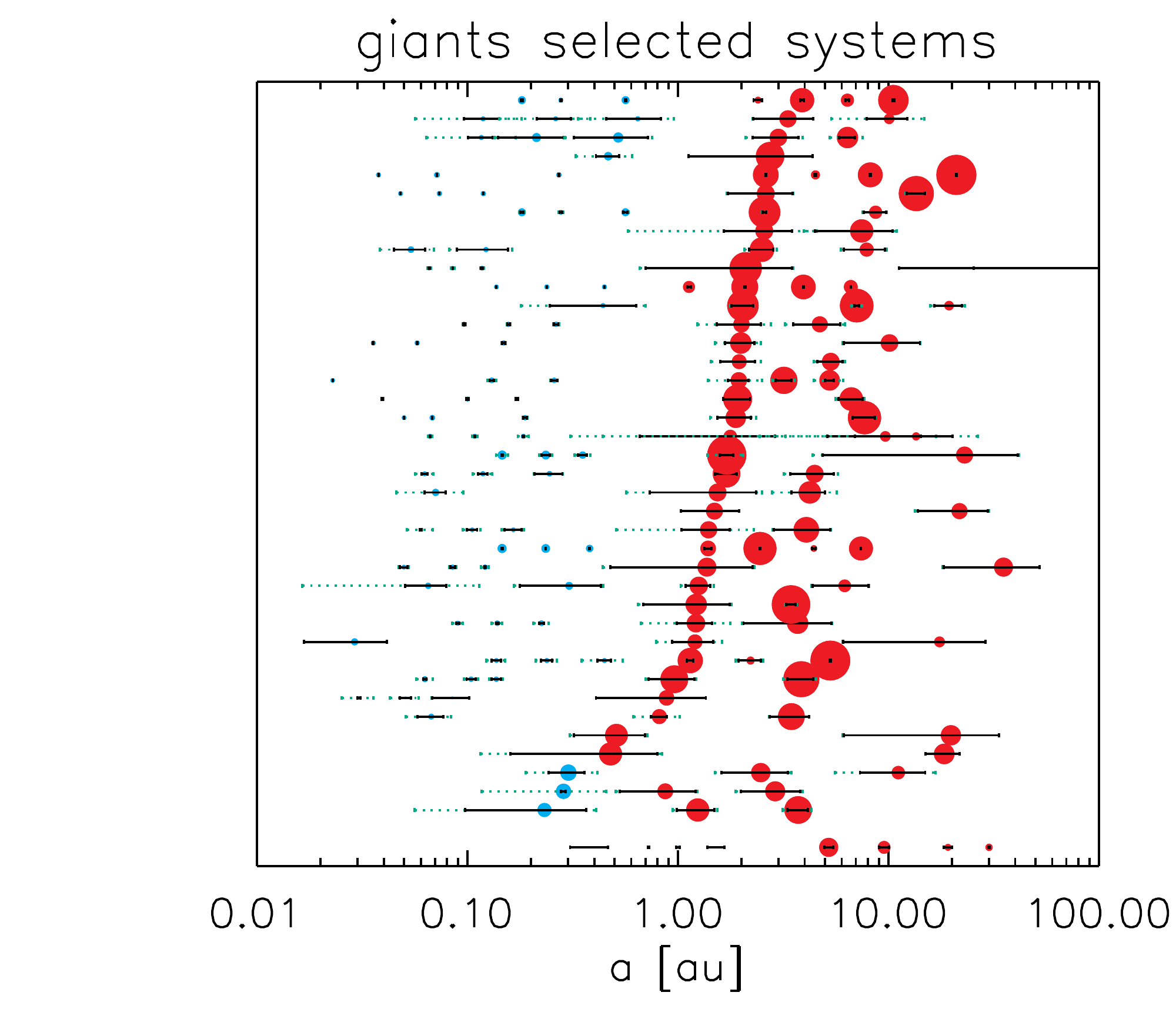}
  \caption{\textbf{Top: }Eccentricity distributions of surviving planets more massive than
    Saturn in our \textsc{Giants} runs, the empirical distribution
    for planets in the same mass range and $P>50$\,days, and
    a subset of the former drawn to be consistent with the latter.
    This subset (\textsc{Giants-selected}) is constructed 
    by dividing the observed and simulated giant planets into eccentricity bins,
    assigning each bin a weight according to how over-represented it
    is in the simulations,
    and randomly drawing simulated systems to give the same eccentricity distribution
    as the observed one. See \S\ref{sec:giants-outer} for details.
    \textbf{Bottom: }Masses, semimajor axes and eccentricities of the \textsc{Giants-selected}
    systems at 10\,Myr. Symbol size is proportional to the cube root of planet mass,
    and the Solar System is shown at the bottom for reference. Solid lines show 
      the final eccentricity, dashed lines the greatest eccentricity attained. Blue 
      planets are initially inner KOIs, while red planets are initially outer giants.}
  \label{fig:edist}
\end{figure}

\begin{figure}
  \includegraphics[width=.5\textwidth]{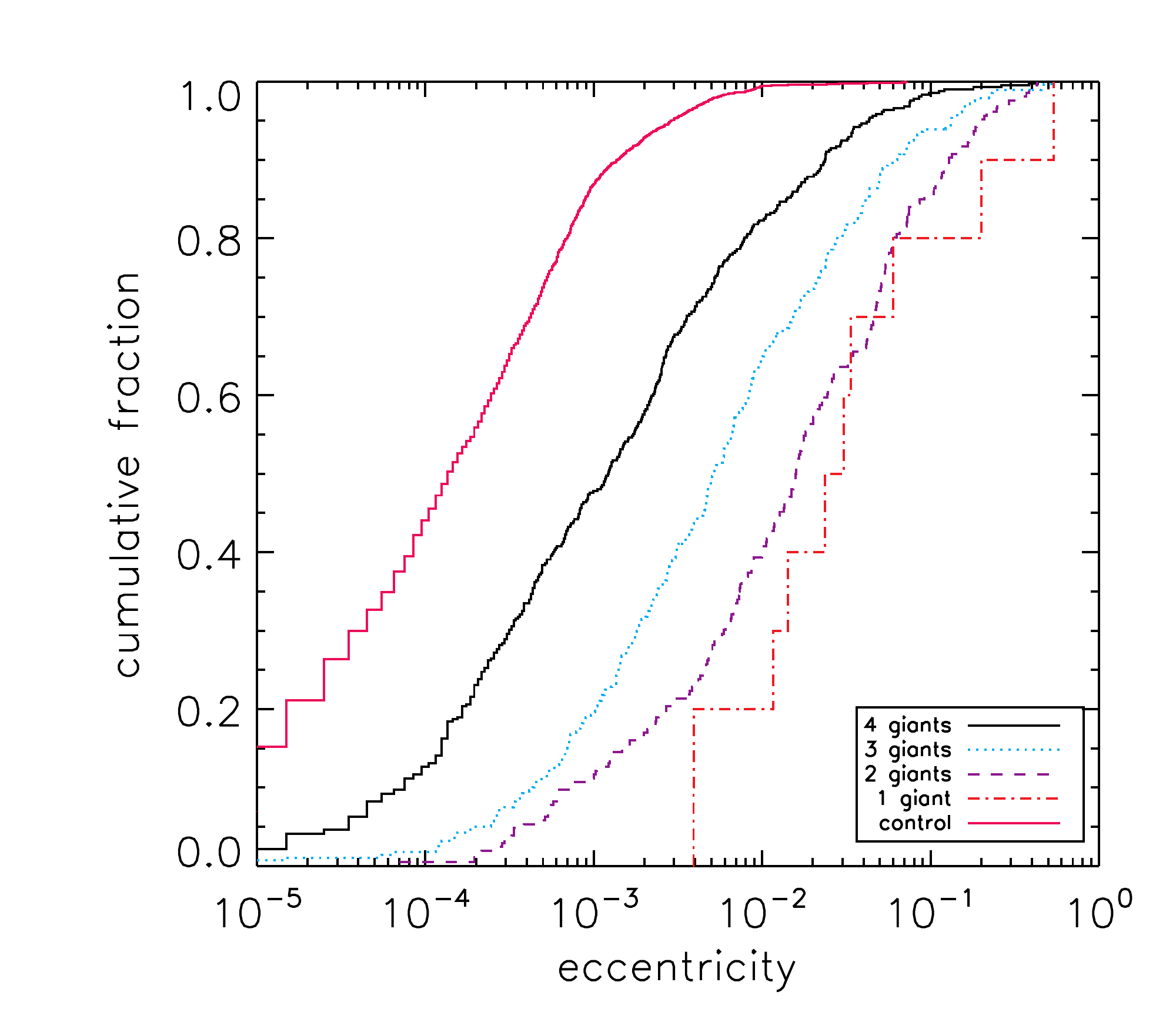}
  \caption{Eccentricities of surviving triple-planet KOI systems
  separated according to the number of surviving giants in the outer system.
  We also include the final eccentricities of KOIs in the \textsc{Control} sample 
  without any additional bodies (\S2). KOIs in systems of stable outer giants 
  have their eccentricities excited by a factor of $\sim10$, but they remain 
  low in absolute terms (median $\sim10^{-3}$). Unstable outer systems excite 
  eccentricities more the more planets they lose.}
  \label{fig:giants-einn}
\end{figure}

\begin{figure}
  \includegraphics[width=.5\textwidth]{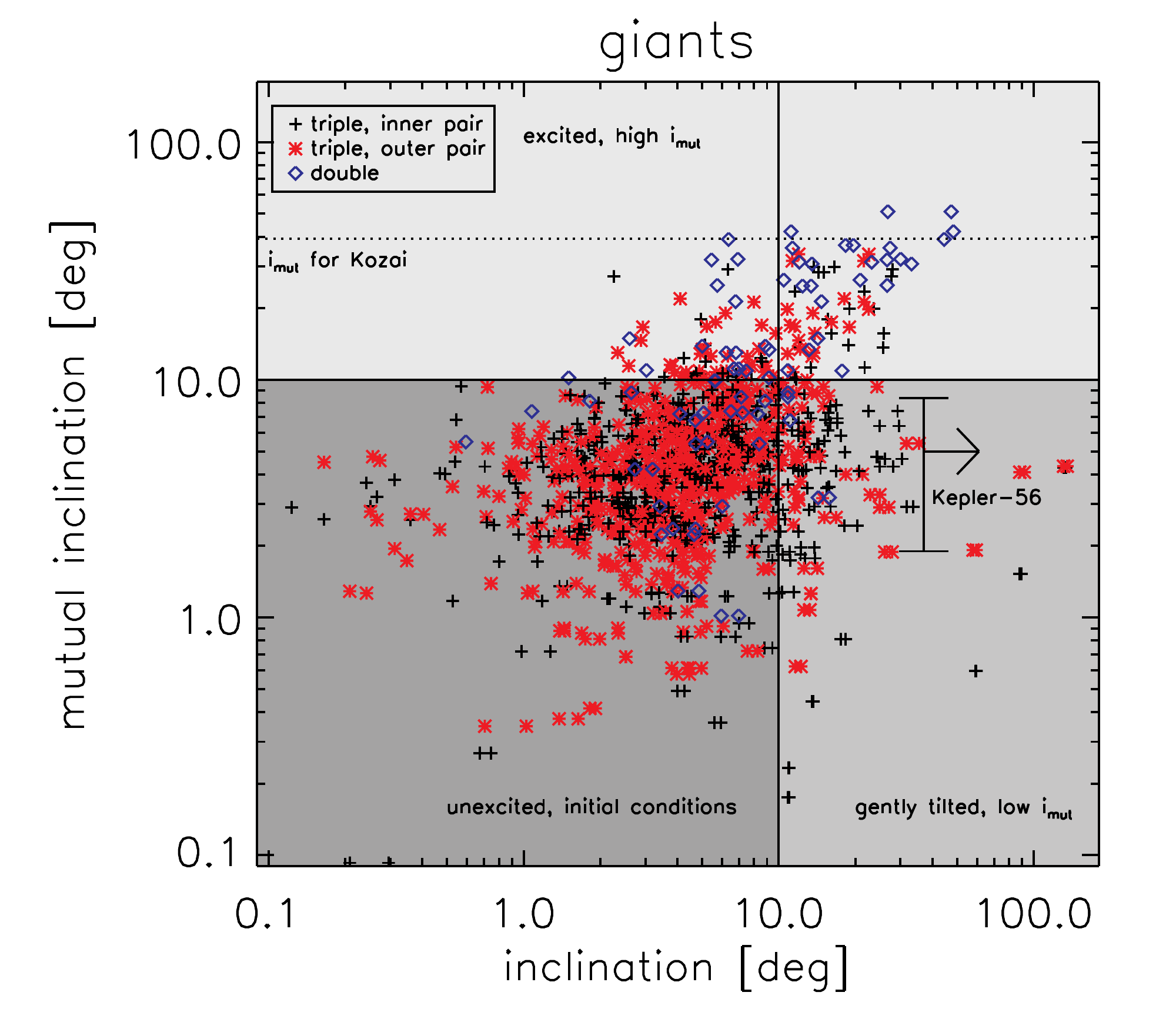}
  \caption{Final inclinations and mutual inclinations of surviving inner 
    double- and triple-planet systems from \textsc{Giants}. 
    On the $x$-axis, the inclination of each planet with respect to 
    the original reference plane is shown. On the $y$-axis, the mutual inclination 
    of each pair of adjacent planets is shown. Hence, each planet pair 
    contributes two points, at identical $y$-values. 
    Systems begin in the 
    lower left quadrant with $i_\mathrm{mut}<10^\circ$. They may subsequently 
    be excited to high mutual inclination (more common when the multiplicity 
    itself is reduced to a double system), or gently tilted to a high inclination 
    with respect to the system's original invariant plane, while maintaining 
    a low mutual inclination. We also mark the system Kepler-56 \citep{Huber+13}.}
  \label{fig:i3-imut3}
\end{figure}

\section{Formation of eccentric warm Jupiters}

\label{sec:warmj}
\begin{figure}
  \includegraphics[width=.5\textwidth]{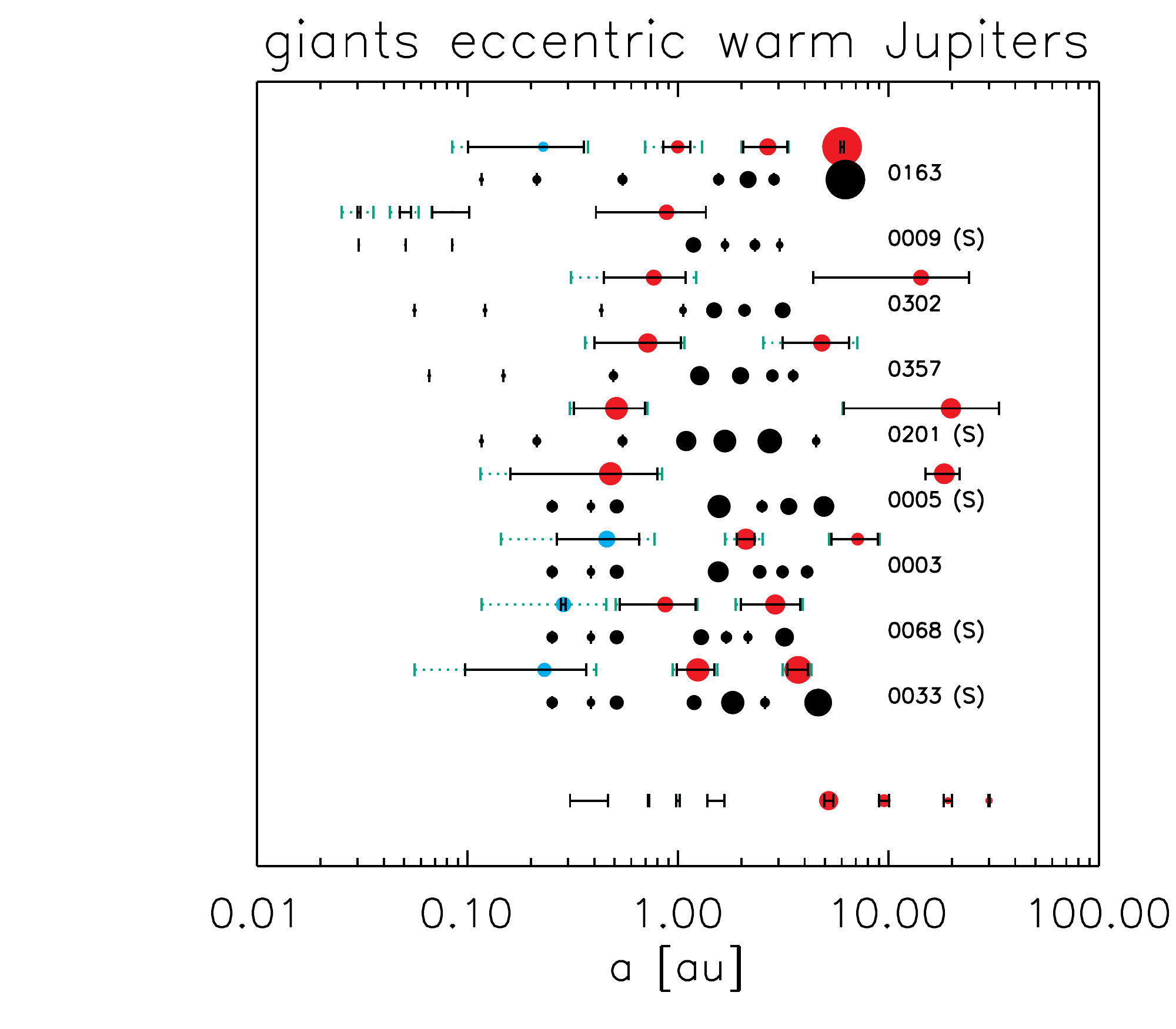}
  \includegraphics[width=.5\textwidth]{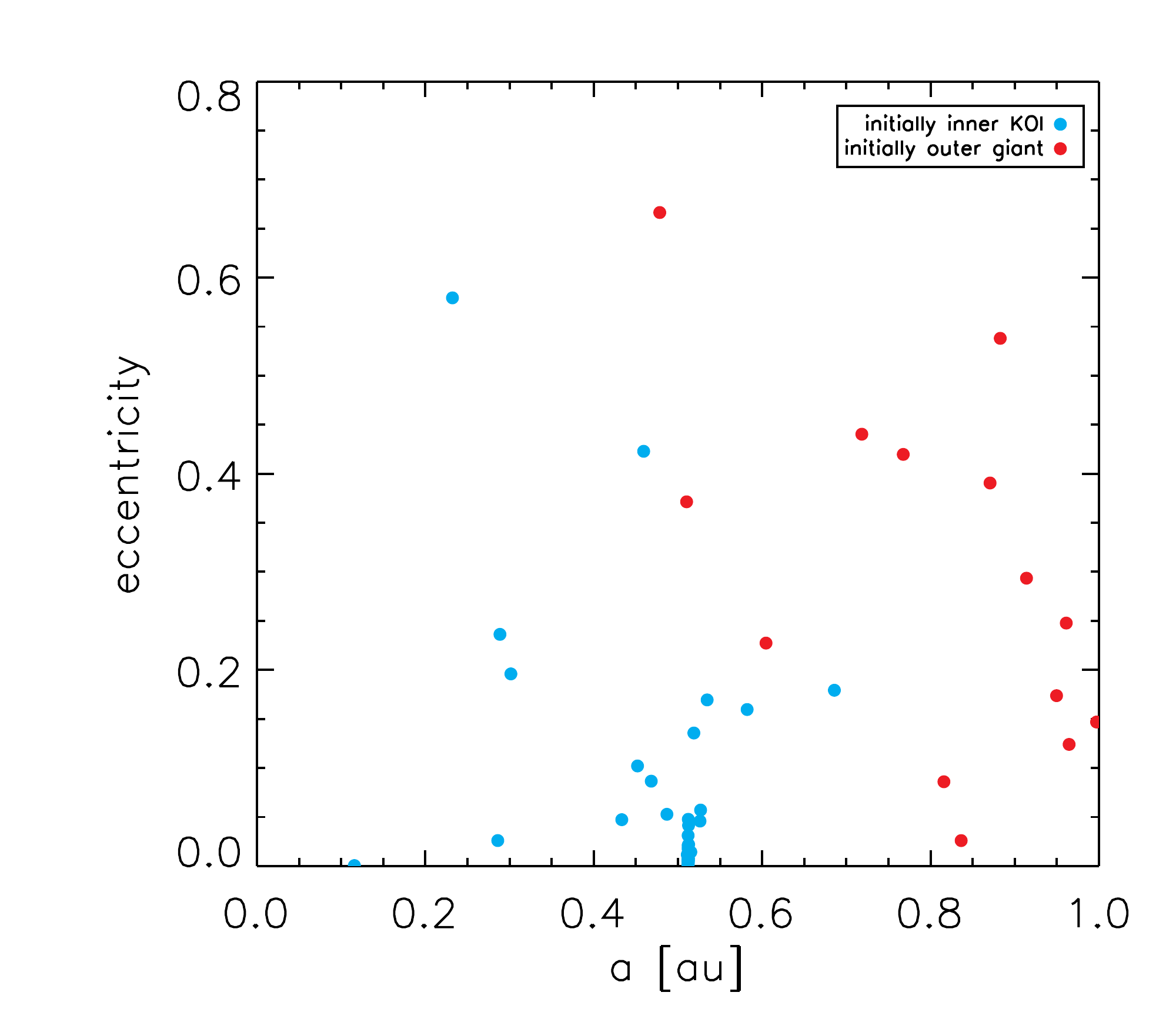}
  \caption{\textbf{Top: }Formation of eccentric warm Jupiters in our \textsc{Giants}
    integrations. We show systems with at least one planet with mass greater
    that that of Saturn, semi-major axis less than 1\,au, and eccentricity
    greater than 0.3 at any point between the time the final planet was
    removed and the end of the integration. The dotted lines show the maximum
    eccentricity of each planet attained during this time, while solid lines show the
    instantaneous eccentricity at the end of the integration. Red planets are
    originally outer planets while blue are originally inner \textit{Kepler} planets. Circle
    radius is proportional to the cube root of mass. Below each final system,
    we show in black the initial configuration. Run IDs are noted; see text for
    discussion of some individual systems. Runs marked ``(S)'' are
    in the \textsc{Giants-selected} sample. The Solar System is shown
    at the bottom for comparison. \textbf{Bottom: }All warm 
    Jupiters (including those with $e<0.3$) at the end of our \textsc{Giants} 
    simulations. Eccentricity is plotted against semimajor axis for each 
    planet with a mass greater than Saturn's and a semimajor axis $<1$\,au. Planets 
    that are originally one of the triple KOIs are shown in blue, while those 
    originally an outer giant are in red.}
  \label{fig:lovisplot-warmj}
\end{figure}

\begin{figure*}
  \includegraphics[width=.48\textwidth]{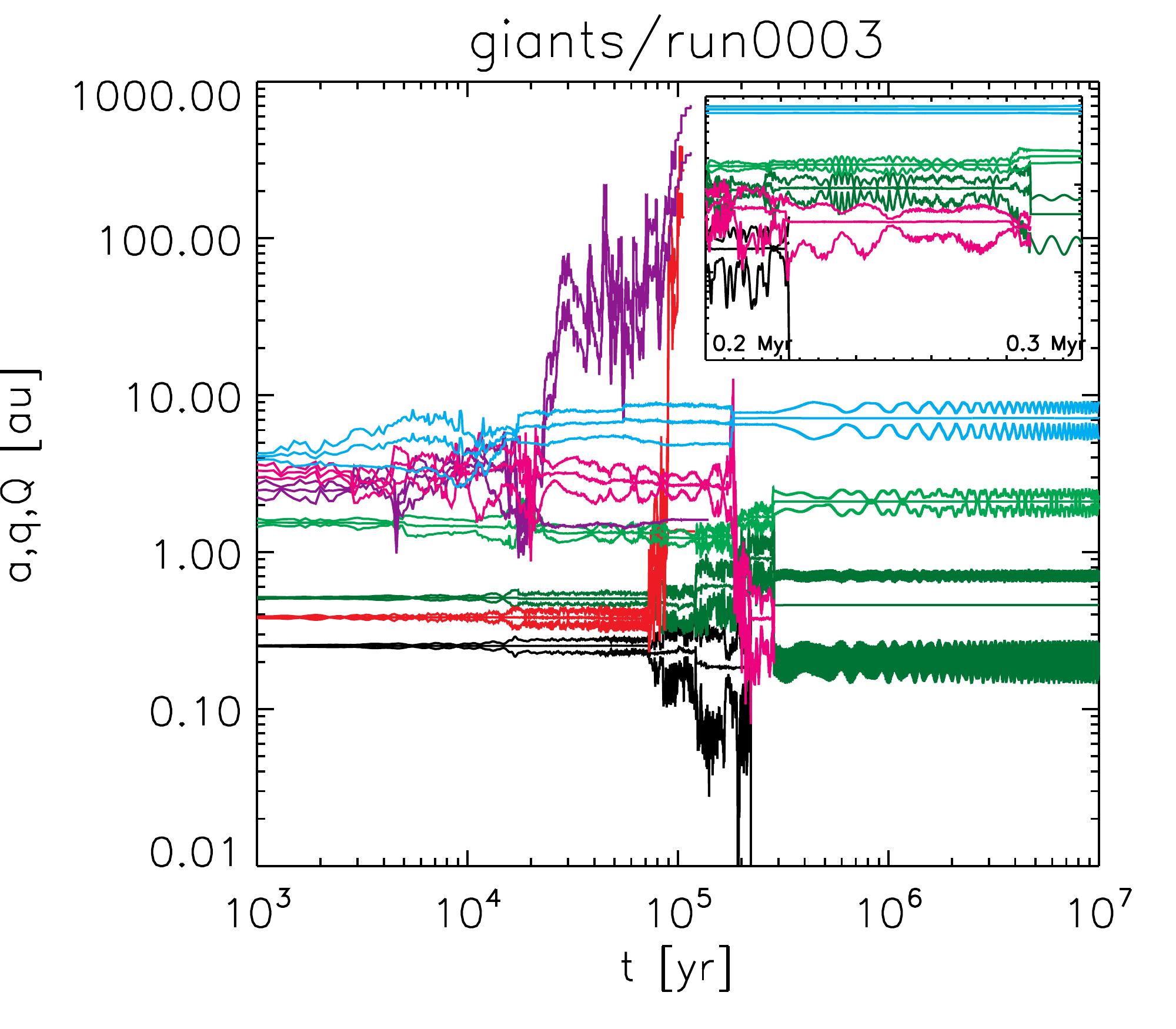}
  \includegraphics[width=.48\textwidth]{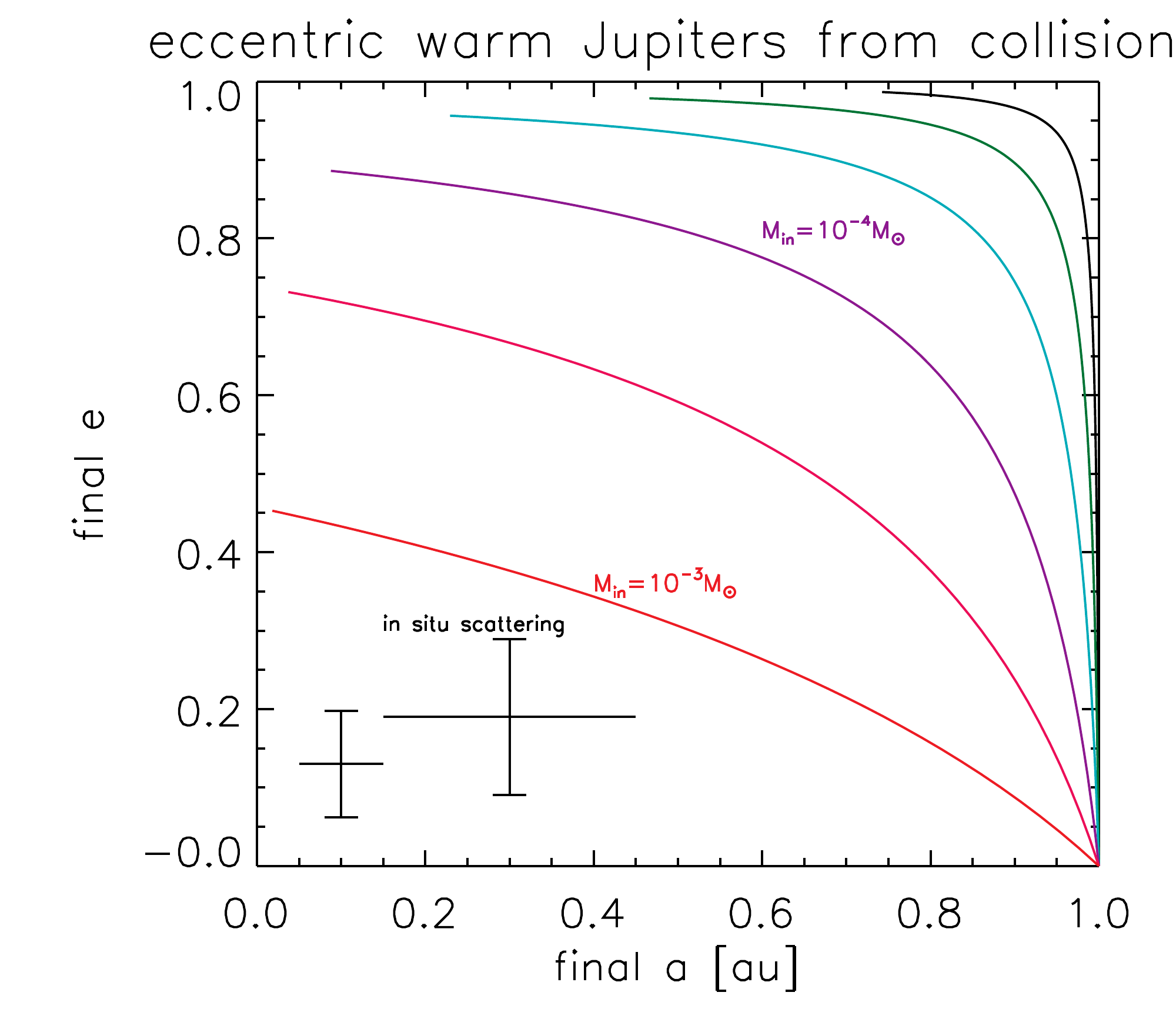}
  \caption{\textbf{Left: }Formation of an eccentric warm
    Jupiter by an inelastic collision between two $\sim$Saturn-mass
    planets, the interior on a low-eccentricity orbit and the
    exterior on a high-eccentricity orbit (inset, at $0.285$\,Myr).
    \textbf{Right: }Toy model of the formation of
    eccentric warm Jupiters through inelastic planet--planet
    collisions. A Jupiter-mass planet is placed at 1\,au
    with another planet interior, with mass ranging
    from Earth-mass to Jupiter mass and semimajor axis
    from $0.01$ to 1\,au. The giant planet is given
    sufficient eccentricity for its pericentre to reach
    the inner planet's orbit. The semimajor axis and
    eccentricity of the merger product are shown,
    assuming that this merger occurs at pericentre
    and conserves mass and angular momentum. Each
    line shows a different mass of the inner planet;
    lines terminate at an inner planet's $a=0.01$\,au. 
    Each line thus shows the locus 
    $(a_\mathrm{f}(a_1;m_1),e_\mathrm{f}(a_1;m_1))$
    for a given $m_1$: see Equations~\ref{eq:ef} 
    and~\ref{eq:af}. 
    Crosses show eccentricities attained from \emph{in-situ} 
    scattering \citep{Petrovich+14}.}
  \label{fig:collide-warmj}
\end{figure*}

An interesting challenge to models of planet migration and dynamics has been
raised by the discovery of a population of eccentric warm Jupiters at
separations of a few tenths of an au. We may divide warm Jupiters 
into three eccentricity ranges:
\begin{enumerate}
\item \emph{Low eccentricity:} These are consistent with disc 
migration and/or \emph{in situ} formation. In reality the two are connected, 
since planets accrete as they migrate: see for example the growth tracks in 
Figure~2 of \cite{Bitsch+15}. These planets can acquire low eccentricities through 
\emph{in situ} scattering, but \cite{Petrovich+14} show that eccentricities 
of above $0.3$ are very hard to attain because planets on close-in orbits 
preferentially collide before they acquire higher eccentricities. We therefore 
take $e=0.3$ as the upper eccentricity bound for this sub-population.
\item \emph{Super eccentricity:} These are warm Jupiters with small 
pericentres moving down tidal circularisation tracks, and would, in time, 
become hot Jupiters \citep{Socrates+12,Dawson+15}. 
The tidal migration rate is a very strong function 
of the planet's pericentre distance, and we may take $q\approx0.04$\,au 
to define an envelope in $a-e$ space below which tidal migration 
cannot populate (without additional forcing; see the next point).
\item \emph{Moderate eccentricity:} Lying in the eccentricity 
range between these two extremes, the moderately-eccentric warm Jupiters 
have eccentricities too high to be produced by \emph{in situ} scattering 
but too low to be following a tidal circularistaion track. Proposed formation 
mechanisms include the ``slow'' regime of Kozai cycles plus tidal friction, 
during which eccentricity oscillations still occur while tidal dissipation 
happens only at eccentricity maxima 
\citep[e.g.,][]{DawsonChiang14,Dong+14,PetrovichTremaine16}, 
and inelastic collisions between 
eccentric giant planets from the outer system and sufficiently massive 
inner planets \citep{Mustill+15}.
\end{enumerate}
Here we focus on the moderately-eccentric population with $e>0.3$ and 
$q>0.04$\,au, elaborating on the inelastic collision mechanism of \cite{Mustill+15} 
and describing two further formation channels for these objects.

Defining such planets as those more massive than Saturn,
with semimajor axis less than 1\,au, and which attain an eccentricity of at
least $0.3$ at some point after the loss of the final planet, we find
eccentric warm Jupiters to be produced in 9 of our \textsc{Giants} simulations, 
and 3 of our \textsc{Binaries} simulations. 

The systems from the \textsc{Giants} simulations are 
displayed in Figure~\ref{fig:lovisplot-warmj}. This formation
rate ($9/400=2.5\pm0.8\%$) is low, but is higher for our
\textsc{Giants-selected} sample at $5/39=14.6\pm5.5\%$; \textsc{selected} systems are
indicated in Figure~\ref{fig:lovisplot-warmj}. Four of our eccentric
warm Jupiter systems (with the smallest semimajor axis of the warm Jupiter: the
lowest four systems in Figure~\ref{fig:lovisplot-warmj}) share a common
KOI inner triple-planet system: KOI620, alias Kepler-51. With our adopted mass--radius relation,
this system has planet masses of $2.1\times10^{-4}$, 
$3.8\times10^{-4}$, and $8.3\times10^{-5}\mathrm{\,M}_\odot$\footnote{Note however that an analysis 
of transit timing variations 
\cite{Masuda14} has yielded exceptionally low masses for these planets
($2.1$, $4.0$ and $7.6\mathrm{\,M}_\oplus$ with 
radii of $7.1$, $9.0$ and $9.7\mathrm{\,R}_\oplus$), 
while RV upper limits from \cite{Santerne+16} are 
consistent with our assigned masses.}. 
Three of the eccentric warm Jupiters in \textsc{Giants}  
were originally members of inner systems, while the remaining 6 were initially 
outer planets. We also show in the bottom panel of 
Figure~\ref{fig:lovisplot-warmj} the eccentricities and semimajor axes 
of all planets more massive than Saturn with $a<1$\,au at the end of 
the simulation. Most of these warm Jupiters that originated as one of the 
triple KOI inner planets retain a low eccentricity, while those 
originating as outer planets have a broader range of eccentricities. This 
underlines the $e>0.3$ criterion we used to distinguish the low-eccentricity 
from the moderate-eccentricity warm Jupiters.

We now discuss the formation of these eccentric warm Jupiters. We identify 
three pathways: inelastic collision between an inner planet and an eccentric outer 
planet as in \cite{Mustill+15} (Section~\ref{sec:collide-warmj}); secular 
forcing, possibly involving freezing into a high-eccentricity state as 
scattering resolves (Section~\ref{sec:secular}); and \emph{in-situ} scattering, 
which may be aided by ``uplift'' as one planet is removed from the inner system by the 
outer planets (Section~\ref{sec:uplift}).

\subsection{Eccentric warm Jupiters from inelastic planet--planet collision}

\label{sec:collide-warmj}

In \cite{Mustill+15} we found that a highly-eccentric Jupiter-mass 
planet experiencing an inelastic collision with an inner Neptune-mass planet can form an 
eccentric warm Jupiter, which is the merger product of the two planets. 
While that study was an idealized case 
of the giant's eccentricity being imposed arbitrarily, rather than 
arising consistently through dynamical evolution, this mechanism 
remains at work when we treat the dynamics consistently in the 
present study, and two of our 
eccentric warm Jupiters form from such inelastic collisions. We show one example 
in the left-hand panel of Figure~\ref{fig:collide-warmj}: after a period 
of scattering, in which two roughly Saturn-mass planets switch places, 
the eccentricity of the outer one of the pair is excited to almost $0.8$, 
and it then collides with the inner, causing its semi-major axis to 
shrink from $\sim0.9$\,au to $0.46$\,au, and leaving its eccentricity stably 
oscillating around $0.6$.

We construct a toy model of this process as follows. Assume that 
a giant planet at a semi-major axis $a_2$ with mass $m_2$ is given sufficient 
eccentricity to collide with a coplanar planet of mass $m_1$ 
on a circular orbit at $a_1<a_2$, and that the planets then collide inelastically before  
their orbits change further, conserving mass and angular 
momentum. The angular momenta of the two planets before the collision are 
\begin{equation}
  L_1=m_1\sqrt{\mathcal{G}M_\star a_1},\qquad L_2=m_2\sqrt{\mathcal{G}M_\star a_1(1+e_2)}.
\end{equation}
The final eccentricity of the merger product is then given by
\begin{equation}
  e_\mathrm{f}=\left(\frac{m_1/m_2 + \sqrt{1+e_2}}{m_1/m_2+1}\right)^2 - 1,\label{eq:ef}
\end{equation}
guaranteed to lie between 0 and the initial eccentricity $e_2$, while 
the final semi-major axis is
\begin{equation}
  a_\mathrm{f}=\frac{a_1}{1-e_\mathrm{f}},\label{eq:af}
\end{equation}
lying between the initial semi-major axes $a_1$ and $a_2$. 
The final $a$ and $e$ of the merger products are shown in 
the right-hand panel of Fig~\ref{fig:collide-warmj}, for $a_2=1$\,au, 
$m_2=10^{-3}\mathrm{\,M}_\odot$, and a range of $a_1$ and $m_1$. An inelastic collision 
of the eccentric Jupiter with a super-Earth at 0.01\,au is sufficient to shrink 
the giant's orbit by a factor of 2, although collision with an object 
more massive than Neptune would be required to simultaneously reduce the eccentricity 
below $0.9$. 

This mechanism was also responsible for producing one of the eccentric warm 
Jupiters in our \textsc{Binaries} simulations: that shown in the left panel 
of Figure~\ref{fig:example}. Here however, the warm Jupiter possesses a very 
small pericentre after the collision, meaning that the warm Jupiter phase would 
only be transient and the planet would in time circularise to become a hot Jupiter.

\subsection{Eccentric warm Jupiters from secular chaos and secular freeze-out}

\label{sec:secular}

\begin{figure*}
  \includegraphics[width=.48\textwidth]{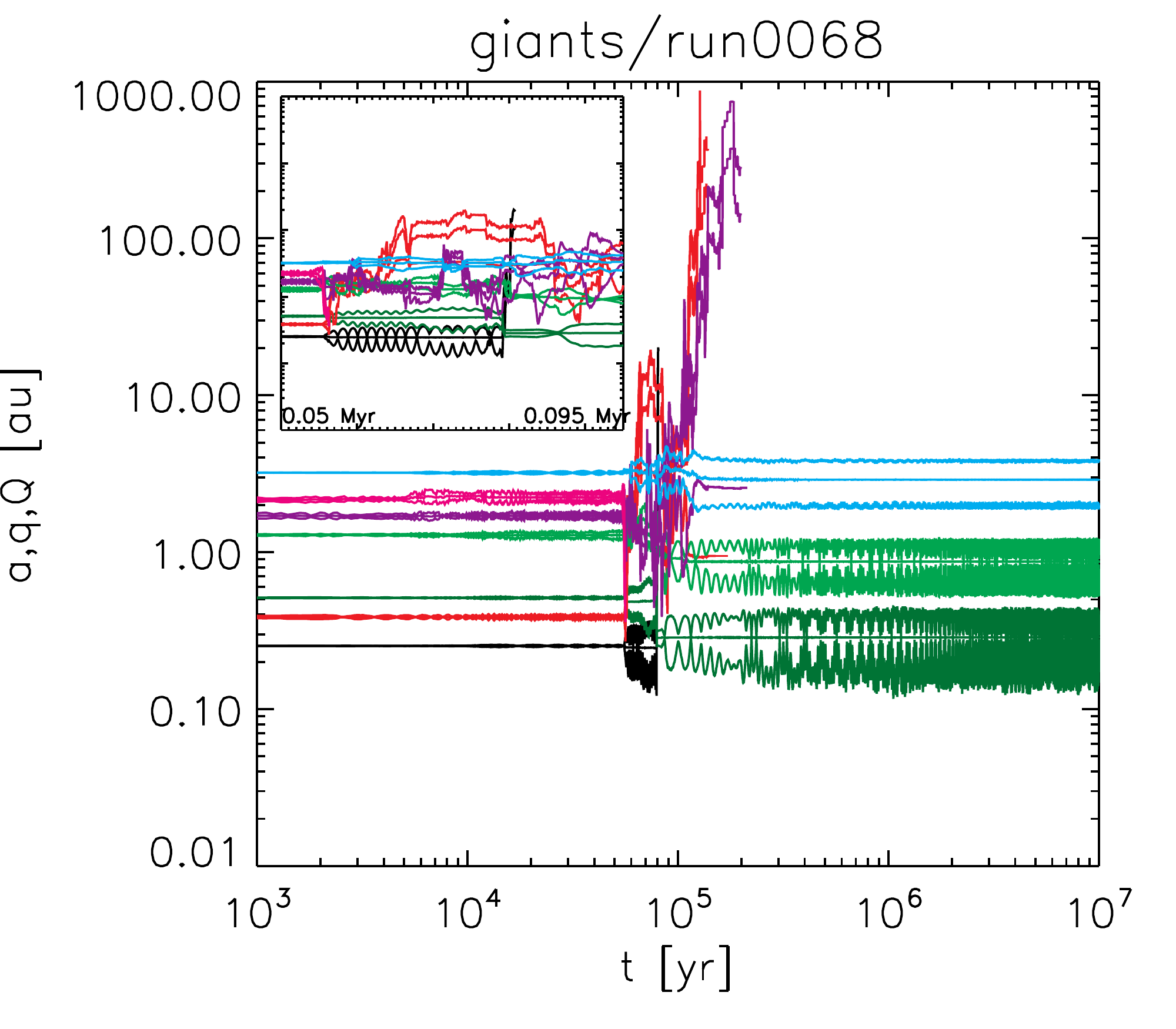}
  \includegraphics[width=.48\textwidth]{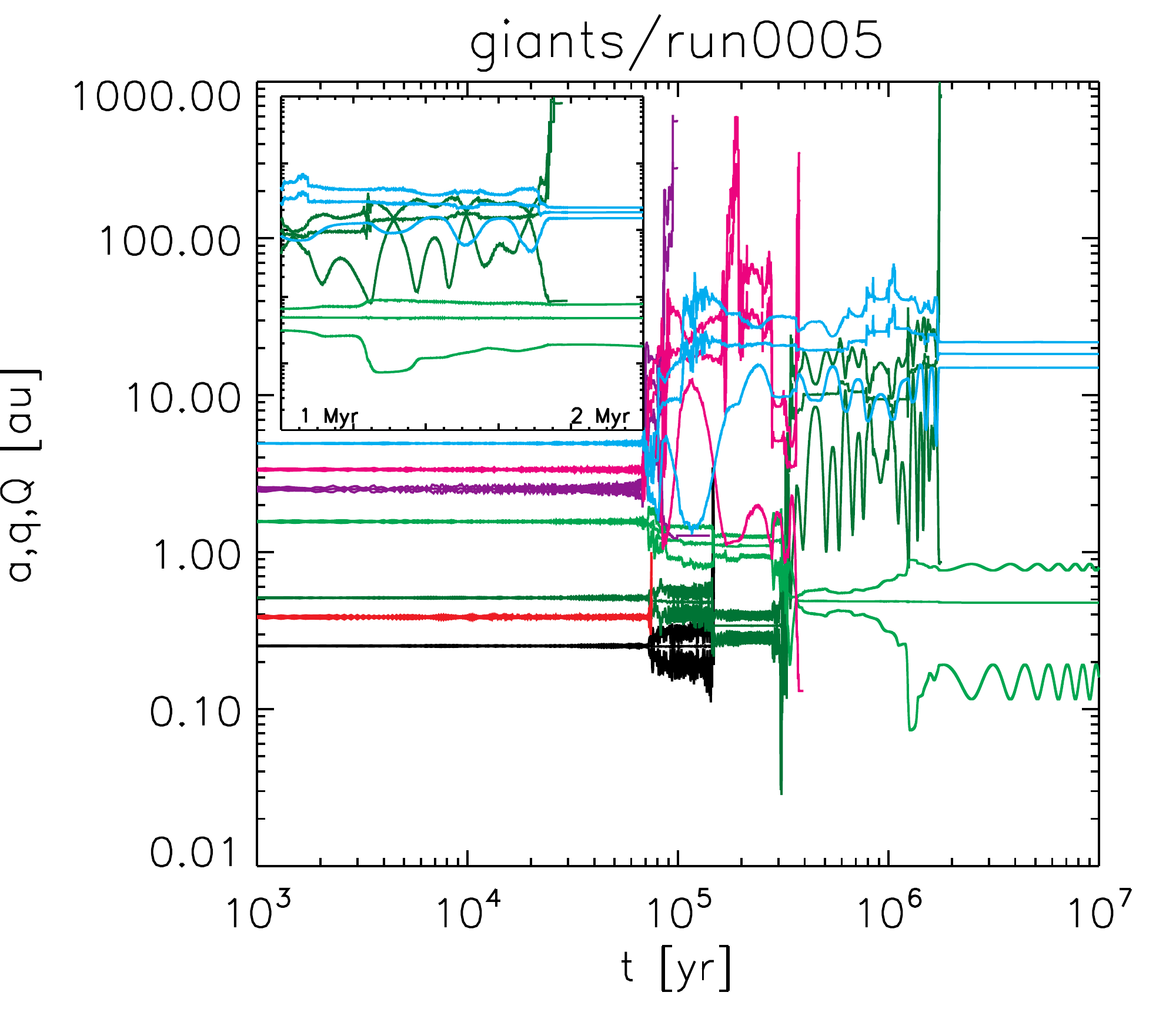}
  \caption{Formation of eccentric warm Jupiters by secular 
    processes. \textbf{Left: }As a result of planet--planet scattering, 
    significant angular momentum defict is generated, allowing 
    the inner planet's eccentricity to be forced up by secular 
    processes (initially at around $0.09$\,Myr; see inset). 
    Secular cycles continue for the remainder of the integration, 
    with irregular modulation. \textbf{Right: }After the initial 
    scattering ends at around $0.4$\,Myr, the warm Jupiter's 
    eccentricity is first forced by the second remaining planet (inset). 
    This forcing ends however when this planet is itself ejected by 
    the third survivor, causing the warm Jupiter to be ``frozen'' 
    into a high-eccentricity state, with the mutual inclination 
    between the two surviving planets at $\sim80^\circ$.}
  \label{fig:run0068}
\end{figure*}

\begin{figure*}
  \includegraphics[width=\textwidth]{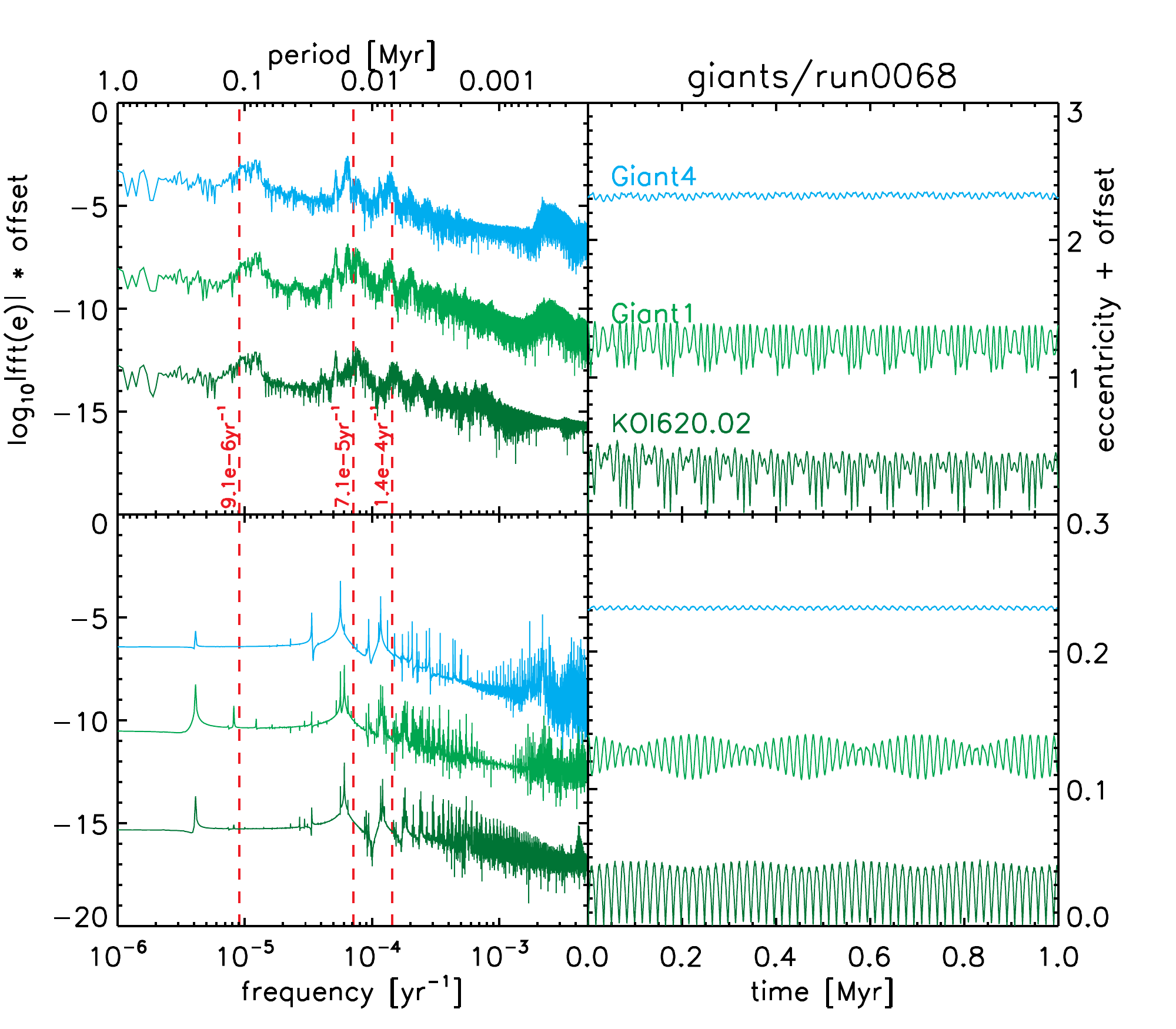}
  \caption{Further evolution the system shown in the left panel of Fig~\ref{fig:run0068}. 
    \textbf{Top: }The system is integrated for a further 10\,Myr; 
    the first 1\,Myr of eccentricity evolution is shown on the 
    right and the FFT of the full 10\,Myr on the left. The strongest peaks 
    are near the frequencies predicted by leading-order Laplace--Lagrange secular 
    theory (red), but the breadth of the peaks is indicative of 
    chaotic behaviour. \textbf{Bottom: }Here the system is initially identical 
    in all orbital elements but the eccentricity is reduced by a factor of 
    10. Peaks in the periodogram are now narrow and distinct, indicative 
    of regular quasiperiodic motion.}
  \label{fig:warmj-ecc}
\end{figure*}

Some warm Jupiters in our simulations are produced by a combination of 
secular processes and scattering. In \textsc{Giants/run0068}, planet--planet 
scattering generates significant angular momentum deficit which results 
in the three surviving planets having significant eccentricities, the inner two 
in particular experiencing large irregular oscillations (Fig~\ref{fig:run0068}). 
The chaotic nature of 
the subsequent secular evolution is revealed in Fig~\ref{fig:warmj-ecc}, where 
we extend the integration run for a further 10\,Myr. We compare the evolution 
of the system continued from the end of our simulation with one identical in all orbital 
elements except that eccentricities are reduced by a factor of 10. In the low-eccentricity 
case, power in the periodogram of the eccentricity evolution is concentrated at 
well-defined peaks close to the frequencies predicted by Laplace--Lagrange secular 
theory, characteristic of regular quasiperiodic motion. In contrast, 
in the high-eccentricity system that arises from our initial integration, these 
peaks are significantly broadened, characteristic of chaotic motion.

A second variant of secular warm Jupiter formation is shown in 
the right-hand panel of Figure~\ref{fig:run0068}. 
Here an unstable three-planet system arises from the initial scattering, with the 
warm Jupiter the innermost of the three. Its eccentricity is initially 
forced by the second of the planets during the latter's phases of high eccentricity, 
a process which ends when the third survivor ejects this planet at around $1.8$\,Myr. The 
warm Jupiter is then frozen into a high-eccentricity state, experiencing weak Kozai 
forcing largely suppressed by general relativistic precession
(mutual inclination of approximately $80^\circ$). This mechanism, 
whereby the warm Jupiter acquires its eccentricity by secular forcing 
from an outer system which is itself unstable and whose evolution ends 
following the ejection of all but one body, we dub ``freeze-out''.

\subsection{Eccentric warm Jupiters from \emph{in-situ} scattering and uplift}

\label{sec:uplift}

\begin{figure}
  \includegraphics[width=.5\textwidth]{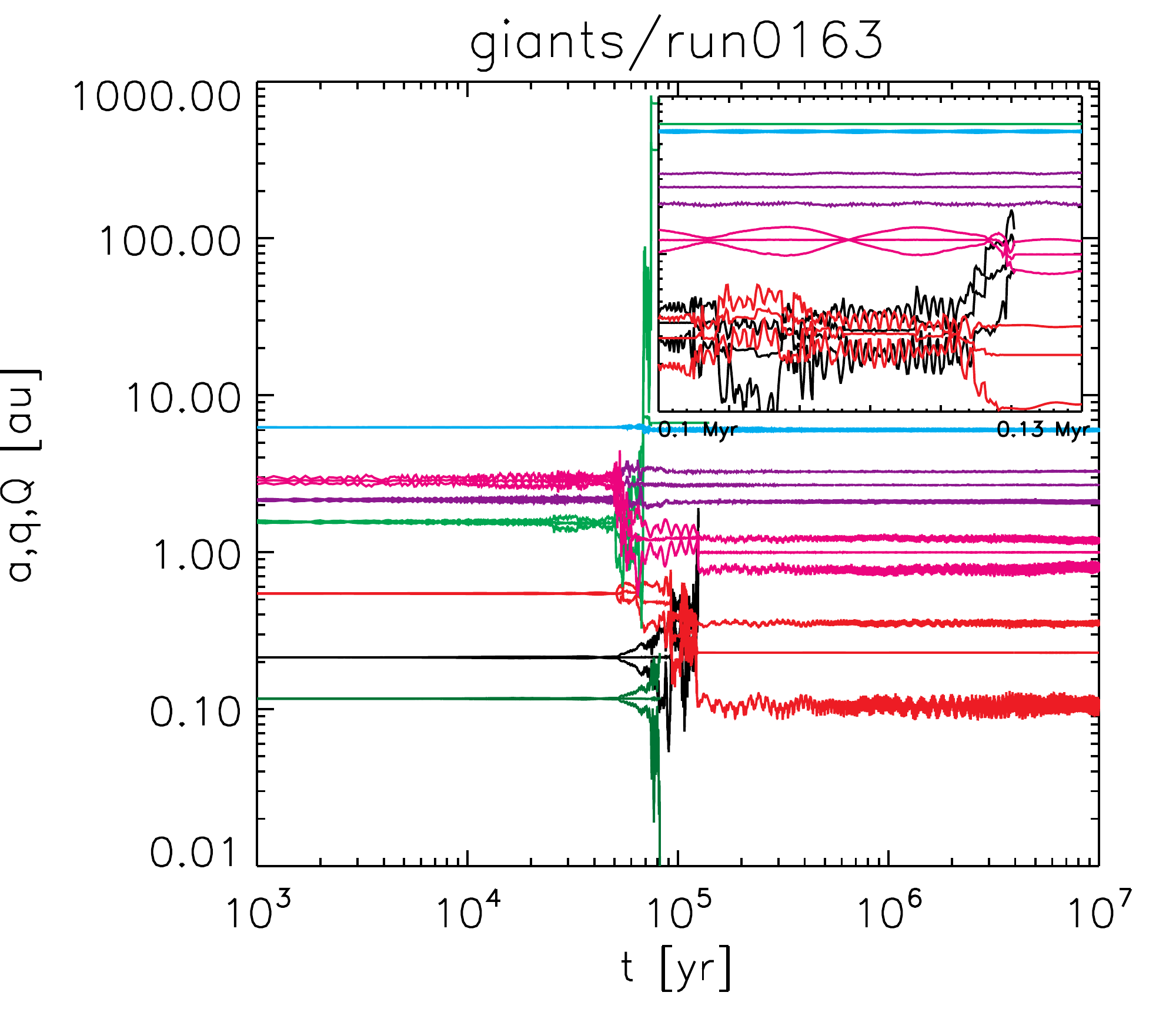}
  \caption{Formation of an eccentric warm Jupiter (pink; 
    second surviving planet) and an eccentric sub-Saturn 
    (red; innermost surviving planet). The initially 
    second and third (black and red) planets undergo 
    a phase of scattering between $0.1$ and $0.12$\,Myr, 
    which ends when the black planet interacts with the pink 
    planet just beyond 1\,au, raising its pericentre to detach 
    it from the innermost planet, a process we call ``uplift'' 
    (inset). The innermost sub-saturn is left with an eccentricity 
    fluctuating between $0.5$ and $0.6$, while the warm 
    Jupiter's eccentricity varies between $0.11$ and just over $0.3$.}
  \label{fig:uplift}
\end{figure}

Planets undergoing scattering at a few tenths of an au are 
inefficient at exciting high eccentricities or causing ejections, 
as their gravitational focusing is reduced owing to the high 
orbital speeds \citep{Petrovich+14}. Instability in such systems 
usually leads to collisions and relatively low eccentricities of the 
surviving planets. Indeed, the only eccentric warm Jupiters we 
formed directly by \emph{in-situ} scattering, which had $a$ just 
under $1$\,au, arose from giant 
outer planets scattering each other, not from scattering in the more massive 
of the inner systems.

However, we find that the addition of 
outer giant planets can enhance the rate of ejections of planets 
from the inner system. Of our 22 \textsc{Giants} systems modelled 
on KOI620 that were unstable, 13 of them ejected an inner planet. 
We ran 100 systems based on KOI620 with no outer planets and small 
separations between the inner planets, and another 100 similar with 
the mutual inclinations reduced to $0.1^\circ$ as in \cite{Petrovich+14}. 
Of these, only 12 out of 100 of the moderately-inclined and 9 out of 
100 of the coplanar systems ejected an inner planet, despite all being 
unstable: the majority of planets lost suffered collisions with 
other bodies. 

The presence of outer giants enhances the ejection rate by a 
process we call ``uplift'': when an inner planet attains a 
moderate eccentricity, it can begin scattering off the outer planets, 
which can raise its pericentre out of the inner system, and ultimately lead to 
its ejection, while quickly decoupling it from the other inner planets 
before a physical collision occurs. An example is shown in 
Figure~\ref{fig:uplift}, where the second and third planets scatter each other 
until one is lifted out of the inner system by one of the outer 
planets, with which it collides shortly afterwards. This leaves the innermost 
surviving planet with a significant eccentricity of $0.5-0.6$; if scattering 
had continued between only the inner two planets; the probable outcome would 
have been a collision resulting in a lower eccentricity for the merger product. 
Although in this case the innermost planet is too low in mass 
($1.4\times10^{-4}\mathrm{\,M}_\odot$) to qualify as 
a warm Jupiter (the warm Jupiter here is actually the second planet, which 
just meets our criterion at $a=0.997$\,au), this process should work 
more efficiently if the inner planet were higher in mass, as scattering 
to the outer system would then be easier, making this a viable route to 
produce eccentric warm Jupiters.

\begin{figure*}
  \includegraphics[width=\textwidth]{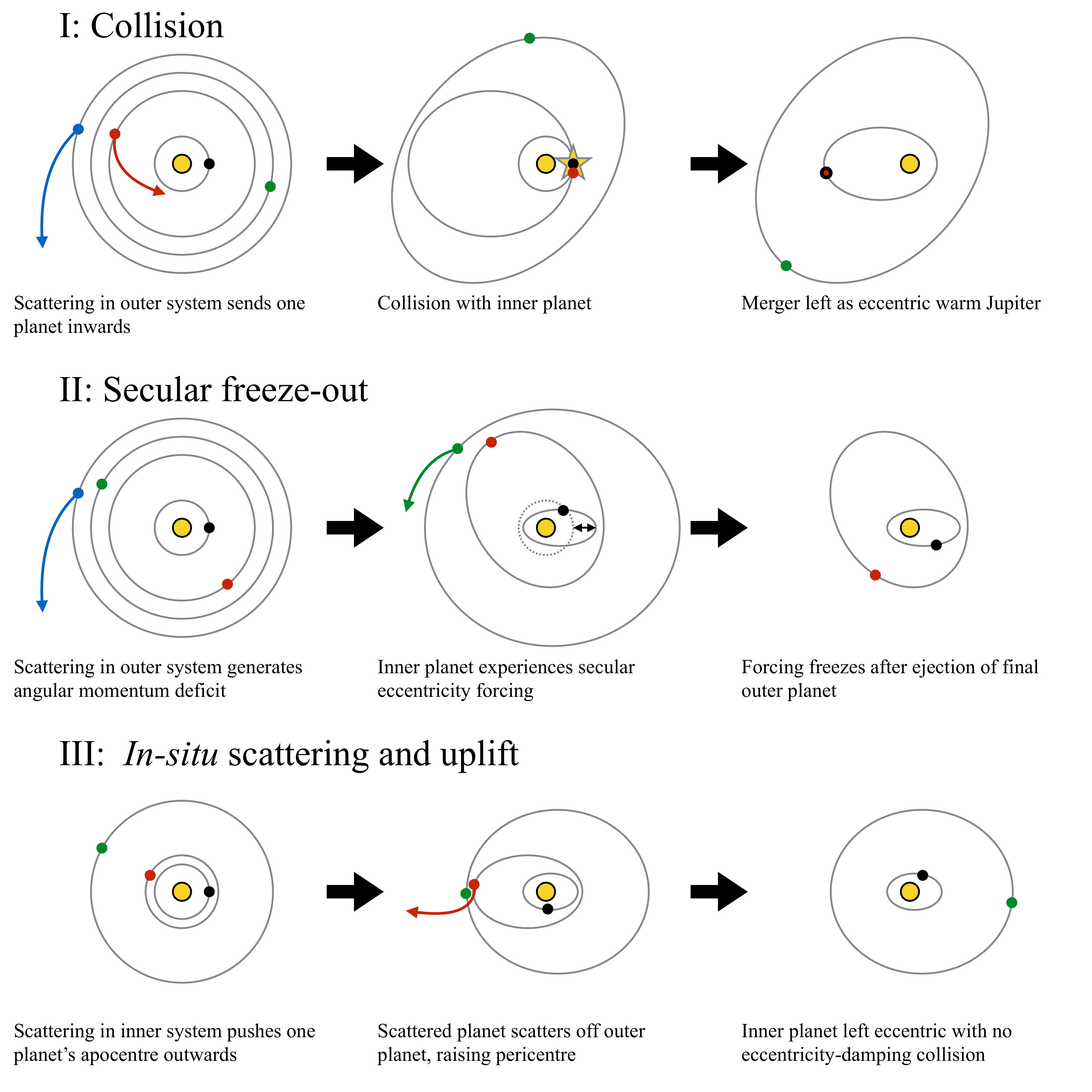}
  \caption{Cartoon summarising routes to the formation 
    of eccentric warm Jupiters. \textbf{Top: }An outer 
    giant is scattered into the inner system, where it 
    collides with a close-in planet to lower its 
    eccentricity (see Fig~\ref{fig:collide-warmj}, 
    Section~\ref{sec:collide-warmj} and 
    \citealt{Mustill+15}). \textbf{Middle: }Scattering 
    in the outer system generates angular momentum 
    deficit (AMD), which periodically excites 
    the inner planet's eccentricity (see 
    Fig~\ref{fig:run0068} and Section~\ref{sec:secular}). 
    If scattering continues in the outer system, the inner 
    planet's eccentricity can be frozen into a high value after 
    planetary ejections cease. \textbf{Bottom: }Scattering in the inner 
    system leads to one planet's apocentre being raised enough 
    to interact with outer giants, which then decouples this 
    planet from the inner system and prevents a collision 
    with the other inner planet, which would generally 
    reduce eccentricity (see Figure~\ref{fig:uplift} and 
    Section~\ref{sec:uplift}).}
  \label{fig:cartoon}
\end{figure*}

\section{Discussion}

\label{sec:discuss}

\subsection{Relation to previous work}

We have studied several aspects of the interactions between 
outer planetary systems beyond 1\,au and inner systems such 
as those discovered by \textit{Kepler}. Our present work 
builds on the study of \cite{Mustill+15}, wherein we showed 
that highly-eccentric giant planets \emph{en route} to 
becoming hot Jupiters would clear inner planets out 
of the inner system. While in our previous study we imposed 
the high eccentricity at the beginning of the integration, 
in our present study we allow the eccentricity to arise 
naturally as a result of the dynamics of the outer system. 
We verify that these highly-eccentric planets also clear 
out their inner systems when we model the dynamics 
consistently: in our \textsc{Binaries} integrations, 
all 11 systems where the outer planet both attained a 
pericentre $<0.05$\,au and survived to the end of the 
integration lost all other planets in the system.

A number of other authors have considered aspects 
of the effects of the dynamics of outer systems on 
inner ones. The study of the effects of planet--planet 
scattering on inner terrestrial planets in the 
habitable zone has a venerable tradition 
\citep[e.g.,][]{VerasArmitage05,VerasArmitage06,Raymond+11,
  Raymond+12,Matsumura+13,Carrera+16,KaibChambers16}. 
Our inner systems, being modelled on \textit{Kepler} 
systems, are often closer to the 
star than the habitable zone, but an extrapolation 
of our \textsc{Giants} results suggests that $\sim40\%$ 
of habitable-zone planets in unstable systems of 
giant planets would belong to inner systems 
that were themselves destabilised 
(Figure~\ref{fig:funstable}). This is somewhat 
less than the hierarchical case (4Gb+4e) studied by 
\cite{Carrera+16} where $70\%$ of habitable-zone planets 
were destabilised; most of this discrepancy is 
accounted for by the 141 of our outer systems that 
had not lost a planet during the course of the integration. A 
recent work by \cite{Huang+16b} studies the effects of 
scattering among giant outer planets on super-Earth 
systems, similar to our \textsc{Giants-selected} runs, 
finding a $70-80\%$ destabilisation rate, almost twice 
the value of ours. This may be attributable to their 
outer systems being closer to the inner systems than 
are ours.

Several groups have studied the effects that outer 
planets might have on the inclinations of inner 
systems. \cite{LaiPu16} find that an inclined 
outer planet can excite the mutual inclinations 
of \textit{Kepler} systems, while \cite{Hansen16} 
finds that this process is greatly enhanced with 
two dynamically-excited outer planets, which can 
land secular resonances in the midst of the 
\textit{Kepler} zone. \cite{GratiaFabrycky17} point out, 
in the context of Kepler-56, that such inclined outer 
planets can lead to an increase of the obliquity 
of an inner system without exciting mutual 
inclinations. Our study complements these works 
by treating consistently the origin of the eccentricities 
and inclinations of the outer planet(s).

We now proceed to discuss in more detail the effects 
on the multiplicities of \textit{Kepler} systems, 
both in their intrinsic numbers of planets 
(Section~\ref{sec:kepler-multi}) and in their mutual 
inclinations (Section~\ref{sec:kepler-inc}); the 
generation of significant obliquities and mutual 
inclinations as seen in Kepler-56 and Kepler-108 
(Section~\ref{sec:inc}); 
and the formation 
of eccentric warm Jupiters (Section~\ref{sec:discuss-warmj}); 
before proceeding to critique our choice of initial 
conditions (Section~\ref{sec:initial}), discuss neglected 
physical processes (Section~\ref{sec:neglect}) and 
consider the effects of long-term dynamical evolution 
(Section~\ref{sec:longterm}).

\subsection{Sculpting the Kepler multiplicity function}

\subsubsection{Intrinsic multiplicities}

\label{sec:kepler-multi}

The multiplicities of systems observed by \textit{Kepler} depends on 
both the intrinsic multiplicity of the system and the mutual inclinations 
between the planets: a system with many planets but high mutual inclinations 
is more likely to only be observed as a single. \cite{Johansen+12} showed that 
the apparent excess of single-transit systems cannot be explained by 
varying the mutual inclinations of intrinsically triple-planet systems. Nor can 
internal dynamics of the triple-planet systems reduce multiplicity, as the surviving 
triple-planet systems are much too stable to be the survivors of a population with 
a continuous range of stability times, in contrast to higher multiplicity 
systems \cite{PuWu15}.

Given the prevalence of binary stellar companions and wide-orbit planets, 
we have explored the dynamical effects of these on \textit{Kepler}-detected inner systems. 
We find that a reduction of the mutliplicity of the inner system occurs 
in $20-25\%$ of systems in our population syntheses. If we restrict our attention 
to a subsample of scattering planets chosen to have an 
eccentricity distribution matching observed exoplanets, the disruption rate 
rises to $\sim40\%$; a similar rate is found for the systems 
with outer planets forced by binary companions to $e>0.5$. 
This is insufficient to explain the large excess of 
single-planet candidates from \textit{Kepler,} especially when we consider that not all 
such systems will possess the necessary outer architectures.

However, violent dynamics in the outer system does make some contribution 
to the multiplicity function of inner systems. Assuming that the 
occurrence rates of system components are independent, 
we make the following estimates: For \textsc{Binaries}, 
around 25\% of stars have a wide binary companion \citep{DucheneKraus13}, 
and microlensing suggests that $\sim50\%$ of stars has a wide-orbit 
Neptune-mass planet or above. If 25\% of such systems destabilise their 
inner systems, that gives a fraction of $\sim3\%$ of inner systems 
destabilised by Kozai cycles induced on outer planets. For \textsc{Giants}, 
taking a giant planet occurrence rate of 20\% as found by 
\cite{Uehara+16}, noting that $\gtrsim75\%$ 
of these may have undergone instability, and $40\%$ of these disrupt 
their inner systems (working with \textsc{Giants-selected}), we now 
end up with 6\% of inner systems having been disrupted by an unstable 
outer system of giant planets.

These rates are clearly too low to reproduce the excess of single-planet 
\textit{Kepler} systems. 
We note that unexplored architectures may raise this rate. In particular, 
we have not explored unstable systems of low-mass outer planets (almost 
all of our \textsc{Giants} runs possess at least one Saturn-mass planet). 
Scattering instabilities in low-mass 
systems take far longer to resolve than in high-mass systems 
\citep{Mustill+16,Veras+16}, and can result in large excursions in 
eccentricity and pericentre \citep{Veras+16}. While super-Earths or 
Neptunes penetrating the inner system directly would be less damaging 
than gas giant planets, and would probably simply result in the ejection 
of the intruder \citep{Mustill+15}, the longer timescales on which 
the scattering occurs would give more time for moving secular 
resonances to act on the inner system. If we assume that systems 
of unstable low-mass planets would be as disruptive as our \textsc{Giants} 
systems, we would estimate (raising the outer planet occurrence rate from 
$20\%$ to $50\%$) that 15\% of inner systems are disrupted 
this way, for a total of 18\% when adding the effects of the \textsc{Binaries} 
run: a significant contribution to the \textit{Kepler} multiplicity 
function, but insufficient by itself to resolve the \textit{Kepler} Dichotomy. 
The ratio of single to multiple planet \textit{Kepler} systems is around 
4:1 \citep{Johansen+12}, but as we find that only 18\% of triple-planet systems 
are expected to be disrupted, this leaves over 75\% of the single \textit{Kepler} 
planets unaccounted for. 
From the outcomes in Table~\ref{tab:nsurv}, we find about $10.6\%$ of 
\textit{Kepler} triple-planet systems would be reduced to single-planet systems and 
around $6.0\%$ would lose all their planets. Unstable giants 
would contribute the most to this destabilisation, but 
binaries contribute more to the zero-planet systems than they do to the 
single-planet systems.

\subsubsection{Mutual inclinations of inner planets}

\label{sec:kepler-inc}

The observed multiplicity also depends on the mutual inclinations 
of the inner planets. Several recent papers have suggested that 
inclined outer planets can induce mutual inclinations in the 
inner systems that could help to generate an overabundance of single 
planet candidates and resolve the Dichotomy. 
\cite{LaiPu16} recently argued that inclined outer companions to \textit{Kepler} multiple 
systems can excite large mutual inclinations through secular perturbations, leading to a large population of 
single-transit systems. We can look to see whether this effect occurs in our $N$-body runs.

Secular inclination forcing in the inner system depends on the strength of the coupling between 
inner planets compared to the forcing from the outer planet. Strong coupling between the inner 
planets means that high mutual inclinations cannot be excited. \cite{LaiPu16} parametrise this 
coupling with a parameter $\epsilon$ (their $\bar\epsilon$, Eq 29), where $\epsilon\ll1$ means strong coupling 
and $\epsilon \gg1$ means weak coupling. Secular resonances occur at $\epsilon\approx1$ where 
very high mutual inclinations can be excited. However, in our integrations all of 
the systems had $\epsilon<1$.

To attain $\epsilon\ge1$ the outer planets would have to be brought closer to the inner systems. 
While this is possible (note the gaps between the inner and outer systems in 
Fig~\ref{fig:a-m}), the 
inclination of the outer body or bodies with respect to the inner system must still be excited. 
It is possible that in this case the inclinations of the inner planets would be directly 
excited by scattering during the excitation of the outer system. 

We do find a small number of mutually-misaligned two-planet survivors from the 
initial triple-planet systems (Figure~\ref{fig:imut}): as in \cite{SpaldingBatygin16}, 
who argued for secular inclination driving as a young star spins down, 
and \cite{Hansen16}, who treated less violent outer system dynamics than we, 
significant inclinations are concomitant with dynamical instability. 
However, these systems are a small fraction of our full set of runs and 
cannot contribute significantly to the \textit{Kepler} Dichotomy.

\subsection{Tilting and strongly misaligning inner systems}

\label{sec:inc}

However, in occasional cases these inclination effects are of interest. 
We draw comparisons to two observed systems of interest: Kepler-56 
\citep{Huber+13} and Kepler-108 \citep{MillsFabrycky16}.

\begin{figure}
  \includegraphics[width=.5\textwidth]{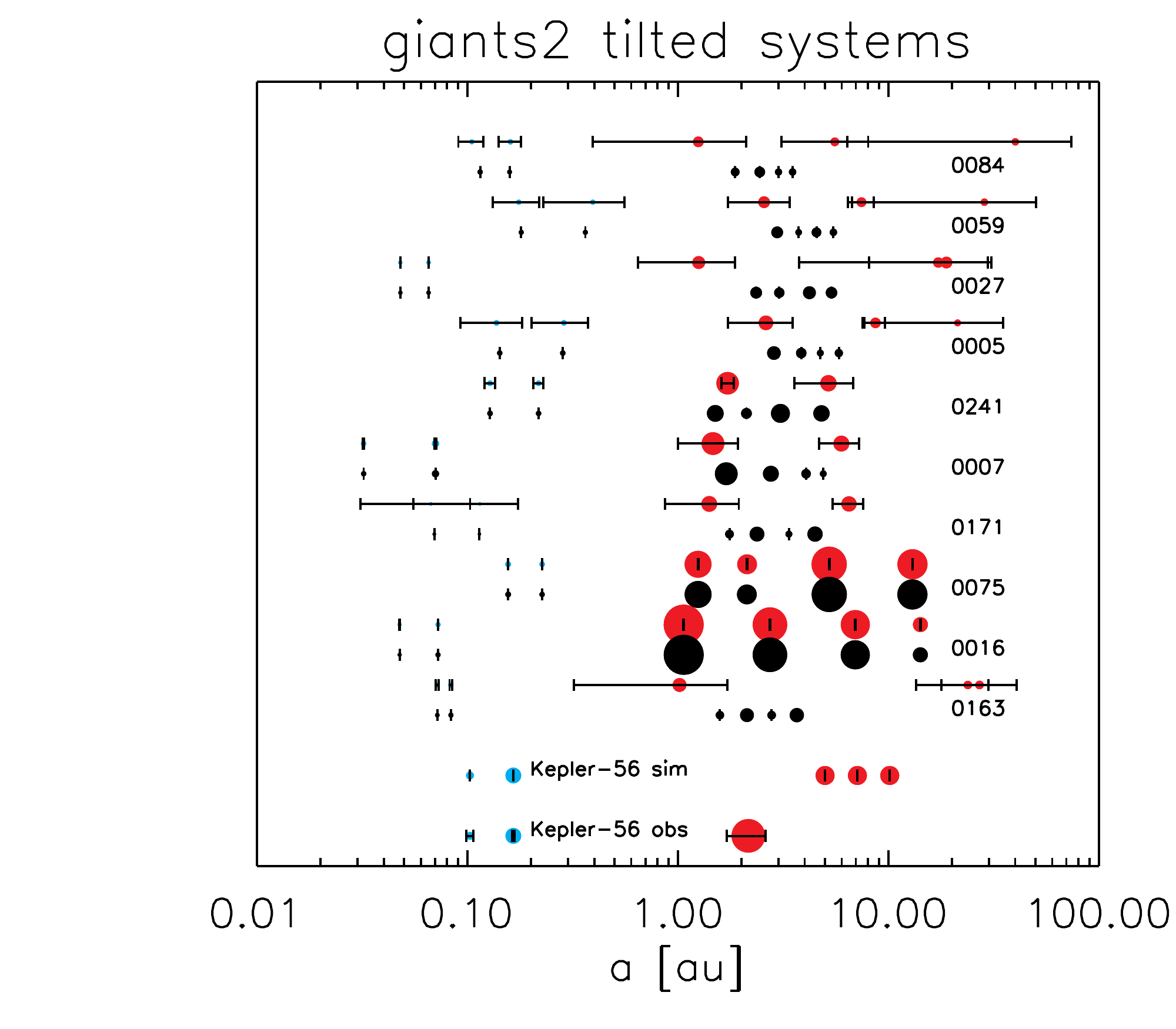}
  \caption{Systems of initially two inner planets resulting in 
    orbits coplanar with each other but inclined away from 
    the original reference plane, similar to Kepler-56, from the 
    \textsc{Giants2} simulations. The bottom rows show 
    the observed Kepler-56 system from \protect\cite{Otor+16}, 
    as well as the initial conditions for the simulations 
    of \protect\cite{MillsFabrycky16}. In the two \emph{stable} 
    simulations with four surviving outer planets, the inner 
    planets lie close to strong secular inclination 
    resonances.}
  \label{fig:kepler56}
\end{figure}

Kepler-56 is a system of two transiting planets \citep{Huber+13} with a third 
planet at detected by RV at $2.1$\,au, and likely no other giant 
planets within 20\,au \citep{Otor+16}. The two inner planets have 
a misalignment with the host star of at least $45^\circ$ \cite{Huber+13}, 
while having a low mutual inclination; the inclination of the outer 
planet is unknown. \cite{Li+14a} showed that the high obliquity 
of the inner planets could be explained by an inclined outer giant 
planet; we have shown that this can indeed occur naturally, albeit 
somewhat rarely, as a result of scattering in the outer system. 
Successful examples (mutual inclination of the inner planets $<10^\circ$, 
and an absolute inclination of the innermost $>20^\circ$, at 
the end of the integration) from our \textsc{Giants2} simulations are shown in 
Figure~\ref{fig:kepler56}, together with the observed system. 
\cite{GratiaFabrycky17} also
successfully achieved the mutual inclinations with scattering 
between three equal-mass giant planets, as marked on the Figure. 
The equal mass case 
results in the most violent instability, and doubtless helps 
to leave the surviving giant with the necessary high inclination. 
\cite{Otor+16} place limits on undetected planets of $0.5\mathrm{\,M_J}$ 
at 10\,au and $2\mathrm{\,M_J}$ at 20\,au, although this assumes a 
circular orbit for such a planet. Surviving outer planets (in our eight 
unstable systems) have masses ranging from $0.05$ to $6\mathrm{\,M_J}$, 
and in no system was only one outer planet left. This suggests 
that one or more sub-Saturn mass planets could exist on wide 
orbits in Kepler-56, although our sample is too small to draw 
firm conclusions.

Kepler-108 is a two-planet system in a binary with a high mutual inclination
of $15-60^\circ$ measured from TTVs, whose dynamics was recently
analysed by \cite{MillsFabrycky16}. They note that the system is at present
too strongly coupled for the binary to have excited the mutual
inclination, and speculate that the binary might have thrown in
an outer planet to excite the inclination. We show that this is
possible but rare; an alternative route is to start with an
extra planet in the inner system as well as the outer, which
then leads to  higher mutual inclination excitation (Figure~\ref{fig:imut}).

\cite{Campante+16} recently found that, among 16 multi-planet 
systems whose stellar obliquities were determined through 
asteroseismology, none had significant misalignment 
between the stellar and orbital angular momenta. While these 
numbers are still small, 
the next generation of space-borne transit observatories 
(TESS, \citealt{Ricker+15}; CHEOPS, \citealt{Broeg+13}; 
and PLATO, \citealt{Rauer+14}) are set to offer both improved cadence 
over \textit{Kepler,} and better amenability to ground-based 
follow-up: we can expect future statistical studies to probe 
the incidence of tilted and misaligned systems, which may, 
in conjunction with further theoretical studies, constrain 
the frequency of the violent outer system dynamics we have considered 
in this paper.

Implicit in this discussion is the assumption that 
there is initially no significant misalignment between the stellar equator 
and the orbital plane of the planets, a situation which will arise 
if the planets form in a disc whose angular momentum vector is 
aligned with the stellar spin axis. This is not necessarily a given, 
particularly for the wide binaries we consider, as a binary 
companion misaligned with a protoplanetary disc can drive significant 
inclination fluctuations in the disc and any planets forming within it 
\citep{Batygin12,PicognaMarzari15}. Observational evidence shows that 
at least some wide binaries, such as the IRS~43 system 
\citep[$a_\mathrm{B}=74$\,au][]{Brinch+16} 
and the HK~Tau system \citep[projected separation $\approx400$\,au][]{JensenAkeson14}, 
possess discs that are misaligned with each other. The angles between these 
discs and their stellar spin axes have not been determined, however. 
An initially misaligned disc may also be present around a single star, 
due to magnetic interactions between the star and the protoplanetary disc 
\citep{Lai+11} or to the infall of material in a turbulent 
environment with angular momentum vectors different from that of the star 
\citep[e.g.,][]{Fielding+15}.

\subsection{Formation of eccentric warm Jupiters}

\label{sec:discuss-warmj}

Excitation of the eccentricities of warm Jupiters through 
\emph{in-situ} scattering is challenging \citep{Petrovich+14}. 
Current explanations favour secular processes, particularly 
``stalled'' Kozai migration where an eccentric warm Jupiter 
is essentially a slowly-migrating proto-hot Jupiter that continues 
to experience Kozai cycles as tidal dissipation acts to shrink its orbit 
\citep{DawsonChiang14,Dong+14,Petrovich15b,PetrovichTremaine16}; where the outer perturber 
is a planet, as in \cite{DawsonChiang14} or 
\cite{PetrovichTremaine16}, a means of initially 
exciting the mutual inclination is required. The perturbers described 
by \cite{DawsonChiang14} are close (few au) and apsidally 
misaligned. Our eccentric warm Jupiters are accompanied by 
zero, one or two outer giant planets, at a range of separations 
(see Fig~\ref{fig:lovisplot-warmj}). More 
extensive and dedicated work will be required to fully 
predict the orbits of outer companions to eccentric 
warm Jupiters formed through the mechanisms we have 
identified, but we note that our systems with wide-orbit 
($\sim20$\,au) or no outer companions may explain those 
eccentric warm Jupiters with no companion as yet 
detected.

We note that only two of the eccentric warm Jupiters we form 
retain their inner companions; both of these warm Jupiters have $a$ 
close to 1\,au. Other warm Jupiters we formed, even with similar $a$, 
destroyed their inner planets, including those within $0.1$\,au 
(third and fourth rows of Fig~\ref{fig:lovisplot-warmj}). 
\cite{Huang+16} showed that some, but not all, 
\textit{Kepler-}detected warm Jupiters have companions, while 
\cite{Dong+14} and \cite{Bryan+16} found that RV-detected warm Jupiters are more 
likely to have Jovian companions the higher their eccentricity. This points 
to many warm Jupiters (considering all eccentricities) having 
undergone low-eccentricity disc migration or \emph{in-situ} formation; 
those warm Jupiters that have experienced high eccentricities during 
their evolution will have cleared out their inner systems, in a manner similar to 
hot Jupiters undergoing high-eccentricity migration \citep{Mustill+15}.

Models of the formation of warm Jupiters can be tested by 
comparing the eccentricity distributions of modelled and 
observed populations \citep[e.g.,][]{PetrovichTremaine16}. 
Here we defer a quantitative comparison to a future study, as 
we suffer from small number statistics in our model population. 
However, we show the semimajor axes and eccentricities of 
all warm Jupiters at the end of our \textsc{Giants} simulations 
in the lower panel of Figure~\ref{fig:lovisplot-warmj}. We find a 
population of low-eccentricity warm Jupiters at $a\approx0.5$\,au 
which remain on orbits close to their initial ones. This is 
qualitatively in agreement with \cite{PetrovichTremaine16} who 
found that high-eccentricity migration produces few low-eccentricity 
warm Jupiters, and suggests that disc migration contributes 
significantly to the low-eccentricity population. Similarly, 
\cite{Antonini+16} argue that most warm Jupiters acquire their orbits 
through disc migration, based on the likely dynamical instability 
of the pre-migration configurations of warm Jupiters with companions. 
We also note that further clues to the origins of warm Jupiters may come 
from their atmospheric compositions: Jovian planets which accrete most 
of their gas at large radii beyond the ice line, before 
later dynamical migration,  
are expected to have higher atmospheric C/O ratios than those 
which migrate through the disc and accrete significant gas 
interior to the ice line \citep{Madhusudhan+16}. 
Measuring a high atmospheric C/O ratio of a warm Jupiter would therefore 
imply a dynamical formation mechanism from outside the ice line 
(inelestic collision or slow Kozai migration), while a low C/O ratio 
would be consistent with disc migration, followed by the secular freeze-out 
or \emph{in situ} scattering plus uplift mechanisms if the eccentricity 
has been excited. This would thus allow some discrimination between different 
dynamical formation models on a per-system basis, complementing the statistical 
approach afforded by the eccentricity distributions.

\subsection{Initial conditions}

\label{sec:initial}

As with any $N$-body study, the initial conditions for our integrations 
require defending. The main issues are (i) the occurrence rate of the 
configurations studied (Kozai in binaries and planet--planet scattering) 
and (ii) how physically motivated are the initial conditions.

We expect to find configurations such as we have studied around a 
significant minority of stars. Around 50\% of Solar-type stars 
are in binaries, and half of these again are wider than 
a few tens of au \citep{DuquennoyMayor91,Raghavan+10,DucheneKraus13}. 
Our binary population thus represents a little under a quarter 
of all Solar-type stars (we miss the very wide binaries beyond 1000\,au, 
which are less dynamically interesting: see 
Figure~\ref{fig:hj-formation-binaries}, but cf \citealt{Kaib+13}). We 
also note that ``binarity'' is not a fixed property of a system, 
and at young ages stars may freely exchange into and out 
of multiple systems in their birth cluster \citep[e.g.][]{Malmberg+07b}. 
The frequency of outer giant planets is harder to constrain. 
We assume, perhaps optimistically, that wide-orbit giant planets are as frequent around 
components of wide binaries as around single stars. \cite{Mayor+11} 
estimated an ocurrence rate of 14\% for planets more massive than 
$50\mathrm{\,M}_\oplus$ within 5\,au. \cite{Wittenmyer+16} 
find a frequency of ``Jupiter analogues'' ($a\in[3,7]$\,au, 
$e<0.3$, and mass greater than Saturn's) of 6\%, while 
\cite{Rowan+16} obtain the slightly lower value of 
3\%, albeit with a slightly more restrictive definition. 
Both transit and RV studies of the inner system reveal that 
the occurrence rate of planets rises strongly with decreasing 
mass, and indeed recent microlensing results put the 
occurrence rate of snow-line planets with a mass ratio greater 
than $10^{-5}$ at 55\% \citep{Shvartzvald+16}. As this 
corresponds to our lower mass limit for the planets in 
\textsc{Binaries}, our simulations may represent 
around 1 in 8 of all inner systems.

This all assumes that the occurrence rates of various components 
are independent. In reality, it has long been suspected 
that binary companions would suppress planet formation, 
\citep[see][for a review]{ThibaultHaghighipour15}, 
although observational evidence of this is 
ambivalent \citep{Wang+14,Deacon+16,Kraus+16}. \cite{Wang+14} 
found a reduction in the frequency of planets detected by 
\textit{Kepler} of $4.5\pm3.2$ for $\sim10$\,au binaries, $2.6\pm1.0$ 
for $\sim100$\,au and $1.7\pm0.5$ for $\sim1000$\,au. 
\cite{Deacon+16} find no evidence for an effect of 
very wide ($\gtrsim3000$\,au) binaries on planet occurrence, 
while \cite{Kraus+16} find a suppression of a factor three 
in occurrence for binaries within $\sim50$\,au, arguing 
that the statistics are not good enough to justify more complex 
models. \cite{Zuckerman14} argued for an impact of binaries within 
$\sim1000$\,au on the formation or long-term stability of 
planetary systems around A-type stars, based on the occurrence 
of metal pollution in their descendant white dwarf atmospheres. 
One could also query whether the planetary systems forming 
in the protoplanetary disc in these binary systems will emerge 
from the disc phase in a coplanar configuration as we have 
assumed. While \cite{Batygin+11} and \cite{Batygin12} 
argued that in many cases of 
interest a disc should precess as a rigid body, 
\cite{PicognaMarzari15} find that a planet will decouple 
from its disc in a relatively short time. 

The initial conditions for our \textsc{Giants} runs 
also require justification. The frequency of multiple 
planet systems at several au is even more poorly constrained than 
the frequency of systems with \emph{any} planet. However, there 
are suggestions that giant planets preferentially form in 
multiples. \cite{Bryan+16} find that 50\% of stars hosting 
one planet (usually a giant) have a wider ($5-20$\,au) 
companion. More importantly, the eccentricity distribution 
of giant planets shows strong indications of scattering 
such as we have considered: \cite{JuricTremaine08} and 
\cite{Raymond+11} found that the eccentricity distribution 
requires a contribution of $\sim80\%$ of unstable systems. 
That is, the majority of giant planets form in unstable 
multiple systems. Do our systems resemble the real ones? 
We do not initially try to tune our systems to be in mean-motion 
resonances, a configuration which na\"{\i}ve models of 
planet migration produce frequently \citep[e.g.,][]{LeePeale02}. 
However, as we discussed above, the \textit{Kepler} planets do \emph{not} 
appear to reside in these resonances, the reasons for which are not 
yet clear. Gas giant systems such as HR~8799 may however 
be genuinely resonant (e.g., \citealt{FabryckyMurray-Clay10}, but see 
\citealt{Gotberg+16} for a dissenting view). 
Hence, although migration into resonances can be 
incorporated into models of scattering 
\citep[e.g.,][]{LibertTsiganis11,Sotiriadis+17}, 
this may not accurately represent the configurations of systems 
at the end of planet formation. Furthermore, 
recall that most multiple giant systems have probably undergone 
instability, and so either have not been protected for the long term by 
resonant lock, or the initial resonant configuration was unstable 
and the initial conditions will be rendered immaterial once 
strong scattering begins. Hence, we believe that although 
in some sense arbitrary, our choice of tight-packed planets 
with no attention paid to orbital phases has little impact 
on the nature of the scattering.

One issue with our simulations appears to be that they are too 
hierarchical: our scattering population is of lower eccentricity 
than the observed population (Figure~\ref{fig:edist}). 
Fortunately our sample size was large enough to draw a sample 
with a correct eccentricity distribution (\textsc{Giants-selected}), 
which in fact contains many of our most interesting runs, being 
more disruptive of the inner systems and forming eccentric 
warm Jupiters at a much higher rate (Figure~\ref{fig:lovisplot-warmj}). 
In future work the low eccentricities could be rectifed by 
imposing a correlation between planet masses. This and the other 
objections in this section could be overcome by combining the $N$-body 
dynamics with a planet formation and migration model. 
This would, however, commit one to a particular formation model, possibly 
locking one out of interesting regions of parameter space, and as 
discussed in the context of resonances the physics of the 
formation/migration is not yet fully understood.

As with outer binary stars, the presence of massive outer planets 
may hinder or suppress planet formation before the systems reach the starting 
point of our simulations. An eccentric massive planet can excite the relative 
velocities of planetesimals or planetary embryos, hindering their continued 
collisional growth \citep[e.g.,][]{MustillWyatt09,BatyginLaughlin15}. In the context 
of the new pebble accretion model for planetary growth 
\citep{JohansenLacerda10,OrmelKlahr10,LambrechtsJohansen12}, planets can 
reduce the radial pebble flux through the protoplanetary disc 
\citep{MorbidelliNesvorny12,LambrechtsJohansen14,Lambrechts+14}, which may reduce 
the availability of solids to form planets in the inner disc. Thus, the 
characteristics, or even the presence, of planets in the inner system 
may depend on the presence or characteristics of planets in the outer system. 
Our simulations (\textsc{Binaries} as well as \textsc{Giants}) may 
thus only be probing a restricted set of valid initial conditions.

\subsection{Further unmodelled processes}

\label{sec:neglect}

\subsubsection{Collisions} 

In common with most $N$-body simulations, we have assumed that planet--planet collisions result in perfect 
mergers and do not make allowance for different collision outcomes (hit-and-run, erosive, disruptive) 
as a function of impact parameter, planet 
size, and collision velocity \citep[see][for an overview]{LeinhardtStewart12}. 

Perfect mergers are a good approximation when the encounter velocity is less than the planets' escape velocity, 
which is the case for our outer giant planets. However, amongst the \textit{Kepler} systems the Keplerian velocities 
are very high ($\sim100$\,km\,s$^{-1}$ at $0.1$ au) and impact velocities can easily exceed the escape velocity, 
which may provide a means to further reduce the number of planets in high-multiplicity systems as planets are 
ground into small fragments following collisions \citep{VolkGladman15}. For our systems, changing the collision 
prescription will not affect the majority of systems which do not experience collisions in the inner system, but 
may strongly affect the multiplicities of systems where collisions \emph{do} occur; 
$40-50\%$ of our inner planets lost are lost to planet--planet collisions, and while they may only 
make a small contribution to the planet population as a whole, the effects may be seen in the 
mass--radius relation for individual objects, as is suggested for example for Kepler-36 
\citep{Quillen+13}. We will explore the effects 
of changing the collision prescription in a future work (Mustill, Davies \& Johansen in prep).

\subsubsection{Remnant planetesimal discs} 
Young planetary systems can 
be expected to host large numbers of planetesimals 
or embryos that have not grown to a detectable size, particularly in the very outer regions where their presence 
is hinted at by the prevalence of dusty debris discs 
\citep[e.g.,][]{Rieke+05,Su+06,Trilling+08,Wyatt08,Eiroa+13,Matthews+14,Thureau+14}. 
The quantity of material may be several tens of Earth masses and may thus have strong dynamical effects on our 
multi-planet systems. It is difficult to say whether the inclusion of small bodies will strengthen or weaken 
the effects of the dynamics of outer systems on inner planets. In the Nice model of the adolescent Solar System, 
the primordial Kuiper Belt is responsible for planet migration that leads to a global instability, but then also 
damps the eccentricities of the planets to their present-day values \citep{Tsiganis+05}. In studies of the HR~8799 
system, \cite{MooreQuillen13} found that inclusion of massive planetesimals populating the debris disc could have a 
stabilising role (causing divergent planet migration to more stable, widely-spaced configurations) or a 
destabilising role (pulling the planets out of a stable resonance). \cite{Raymond+09,Raymond+10} 
found a similar diversity of outcomes. As we have set up our multi-planet systems to be mostly 
unstable with no resonant protection, massive planetesimal discs would probably 
stabilise the systems and make them slightly less damaging to the inner planets.

\subsubsection{Tides}

Over long time-scales, tidal forces act on planets' orbits, 
typically causing a reduction in semimajor axis and eccentricity. 
The tides raised on the star by the planet are, save for hot Jupiters, 
negligible until the star leaves the main sequence \citep{Villaver+14}, 
and for our systems the dominant tidal effect is damping of 
the eccentricity and semimajor axis by tidal dissipation 
within the planet itself. However, for low values of the eccentricity 
even planetary tides act on time-scales long compared to our 
simulations: for a planet of radius $10^{-4}$\,au ($2.3\mathrm{\,R}_\oplus$), 
mass $3\times 10^{-5}\mathrm{\,M}_\odot$ ($10\mathrm{\,M}_\oplus$), quality 
factor $Q_\mathrm{p}=100$ and semimajor axis of 0.1\,au, the eccentricity 
damping timescale from Equation 4 of \cite{Jackson+08} is around $10^9$\,yrs. 
Hence, all but the closest or most eccentric of planets will experience 
negligible tidal effects over the course of our integrations.

\subsection{Long-term evolution}

\label{sec:longterm}

Our simulations are run for 10\,Myr. While short compared to typical 
ages of MS stars, this already represents a significant number of 
dynamical timescales for the inner system. Would we expect further 
evolution on Gyr timescales to change our results?

Regarding the \textsc{Binaries} simulations, the Kozai timescale 
increases as the cube of the semi-major axis ratio, and we do not 
acheive the necessary integration times to drive Kozai cycles 
in our widest binaries (see Fig~\ref{fig:hj-formation-binaries}). 
However, in these systems the presence of the inner system 
suppresses Kozai cycles because of the induced extra precession.
Hence, we do not expect that extending the duration of our integrations 
would lead to significant further dynamical evolution in most of these systems. 
In the \textsc{Giants} simulations, however, we expect that we 
have missed some instabilities amongst the outer planets due to 
the short duration of our simulations: stability times are a very strong 
function of initial spacing \citep[e.g.,][]{Chambers+96,FaberQuillen07,Mustill+14}. 
However, once strong scattering 
starts we do not expect significant differences to arise between 
more or less tightly-spaced systems, as orbital elements quickly become 
randomised\footnote{\cite{Kaib+13} found that more widely-spaced simulated 
systems have lower eccentricity, but attributed this to the larger 
fraction of stable systems amongst them.}. Our \textsc{Giants-unstable} 
and \textsc{Giants-selected} subsets can therefore be taken as the 
most detrimental effects of scattering in the outer system.

One mechanism of eccentricity excitation we have not considered 
in the outer system is secular chaos among outer giant planets 
\cite{WuLithwick11,LithwickWu14}. This can cause order-unity 
fluctuations in eccentricity if the system possesses sufficient 
angular momentum deficit and the mode coupling is suitable. We 
expect that the effects of this will be broadly similar to the Kozai 
cycles we have studied in this paper, with a relatively smooth 
decrease in pericentre, compared to the more impulsive changes 
to pericentre that often arise during planet--planet scattering.

In addition to the dynamics of the outer system, long-term instabilities 
may also be incipient in some surviving inner systems where the 
instability in the outer system has already resolved itself. We have 
already mentioned chaotic behaviour in the inner system in the context 
of formaing eccentric warm Jupiters, where a single surviving planet's 
eccentricity appears to vary chaotically (Figures~\ref{fig:run0068} 
and~\ref{fig:warmj-ecc}). 
Unfortunately, predicting whether a given chaotic system is Hill- or 
Lagrange-unstable is difficult 
\citep[but see e.g.][for Mercury in the Solar System]{Batygin+15}, and 
predicting the outcomes of instability---scattering outcomes, multiplicities and 
eccentricity and inclination excitation---requires $N$-body integrations, 
vastly expensive on timescales of Gyr when the inner planets have such short 
orbital periods as those discovered by \textit{Kepler}.

\section{Conclusions}

\label{sec:conclude}

We have run $N$-body simulations to study the effects of 
the dynamics of outer systems---experiencing 
Kozai perturbations and planet--planet scattering---on 
close-in inner systems such as those detected by 
\textit{Kepler}. Our main simulation sets are \textsc{Binaries}, 
where we add an extra outer planet and a stellar binary 
companion, and \textsc{Giants}, where we add a close-packed 
system of four outer planets. We address the issues of: the 
contribution of the ensuing perturbations to the 
``\textit{Kepler} Dichotomy'' of an excess of single-transit 
systems, by excitation of mutual inclinations or 
outright destabilisation and loss of planets; 
the excitation of extreme mutual inclinations as in 
Kepler-108, or obliquities as in Kepler-56; and 
the formation of eccentric warm Jupiters. Our key findings are:
\begin{itemize}
  \item In the most destructive cases, $40-50\%$ of 
    inner systems lose one or more planets within 10\,Myr 
    as a result of dynamics in the outer system. This applies 
    to systems where Kozai cycles excite a large ($>0.5$) eccentricity 
    on the outer planet, and to a subset of planet--planet scattering 
    simulations that reproduces the observed eccentricity distribution 
    for giant exoplanets.
  \item Over our entire set of simulation runs, including quiescent 
    outer systems where Kozai cycles were not excited due to 
    low inclination or extra precession, and where planet--planet 
    scattering was weak or non-existent, this destabilisation 
    fraction falls to $20-25\%$.
  \item In the inner systems that keep all their inner planets, 
    mutual inclinations are not excited significantly. This is true 
    both for inner systems starting with three and with two planets. 
    Triple-planet systems that are reduced to 
    double-planet systems experience more excitation however.
  \item These rates make some contribution to the 
    \textit{Kepler} Dichotomy, but the majority must be 
    explained through other means: with plausible estimates of 
    the occurrence of suitable outer architectures, we find that 
    $\approx18\%$ of \textit{Kepler} triple-planet systems would lose 
    one or more planets, with $\approx10\%$ of triple-planet systems being 
    reduced to singles, meaning that at least $75\%$ of the single-planet 
    \textit{Kepler} systems do not arise from the dynamical mechanisms 
    that we have studied. As the internal 
    evolution of inner systems is inefficient 
    at reducing multiplicities to zero or unity, formation or a high 
    false positive rate amongst the single-planet candidates 
    may play dominant roles.
  \item Similarly, there is a small contribution to the 
    population of stars with \emph{no} inner planetary system, 
    with $\approx5\%$ of triple-planet systems being reduced to ``zeros''.
  \item Although inclination effects are relatively unimportant 
    for the population of \textsc{Kepler} planets as a whole, 
    occasional interesting systems emerge. We find both 
    tilted but coplanar systems such as Kepler-56, as well 
    as highly-misaligned two-planet systems such as Kepler-108. 
  \item We identify three routes to the formation of eccentric 
    warm Jupiters: \emph{in-situ} scattering (possibly helped 
    by ``uplift'' from outer system); secular eccentricity 
    oscillations which can be ``frozen out'' if an outer planet 
    is ejected; and direct inelastic collision between an outer and an inner 
    planet as in \cite{Mustill+15}. Eccentric warm Jupiters form in 
    15\% of our scattering simulations which reproduce the observed 
    eccentricity distribution of more distant giant planets.
\end{itemize}

\section*{Acknowledgements}

The authors are supported by the project grant ``IMPACT'' 
from the Knut and Alice Wallenberg Foundation. 
A.J.\ was supported by the European Research Council under ERC                                                                                                                  
Starting Grant agreement 278675-PEBBLE2PLANET, the Swedish Research
Council (grant 2014-5775) and the Knut and Alice Wallenberg Foundation.
We thank Fabio Antonini, Daniel Carrera, Rebekah Dawson, 
Dan Fabrycky, Brad Hansen, Kaitlin Kratter, Jasques Laskar, 
Rosemary Mardling, Hagai Perets, Cristobal Petrovich, 
Sean Raymond, and the anonymous referee
for useful discussions and suggestions. 
This research has made use of the Exoplanet Orbit Database 
and the Exoplanet Data Explorer at exoplanets.org.

\bibliographystyle{mnras}
\bibliography{3p+p-outer}

\begin{thebibliography}{}
\makeatletter
\relax
\def\mn@urlcharsother{\let\do\@makeother \do\$\do\&\do\#\do\^\do\_\do\%\do\~}
\def\mn@doi{\begingroup\mn@urlcharsother \@ifnextchar [ {\mn@doi@}
  {\mn@doi@[]}}
\def\mn@doi@[#1]#2{\def\@tempa{#1}\ifx\@tempa\@empty \href
  {http://dx.doi.org/#2} {doi:#2}\else \href {http://dx.doi.org/#2} {#1}\fi
  \endgroup}
\def\mn@eprint#1#2{\mn@eprint@#1:#2::\@nil}
\def\mn@eprint@arXiv#1{\href {http://arxiv.org/abs/#1} {{\tt arXiv:#1}}}
\def\mn@eprint@dblp#1{\href {http://dblp.uni-trier.de/rec/bibtex/#1.xml}
  {dblp:#1}}
\def\mn@eprint@#1:#2:#3:#4\@nil{\def\@tempa {#1}\def\@tempb {#2}\def\@tempc
  {#3}\ifx \@tempc \@empty \let \@tempc \@tempb \let \@tempb \@tempa \fi \ifx
  \@tempb \@empty \def\@tempb {arXiv}\fi \@ifundefined
  {mn@eprint@\@tempb}{\@tempb:\@tempc}{\expandafter \expandafter \csname
  mn@eprint@\@tempb\endcsname \expandafter{\@tempc}}}

\bibitem[\protect\citeauthoryear{{Anderson}, {Storch}  \& {Lai}}{{Anderson}
  et~al.}{2016}]{Anderson+16}
{Anderson} K.~R.,  {Storch} N.~I.,   {Lai} D.,  2016, \mn@doi [\mnras]
  {10.1093/mnras/stv2906}, \href
  {http://adsabs.harvard.edu/abs/2016MNRAS.456.3671A} {456, 3671}

\bibitem[\protect\citeauthoryear{{Antonini}, {Hamers}  \&
  {Lithwick}}{{Antonini} et~al.}{2016}]{Antonini+16}
{Antonini} F.,  {Hamers} A.~S.,   {Lithwick} Y.,  2016, \mn@doi [\aj]
  {10.3847/0004-6256/152/6/174}, \href
  {http://adsabs.harvard.edu/abs/2016AJ....152..174A} {152, 174}

\bibitem[\protect\citeauthoryear{{Ballard} \& {Johnson}}{{Ballard} \&
  {Johnson}}{2016}]{BallardJohnson16}
{Ballard} S.,  {Johnson} J.~A.,  2016, \mn@doi [\apj]
  {10.3847/0004-637X/816/2/66}, \href
  {http://adsabs.harvard.edu/abs/2016ApJ...816...66B} {816, 66}

\bibitem[\protect\citeauthoryear{{Bate}}{{Bate}}{2012}]{Bate12}
{Bate} M.~R.,  2012, \mn@doi [\mnras] {10.1111/j.1365-2966.2011.19955.x}, \href
  {http://adsabs.harvard.edu/abs/2012MNRAS.419.3115B} {419, 3115}

\bibitem[\protect\citeauthoryear{{Batygin}}{{Batygin}}{2012}]{Batygin12}
{Batygin} K.,  2012, \mn@doi [\nat] {10.1038/nature11560}, \href
  {http://adsabs.harvard.edu/abs/2012Natur.491..418B} {491, 418}

\bibitem[\protect\citeauthoryear{{Batygin} \& {Laughlin}}{{Batygin} \&
  {Laughlin}}{2015}]{BatyginLaughlin15}
{Batygin} K.,  {Laughlin} G.,  2015, \mn@doi [Proceedings of the National
  Academy of Science] {10.1073/pnas.1423252112}, \href
  {http://adsabs.harvard.edu/abs/2015PNAS..112.4214B} {112, 4214}

\bibitem[\protect\citeauthoryear{{Batygin}, {Morbidelli}  \&
  {Tsiganis}}{{Batygin} et~al.}{2011}]{Batygin+11}
{Batygin} K.,  {Morbidelli} A.,   {Tsiganis} K.,  2011, \mn@doi [\aap]
  {10.1051/0004-6361/201117193}, \href
  {http://adsabs.harvard.edu/abs/2011A%26A...533A...7B} {533, A7}

\bibitem[\protect\citeauthoryear{{Batygin}, {Morbidelli}  \&
  {Holman}}{{Batygin} et~al.}{2015}]{Batygin+15}
{Batygin} K.,  {Morbidelli} A.,   {Holman} M.~J.,  2015, \mn@doi [\apj]
  {10.1088/0004-637X/799/2/120}, \href
  {http://adsabs.harvard.edu/abs/2015ApJ...799..120B} {799, 120}

\bibitem[\protect\citeauthoryear{{Beaug{\'e}} \& {Nesvorn{\'y}}}{{Beaug{\'e}}
  \& {Nesvorn{\'y}}}{2012}]{BeaugeNesvorny12}
{Beaug{\'e}} C.,  {Nesvorn{\'y}} D.,  2012, \mn@doi [ApJ]
  {10.1088/0004-637X/751/2/119}, \href
  {http://adsabs.harvard.edu/abs/2012ApJ...751..119B} {751, 119}

\bibitem[\protect\citeauthoryear{{Becker} \& {Adams}}{{Becker} \&
  {Adams}}{2016}]{BeckerAdams16}
{Becker} J.~C.,  {Adams} F.~C.,  2016, \mn@doi [\mnras]
  {10.1093/mnras/stv2444}, \href
  {http://adsabs.harvard.edu/abs/2016MNRAS.455.2980B} {455, 2980}

\bibitem[\protect\citeauthoryear{{Bitsch}, {Lambrechts}  \&
  {Johansen}}{{Bitsch} et~al.}{2015}]{Bitsch+15}
{Bitsch} B.,  {Lambrechts} M.,   {Johansen} A.,  2015, \mn@doi [\aap]
  {10.1051/0004-6361/201526463}, \href
  {http://adsabs.harvard.edu/abs/2015A%26A...582A.112B} {582, A112}

\bibitem[\protect\citeauthoryear{{Bou{\'e}} \& {Fabrycky}}{{Bou{\'e}} \&
  {Fabrycky}}{2014}]{BoueFabrycky14}
{Bou{\'e}} G.,  {Fabrycky} D.~C.,  2014, \mn@doi [\apj]
  {10.1088/0004-637X/789/2/111}, \href
  {http://adsabs.harvard.edu/abs/2014ApJ...789..111B} {789, 111}

\bibitem[\protect\citeauthoryear{{Brinch}, {J{\o}rgensen}, {Hogerheijde},
  {Nelson}  \& {Gressel}}{{Brinch} et~al.}{2016}]{Brinch+16}
{Brinch} C.,  {J{\o}rgensen} J.~K.,  {Hogerheijde} M.~R.,  {Nelson} R.~P.,
  {Gressel} O.,  2016, \mn@doi [\apjl] {10.3847/2041-8205/830/1/L16}, \href
  {http://adsabs.harvard.edu/abs/2016ApJ...830L..16B} {830, L16}

\bibitem[\protect\citeauthoryear{{Broeg} et~al.,}{{Broeg}
  et~al.}{2013}]{Broeg+13}
{Broeg} C.,  et~al., 2013, in European Physical Journal Web of Conferences. p.
  03005 (\mn@eprint {arXiv} {1305.2270}), \mn@doi{10.1051/epjconf/20134703005}

\bibitem[\protect\citeauthoryear{{Bryan} et~al.,}{{Bryan}
  et~al.}{2016}]{Bryan+16}
{Bryan} M.~L.,  et~al., 2016, \mn@doi [\apj] {10.3847/0004-637X/821/2/89},
  \href {http://adsabs.harvard.edu/abs/2016ApJ...821...89B} {821, 89}

\bibitem[\protect\citeauthoryear{{Campante} et~al.,}{{Campante}
  et~al.}{2015}]{Campante+15}
{Campante} T.~L.,  et~al., 2015, \mn@doi [\apj] {10.1088/0004-637X/799/2/170},
  \href {http://cdsads.u-strasbg.fr/abs/2015ApJ...799..170C} {799, 170}

\bibitem[\protect\citeauthoryear{{Campante} et~al.,}{{Campante}
  et~al.}{2016}]{Campante+16}
{Campante} T.~L.,  et~al., 2016, \mn@doi [\apj] {10.3847/0004-637X/819/1/85},
  \href {http://adsabs.harvard.edu/abs/2016ApJ...819...85C} {819, 85}

\bibitem[\protect\citeauthoryear{{Carrera}, {Davies}  \& {Johansen}}{{Carrera}
  et~al.}{2016}]{Carrera+16}
{Carrera} D.,  {Davies} M.~B.,   {Johansen} A.,  2016, \mn@doi [\mnras]
  {10.1093/mnras/stw2218}, \href
  {http://adsabs.harvard.edu/abs/2016MNRAS.463.3226C} {463, 3226}

\bibitem[\protect\citeauthoryear{{Chambers}}{{Chambers}}{1999}]{Chambers99}
{Chambers} J.~E.,  1999, \mn@doi [MNRAS] {10.1046/j.1365-8711.1999.02379.x},
  \href {http://adsabs.harvard.edu/abs/1999MNRAS.304..793C} {304, 793}

\bibitem[\protect\citeauthoryear{{Chambers}, {Wetherill}  \& {Boss}}{{Chambers}
  et~al.}{1996}]{Chambers+96}
{Chambers} J.~E.,  {Wetherill} G.~W.,   {Boss} A.~P.,  1996, \mn@doi [\icarus]
  {10.1006/icar.1996.0019}, \href
  {http://adsabs.harvard.edu/abs/1996Icar..119..261C} {119, 261}

\bibitem[\protect\citeauthoryear{{Chatterjee}, {Ford}, {Matsumura}  \&
  {Rasio}}{{Chatterjee} et~al.}{2008}]{Chatterjee+08}
{Chatterjee} S.,  {Ford} E.~B.,  {Matsumura} S.,   {Rasio} F.~A.,  2008,
  \mn@doi [ApJ] {10.1086/590227}, \href
  {http://adsabs.harvard.edu/abs/2008ApJ...686..580C} {686, 580}

\bibitem[\protect\citeauthoryear{{Coleman} \& {Nelson}}{{Coleman} \&
  {Nelson}}{2016}]{ColemanNelson16}
{Coleman} G.~A.~L.,  {Nelson} R.~P.,  2016, \mn@doi [\mnras]
  {10.1093/mnras/stw149}, \href
  {http://adsabs.harvard.edu/abs/2016MNRAS.457.2480C} {457, 2480}

\bibitem[\protect\citeauthoryear{{Coughlin} et~al.,}{{Coughlin}
  et~al.}{2016}]{Coughlin+16}
{Coughlin} J.~L.,  et~al., 2016, \mn@doi [\apjs] {10.3847/0067-0049/224/1/12},
  \href {http://adsabs.harvard.edu/abs/2016ApJS..224...12C} {224, 12}

\bibitem[\protect\citeauthoryear{{Cumming}, {Butler}, {Marcy}, {Vogt}, {Wright}
   \& {Fischer}}{{Cumming} et~al.}{2008}]{Cumming+08}
{Cumming} A.,  {Butler} R.~P.,  {Marcy} G.~W.,  {Vogt} S.~S.,  {Wright} J.~T.,
   {Fischer} D.~A.,  2008, \mn@doi [\pasp] {10.1086/588487}, \href
  {http://adsabs.harvard.edu/abs/2008PASP..120..531C} {120, 531}

\bibitem[\protect\citeauthoryear{{Dawson} \& {Chiang}}{{Dawson} \&
  {Chiang}}{2014}]{DawsonChiang14}
{Dawson} R.~I.,  {Chiang} E.,  2014, \mn@doi [Science]
  {10.1126/science.1256943}, \href
  {http://adsabs.harvard.edu/abs/2014Sci...346..212D} {346, 212}

\bibitem[\protect\citeauthoryear{{Dawson}, {Lee}  \& {Chiang}}{{Dawson}
  et~al.}{2015}]{Dawson+15}
{Dawson} R.~I.,  {Lee} E.~J.,   {Chiang} E.,  2015, preprint, \href
  {http://adsabs.harvard.edu/abs/2015arXiv151204951D} {} (\mn@eprint {arXiv}
  {1512.04951})

\bibitem[\protect\citeauthoryear{{Deacon} et~al.,}{{Deacon}
  et~al.}{2016}]{Deacon+16}
{Deacon} N.~R.,  et~al., 2016, \mn@doi [\mnras] {10.1093/mnras/stv2132}, \href
  {http://adsabs.harvard.edu/abs/2016MNRAS.455.4212D} {455, 4212}

\bibitem[\protect\citeauthoryear{{Dong}, {Katz}  \& {Socrates}}{{Dong}
  et~al.}{2014}]{Dong+14}
{Dong} S.,  {Katz} B.,   {Socrates} A.,  2014, \mn@doi [ApJL]
  {10.1088/2041-8205/781/1/L5}, \href
  {http://adsabs.harvard.edu/abs/2014ApJ...781L...5D} {781, L5}

\bibitem[\protect\citeauthoryear{{Duch{\^e}ne} \& {Kraus}}{{Duch{\^e}ne} \&
  {Kraus}}{2013}]{DucheneKraus13}
{Duch{\^e}ne} G.,  {Kraus} A.,  2013, \mn@doi [\araa]
  {10.1146/annurev-astro-081710-102602}, \href
  {http://adsabs.harvard.edu/abs/2013ARA%26A..51..269D} {51, 269}

\bibitem[\protect\citeauthoryear{{Duquennoy} \& {Mayor}}{{Duquennoy} \&
  {Mayor}}{1991}]{DuquennoyMayor91}
{Duquennoy} A.,  {Mayor} M.,  1991, \aap, \href
  {http://adsabs.harvard.edu/abs/1991A%26A...248..485D} {248, 485}

\bibitem[\protect\citeauthoryear{{Eiroa} et~al.,}{{Eiroa}
  et~al.}{2013}]{Eiroa+13}
{Eiroa} C.,  et~al., 2013, \mn@doi [\aap] {10.1051/0004-6361/201321050}, \href
  {http://adsabs.harvard.edu/abs/2013A%26A...555A..11E} {555, A11}

\bibitem[\protect\citeauthoryear{{Faber} \& {Quillen}}{{Faber} \&
  {Quillen}}{2007}]{FaberQuillen07}
{Faber} P.,  {Quillen} A.~C.,  2007, \mn@doi [\mnras]
  {10.1111/j.1365-2966.2007.12490.x}, \href
  {http://adsabs.harvard.edu/abs/2007MNRAS.382.1823F} {382, 1823}

\bibitem[\protect\citeauthoryear{{Fabrycky} \& {Murray-Clay}}{{Fabrycky} \&
  {Murray-Clay}}{2010}]{FabryckyMurray-Clay10}
{Fabrycky} D.~C.,  {Murray-Clay} R.~A.,  2010, \mn@doi [\apj]
  {10.1088/0004-637X/710/2/1408}, \href
  {http://adsabs.harvard.edu/abs/2010ApJ...710.1408F} {710, 1408}

\bibitem[\protect\citeauthoryear{{Fabrycky} \& {Tremaine}}{{Fabrycky} \&
  {Tremaine}}{2007}]{FabryckyTremaine07}
{Fabrycky} D.,  {Tremaine} S.,  2007, \mn@doi [\apj] {10.1086/521702}, \href
  {http://adsabs.harvard.edu/abs/2007ApJ...669.1298F} {669, 1298}

\bibitem[\protect\citeauthoryear{{Fabrycky} et~al.,}{{Fabrycky}
  et~al.}{2014}]{Fabrycky+14}
{Fabrycky} D.~C.,  et~al., 2014, \mn@doi [\apj] {10.1088/0004-637X/790/2/146},
  \href {http://adsabs.harvard.edu/abs/2014ApJ...790..146F} {790, 146}

\bibitem[\protect\citeauthoryear{{Fielding}, {McKee}, {Socrates}, {Cunningham}
  \& {Klein}}{{Fielding} et~al.}{2015}]{Fielding+15}
{Fielding} D.~B.,  {McKee} C.~F.,  {Socrates} A.,  {Cunningham} A.~J.,
  {Klein} R.~I.,  2015, \mn@doi [\mnras] {10.1093/mnras/stv836}, \href
  {http://adsabs.harvard.edu/abs/2015MNRAS.450.3306F} {450, 3306}

\bibitem[\protect\citeauthoryear{{Ford}, {Kozinsky}  \& {Rasio}}{{Ford}
  et~al.}{2000}]{Ford+00}
{Ford} E.~B.,  {Kozinsky} B.,   {Rasio} F.~A.,  2000, \mn@doi [\apj]
  {10.1086/308815}, \href {http://adsabs.harvard.edu/abs/2000ApJ...535..385F}
  {535, 385}

\bibitem[\protect\citeauthoryear{{Foreman-Mackey}, {Morton}, {Hogg}, {Agol}  \&
  {Sch{\"o}lkopf}}{{Foreman-Mackey} et~al.}{2016}]{Foreman-Mackey+16}
{Foreman-Mackey} D.,  {Morton} T.~D.,  {Hogg} D.~W.,  {Agol} E.,
  {Sch{\"o}lkopf} B.,  2016, preprint, \href
  {http://adsabs.harvard.edu/abs/2016arXiv160708237F} {} (\mn@eprint {arXiv}
  {1607.08237})

\bibitem[\protect\citeauthoryear{{Fressin} et~al.,}{{Fressin}
  et~al.}{2013}]{Fressin+13}
{Fressin} F.,  et~al., 2013, \mn@doi [ApJ] {10.1088/0004-637X/766/2/81}, \href
  {http://adsabs.harvard.edu/abs/2013ApJ...766...81F} {766, 81}

\bibitem[\protect\citeauthoryear{{Galicher} et~al.,}{{Galicher}
  et~al.}{2016}]{Galicher+16}
{Galicher} R.,  et~al., 2016, \mn@doi [\aap] {10.1051/0004-6361/201527828},
  \href {http://adsabs.harvard.edu/abs/2016A%26A...594A..63G} {594, A63}

\bibitem[\protect\citeauthoryear{{G{\"o}tberg}, {Davies}, {Mustill}, {Johansen}
   \& {Church}}{{G{\"o}tberg} et~al.}{2016}]{Gotberg+16}
{G{\"o}tberg} Y.,  {Davies} M.~B.,  {Mustill} A.~J.,  {Johansen} A.,   {Church}
  R.~P.,  2016, \mn@doi [\aap] {10.1051/0004-6361/201526309}, \href
  {http://adsabs.harvard.edu/abs/2016A%26A...592A.147G} {592, A147}

\bibitem[\protect\citeauthoryear{{Gratia} \& {Fabrycky}}{{Gratia} \&
  {Fabrycky}}{2017}]{GratiaFabrycky17}
{Gratia} P.,  {Fabrycky} D.,  2017, \mn@doi [\mnras] {10.1093/mnras/stw2180},
  \href {http://adsabs.harvard.edu/abs/2017MNRAS.464.1709G} {464, 1709}

\bibitem[\protect\citeauthoryear{{Han}, {Wang}, {Wright}, {Feng}, {Zhao},
  {Fakhouri}, {Brown}  \& {Hancock}}{{Han} et~al.}{2014}]{Han+14}
{Han} E.,  {Wang} S.~X.,  {Wright} J.~T.,  {Feng} Y.~K.,  {Zhao} M.,
  {Fakhouri} O.,  {Brown} J.~I.,   {Hancock} C.,  2014, \mn@doi [\pasp]
  {10.1086/678447}, \href {http://adsabs.harvard.edu/abs/2014PASP..126..827H}
  {126, 827}

\bibitem[\protect\citeauthoryear{{Hansen}}{{Hansen}}{2016}]{Hansen16}
{Hansen} B.~M.~S.,  2016, preprint, \href
  {http://adsabs.harvard.edu/abs/2016arXiv160806300H} {} (\mn@eprint {arXiv}
  {1608.06300})

\bibitem[\protect\citeauthoryear{{Howard} et~al.,}{{Howard}
  et~al.}{2012}]{Howard+12}
{Howard} A.~W.,  et~al., 2012, \mn@doi [\apjs] {10.1088/0067-0049/201/2/15},
  \href {http://adsabs.harvard.edu/abs/2012ApJS..201...15H} {201, 15}

\bibitem[\protect\citeauthoryear{{Huang}, {Petrovich}  \& {Deibert}}{{Huang}
  et~al.}{2016a}]{Huang+16b}
{Huang} C.~X.,  {Petrovich} C.,   {Deibert} E.,  2016a, preprint, \href
  {http://adsabs.harvard.edu/abs/2016arXiv160908110H} {} (\mn@eprint {arXiv}
  {1609.08110})

\bibitem[\protect\citeauthoryear{{Huang}, {Wu}  \& {Triaud}}{{Huang}
  et~al.}{2016b}]{Huang+16}
{Huang} C.,  {Wu} Y.,   {Triaud} A.~H.~M.~J.,  2016b, \mn@doi [\apj]
  {10.3847/0004-637X/825/2/98}, \href
  {http://adsabs.harvard.edu/abs/2016ApJ...825...98H} {825, 98}

\bibitem[\protect\citeauthoryear{{Huber} et~al.,}{{Huber}
  et~al.}{2013}]{Huber+13}
{Huber} D.,  et~al., 2013, \mn@doi [Science] {10.1126/science.1242066}, \href
  {http://adsabs.harvard.edu/abs/2013Sci...342..331H} {342, 331}

\bibitem[\protect\citeauthoryear{{Innanen}, {Zheng}, {Mikkola}  \&
  {Valtonen}}{{Innanen} et~al.}{1997}]{Innanen+97}
{Innanen} K.~A.,  {Zheng} J.~Q.,  {Mikkola} S.,   {Valtonen} M.~J.,  1997,
  \mn@doi [AJ] {10.1086/118405}, \href
  {http://adsabs.harvard.edu/abs/1997AJ....113.1915I} {113, 1915}

\bibitem[\protect\citeauthoryear{{Jackson}, {Greenberg}  \& {Barnes}}{{Jackson}
  et~al.}{2008}]{Jackson+08}
{Jackson} B.,  {Greenberg} R.,   {Barnes} R.,  2008, \mn@doi [\apj]
  {10.1086/529187}, \href {http://adsabs.harvard.edu/abs/2008ApJ...678.1396J}
  {678, 1396}

\bibitem[\protect\citeauthoryear{{Jensen} \& {Akeson}}{{Jensen} \&
  {Akeson}}{2014}]{JensenAkeson14}
{Jensen} E.~L.~N.,  {Akeson} R.,  2014, \mn@doi [\nat] {10.1038/nature13521},
  \href {http://adsabs.harvard.edu/abs/2014Natur.511..567J} {511, 567}

\bibitem[\protect\citeauthoryear{{Johansen} \& {Lacerda}}{{Johansen} \&
  {Lacerda}}{2010}]{JohansenLacerda10}
{Johansen} A.,  {Lacerda} P.,  2010, \mn@doi [\mnras]
  {10.1111/j.1365-2966.2010.16309.x}, \href
  {http://adsabs.harvard.edu/abs/2010MNRAS.404..475J} {404, 475}

\bibitem[\protect\citeauthoryear{{Johansen}, {Davies}, {Church}  \&
  {Holmelin}}{{Johansen} et~al.}{2012}]{Johansen+12}
{Johansen} A.,  {Davies} M.~B.,  {Church} R.~P.,   {Holmelin} V.,  2012,
  \mn@doi [ApJ] {10.1088/0004-637X/758/1/39}, \href
  {http://adsabs.harvard.edu/abs/2012ApJ...758...39J} {758, 39}

\bibitem[\protect\citeauthoryear{{Juri{\'c}} \& {Tremaine}}{{Juri{\'c}} \&
  {Tremaine}}{2008}]{JuricTremaine08}
{Juri{\'c}} M.,  {Tremaine} S.,  2008, \mn@doi [\apj] {10.1086/590047}, \href
  {http://adsabs.harvard.edu/abs/2008ApJ...686..603J} {686, 603}

\bibitem[\protect\citeauthoryear{{Kaib} \& {Chambers}}{{Kaib} \&
  {Chambers}}{2016}]{KaibChambers16}
{Kaib} N.~A.,  {Chambers} J.~E.,  2016, \mn@doi [\mnras]
  {10.1093/mnras/stv2554}, \href
  {http://adsabs.harvard.edu/abs/2016MNRAS.455.3561K} {455, 3561}

\bibitem[\protect\citeauthoryear{{Kaib}, {Raymond}  \& {Duncan}}{{Kaib}
  et~al.}{2011}]{Kaib+11}
{Kaib} N.~A.,  {Raymond} S.~N.,   {Duncan} M.~J.,  2011, \mn@doi [\apjl]
  {10.1088/2041-8205/742/2/L24}, \href
  {http://adsabs.harvard.edu/abs/2011ApJ...742L..24K} {742, L24}

\bibitem[\protect\citeauthoryear{{Kaib}, {Raymond}  \& {Duncan}}{{Kaib}
  et~al.}{2013}]{Kaib+13}
{Kaib} N.~A.,  {Raymond} S.~N.,   {Duncan} M.,  2013, \mn@doi [\nat]
  {10.1038/nature11780}, \href
  {http://adsabs.harvard.edu/abs/2013Natur.493..381K} {493, 381}

\bibitem[\protect\citeauthoryear{{Kipping} et~al.,}{{Kipping}
  et~al.}{2016}]{Kipping+16}
{Kipping} D.~M.,  et~al., 2016, \mn@doi [\apj] {10.3847/0004-637X/820/2/112},
  \href {http://adsabs.harvard.edu/abs/2016ApJ...820..112K} {820, 112}

\bibitem[\protect\citeauthoryear{{Knutson} et~al.,}{{Knutson}
  et~al.}{2014}]{Knutson+14}
{Knutson} H.~A.,  et~al., 2014, \mn@doi [ApJ] {10.1088/0004-637X/785/2/126},
  \href {http://adsabs.harvard.edu/abs/2014ApJ...785..126K} {785, 126}

\bibitem[\protect\citeauthoryear{{Kozai}}{{Kozai}}{1962}]{Kozai62}
{Kozai} Y.,  1962, \mn@doi [AJ] {10.1086/108790}, \href
  {http://adsabs.harvard.edu/abs/1962AJ.....67..591K} {67, 591}

\bibitem[\protect\citeauthoryear{{Kraus}, {Ireland}, {Huber}, {Mann}  \&
  {Dupuy}}{{Kraus} et~al.}{2016}]{Kraus+16}
{Kraus} A.~L.,  {Ireland} M.~J.,  {Huber} D.,  {Mann} A.~W.,   {Dupuy} T.~J.,
  2016, \mn@doi [\aj] {10.3847/0004-6256/152/1/8}, \href
  {http://adsabs.harvard.edu/abs/2016AJ....152....8K} {152, 8}

\bibitem[\protect\citeauthoryear{{Lai} \& {Pu}}{{Lai} \& {Pu}}{2016}]{LaiPu16}
{Lai} D.,  {Pu} B.,  2016, preprint, \href
  {http://adsabs.harvard.edu/abs/2016arXiv160608855L} {} (\mn@eprint {arXiv}
  {1606.08855})

\bibitem[\protect\citeauthoryear{{Lai}, {Foucart}  \& {Lin}}{{Lai}
  et~al.}{2011}]{Lai+11}
{Lai} D.,  {Foucart} F.,   {Lin} D.~N.~C.,  2011, \mn@doi [MNRAS]
  {10.1111/j.1365-2966.2010.18127.x}, \href
  {http://adsabs.harvard.edu/abs/2011MNRAS.412.2790L} {412, 2790}

\bibitem[\protect\citeauthoryear{{Lambrechts} \& {Johansen}}{{Lambrechts} \&
  {Johansen}}{2012}]{LambrechtsJohansen12}
{Lambrechts} M.,  {Johansen} A.,  2012, \mn@doi [\aap]
  {10.1051/0004-6361/201219127}, \href
  {http://adsabs.harvard.edu/abs/2012A%26A...544A..32L} {544, A32}

\bibitem[\protect\citeauthoryear{{Lambrechts} \& {Johansen}}{{Lambrechts} \&
  {Johansen}}{2014}]{LambrechtsJohansen14}
{Lambrechts} M.,  {Johansen} A.,  2014, \mn@doi [\aap]
  {10.1051/0004-6361/201424343}, \href
  {http://adsabs.harvard.edu/abs/2014A%26A...572A.107L} {572, A107}

\bibitem[\protect\citeauthoryear{{Lambrechts}, {Johansen}  \&
  {Morbidelli}}{{Lambrechts} et~al.}{2014}]{Lambrechts+14}
{Lambrechts} M.,  {Johansen} A.,   {Morbidelli} A.,  2014, \mn@doi [\aap]
  {10.1051/0004-6361/201423814}, \href
  {http://adsabs.harvard.edu/abs/2014A%26A...572A..35L} {572, A35}

\bibitem[\protect\citeauthoryear{{Lee} \& {Peale}}{{Lee} \&
  {Peale}}{2002}]{LeePeale02}
{Lee} M.~H.,  {Peale} S.~J.,  2002, \mn@doi [\apj] {10.1086/338504}, \href
  {http://adsabs.harvard.edu/abs/2002ApJ...567..596L} {567, 596}

\bibitem[\protect\citeauthoryear{{Leinhardt} \& {Stewart}}{{Leinhardt} \&
  {Stewart}}{2012}]{LeinhardtStewart12}
{Leinhardt} Z.~M.,  {Stewart} S.~T.,  2012, \mn@doi [ApJ]
  {10.1088/0004-637X/745/1/79}, \href
  {http://adsabs.harvard.edu/abs/2012ApJ...745...79L} {745, 79}

\bibitem[\protect\citeauthoryear{{Li}, {Naoz}, {Valsecchi}, {Johnson}  \&
  {Rasio}}{{Li} et~al.}{2014}]{Li+14a}
{Li} G.,  {Naoz} S.,  {Valsecchi} F.,  {Johnson} J.~A.,   {Rasio} F.~A.,  2014,
  \mn@doi [\apj] {10.1088/0004-637X/794/2/131}, \href
  {http://adsabs.harvard.edu/abs/2014ApJ...794..131L} {794, 131}

\bibitem[\protect\citeauthoryear{{Libert} \& {Tsiganis}}{{Libert} \&
  {Tsiganis}}{2011}]{LibertTsiganis11}
{Libert} A.-S.,  {Tsiganis} K.,  2011, \mn@doi [\mnras]
  {10.1111/j.1365-2966.2010.18059.x}, \href
  {http://adsabs.harvard.edu/abs/2011MNRAS.412.2353L} {412, 2353}

\bibitem[\protect\citeauthoryear{{Lidov}}{{Lidov}}{1962}]{Lidov62}
{Lidov} M.~L.,  1962, \mn@doi [Planetary \& Space Science]
  {10.1016/0032-0633(62)90129-0}, \href
  {http://adsabs.harvard.edu/abs/1962P%26SS....9..719L} {9, 719}

\bibitem[\protect\citeauthoryear{{Lissauer} et~al.,}{{Lissauer}
  et~al.}{2011}]{Lissauer+11}
{Lissauer} J.~J.,  et~al., 2011, \mn@doi [\apjs] {10.1088/0067-0049/197/1/8},
  \href {http://adsabs.harvard.edu/abs/2011ApJS..197....8L} {197, 8}

\bibitem[\protect\citeauthoryear{{Lissauer} et~al.,}{{Lissauer}
  et~al.}{2014}]{Lissauer+14}
{Lissauer} J.~J.,  et~al., 2014, \mn@doi [\apj] {10.1088/0004-637X/784/1/44},
  \href {http://adsabs.harvard.edu/abs/2014ApJ...784...44L} {784, 44}

\bibitem[\protect\citeauthoryear{{Lithwick} \& {Wu}}{{Lithwick} \&
  {Wu}}{2014}]{LithwickWu14}
{Lithwick} Y.,  {Wu} Y.,  2014, \mn@doi [Proceedings of the National Academy of
  Science] {10.1073/pnas.1308261110}, \href
  {http://adsabs.harvard.edu/abs/2014PNAS..11112610L} {111, 12610}

\bibitem[\protect\citeauthoryear{{Liu}, {Mu{\~n}oz}  \& {Lai}}{{Liu}
  et~al.}{2015}]{Liu+15}
{Liu} B.,  {Mu{\~n}oz} D.~J.,   {Lai} D.,  2015, \mn@doi [\mnras]
  {10.1093/mnras/stu2396}, \href
  {http://adsabs.harvard.edu/abs/2015MNRAS.447..747L} {447, 747}

\bibitem[\protect\citeauthoryear{{Madhusudhan}, {Bitsch}, {Johansen}  \&
  {Eriksson}}{{Madhusudhan} et~al.}{2016}]{Madhusudhan+16}
{Madhusudhan} N.,  {Bitsch} B.,  {Johansen} A.,   {Eriksson} L.,  2016,
  preprint, \href {http://adsabs.harvard.edu/abs/2016arXiv161103083M} {}
  (\mn@eprint {arXiv} {1611.03083})

\bibitem[\protect\citeauthoryear{{Malmberg}, {Davies}  \&
  {Chambers}}{{Malmberg} et~al.}{2007a}]{Malmberg+07}
{Malmberg} D.,  {Davies} M.~B.,   {Chambers} J.~E.,  2007a, \mn@doi [\mnras]
  {10.1111/j.1745-3933.2007.00291.x}, \href
  {http://adsabs.harvard.edu/abs/2007MNRAS.377L...1M} {377, L1}

\bibitem[\protect\citeauthoryear{{Malmberg}, {de Angeli}, {Davies}, {Church},
  {Mackey}  \& {Wilkinson}}{{Malmberg} et~al.}{2007b}]{Malmberg+07b}
{Malmberg} D.,  {de Angeli} F.,  {Davies} M.~B.,  {Church} R.~P.,  {Mackey} D.,
    {Wilkinson} M.~I.,  2007b, \mn@doi [\mnras]
  {10.1111/j.1365-2966.2007.11885.x}, \href
  {http://adsabs.harvard.edu/abs/2007MNRAS.378.1207M} {378, 1207}

\bibitem[\protect\citeauthoryear{{Masuda}}{{Masuda}}{2014}]{Masuda14}
{Masuda} K.,  2014, \mn@doi [\apj] {10.1088/0004-637X/783/1/53}, \href
  {http://adsabs.harvard.edu/abs/2014ApJ...783...53M} {783, 53}

\bibitem[\protect\citeauthoryear{{Matsumura}, {Ida}  \& {Nagasawa}}{{Matsumura}
  et~al.}{2013}]{Matsumura+13}
{Matsumura} S.,  {Ida} S.,   {Nagasawa} M.,  2013, \mn@doi [ApJ]
  {10.1088/0004-637X/767/2/129}, \href
  {http://adsabs.harvard.edu/abs/2013ApJ...767..129M} {767, 129}

\bibitem[\protect\citeauthoryear{{Matthews}, {Krivov}, {Wyatt}, {Bryden}  \&
  {Eiroa}}{{Matthews} et~al.}{2014}]{Matthews+14}
{Matthews} B.~C.,  {Krivov} A.~V.,  {Wyatt} M.~C.,  {Bryden} G.,   {Eiroa} C.,
  2014, \mn@doi [Protostars and Planets VI]
  {10.2458/azu_uapress_9780816531240-ch023}, \href
  {http://adsabs.harvard.edu/abs/2014prpl.conf..521M} {pp 521--544}

\bibitem[\protect\citeauthoryear{{Mayor} et~al.,}{{Mayor}
  et~al.}{2011}]{Mayor+11}
{Mayor} M.,  et~al., 2011, preprint, \href
  {http://adsabs.harvard.edu/abs/2011arXiv1109.2497M} {} (\mn@eprint {arXiv}
  {1109.2497})

\bibitem[\protect\citeauthoryear{{Mills} \& {Fabrycky}}{{Mills} \&
  {Fabrycky}}{2016}]{MillsFabrycky16}
{Mills} S.~M.,  {Fabrycky} D.~C.,  2016, preprint, \href
  {http://adsabs.harvard.edu/abs/2016arXiv160604485M} {} (\mn@eprint {arXiv}
  {1606.04485})

\bibitem[\protect\citeauthoryear{{Moore} \& {Quillen}}{{Moore} \&
  {Quillen}}{2013}]{MooreQuillen13}
{Moore} A.,  {Quillen} A.~C.,  2013, \mn@doi [MNRAS] {10.1093/mnras/sts625},
  \href {http://adsabs.harvard.edu/abs/2013MNRAS.430..320M} {430, 320}

\bibitem[\protect\citeauthoryear{{Morbidelli} \& {Nesvorny}}{{Morbidelli} \&
  {Nesvorny}}{2012}]{MorbidelliNesvorny12}
{Morbidelli} A.,  {Nesvorny} D.,  2012, \mn@doi [\aap]
  {10.1051/0004-6361/201219824}, \href
  {http://adsabs.harvard.edu/abs/2012A%26A...546A..18M} {546, A18}

\bibitem[\protect\citeauthoryear{{Morton} \& {Winn}}{{Morton} \&
  {Winn}}{2014}]{MortonWinn14}
{Morton} T.~D.,  {Winn} J.~N.,  2014, \mn@doi [\apj]
  {10.1088/0004-637X/796/1/47}, \href
  {http://adsabs.harvard.edu/abs/2014ApJ...796...47M} {796, 47}

\bibitem[\protect\citeauthoryear{{Mu{\~n}oz}, {Lai}  \& {Liu}}{{Mu{\~n}oz}
  et~al.}{2016}]{Munoz+16}
{Mu{\~n}oz} D.~J.,  {Lai} D.,   {Liu} B.,  2016, \mn@doi [\mnras]
  {10.1093/mnras/stw983}, \href
  {http://adsabs.harvard.edu/abs/2016MNRAS.460.1086M} {460, 1086}

\bibitem[\protect\citeauthoryear{{Mustill} \& {Wyatt}}{{Mustill} \&
  {Wyatt}}{2009}]{MustillWyatt09}
{Mustill} A.~J.,  {Wyatt} M.~C.,  2009, \mn@doi [\mnras]
  {10.1111/j.1365-2966.2009.15360.x}, \href
  {http://adsabs.harvard.edu/abs/2009MNRAS.399.1403M} {399, 1403}

\bibitem[\protect\citeauthoryear{{Mustill}, {Veras}  \& {Villaver}}{{Mustill}
  et~al.}{2014}]{Mustill+14}
{Mustill} A.~J.,  {Veras} D.,   {Villaver} E.,  2014, \mn@doi [\mnras]
  {10.1093/mnras/stt1973}, \href
  {http://adsabs.harvard.edu/abs/2014MNRAS.437.1404M} {437, 1404}

\bibitem[\protect\citeauthoryear{{Mustill}, {Davies}  \& {Johansen}}{{Mustill}
  et~al.}{2015}]{Mustill+15}
{Mustill} A.~J.,  {Davies} M.~B.,   {Johansen} A.,  2015, \mn@doi [\apj]
  {10.1088/0004-637X/808/1/14}, \href
  {http://adsabs.harvard.edu/abs/2015ApJ...808...14M} {808, 14}

\bibitem[\protect\citeauthoryear{{Mustill}, {Raymond}  \& {Davies}}{{Mustill}
  et~al.}{2016}]{Mustill+16}
{Mustill} A.~J.,  {Raymond} S.~N.,   {Davies} M.~B.,  2016, \mn@doi [\mnras]
  {10.1093/mnrasl/slw075}, \href
  {http://adsabs.harvard.edu/abs/2016MNRAS.460L.109M} {460, L109}

\bibitem[\protect\citeauthoryear{{Nagasawa}, {Ida}  \& {Bessho}}{{Nagasawa}
  et~al.}{2008}]{Nagasawa+08}
{Nagasawa} M.,  {Ida} S.,   {Bessho} T.,  2008, \mn@doi [\apj]
  {10.1086/529369}, \href {http://adsabs.harvard.edu/abs/2008ApJ...678..498N}
  {678, 498}

\bibitem[\protect\citeauthoryear{{Naoz}}{{Naoz}}{2016}]{Naoz16}
{Naoz} S.,  2016, \mn@doi [\araa] {10.1146/annurev-astro-081915-023315}, \href
  {http://adsabs.harvard.edu/abs/2016ARA%26A..54..441N} {54, 441}

\bibitem[\protect\citeauthoryear{{Ngo} et~al.,}{{Ngo} et~al.}{2015}]{Ngo+15}
{Ngo} H.,  et~al., 2015, \mn@doi [\apj] {10.1088/0004-637X/800/2/138}, \href
  {http://adsabs.harvard.edu/abs/2015ApJ...800..138N} {800, 138}

\bibitem[\protect\citeauthoryear{{Ngo} et~al.,}{{Ngo} et~al.}{2016}]{Ngo+16}
{Ngo} H.,  et~al., 2016, \mn@doi [\apj] {10.3847/0004-637X/827/1/8}, \href
  {http://adsabs.harvard.edu/abs/2016ApJ...827....8N} {827, 8}

\bibitem[\protect\citeauthoryear{{Ormel} \& {Klahr}}{{Ormel} \&
  {Klahr}}{2010}]{OrmelKlahr10}
{Ormel} C.~W.,  {Klahr} H.~H.,  2010, \mn@doi [\aap]
  {10.1051/0004-6361/201014903}, \href
  {http://adsabs.harvard.edu/abs/2010A%26A...520A..43O} {520, A43}

\bibitem[\protect\citeauthoryear{{Osborn} et~al.,}{{Osborn}
  et~al.}{2016}]{Osborn+16}
{Osborn} H.~P.,  et~al., 2016, \mn@doi [\mnras] {10.1093/mnras/stw137}, \href
  {http://adsabs.harvard.edu/abs/2016MNRAS.457.2273O} {457, 2273}

\bibitem[\protect\citeauthoryear{{Otor} et~al.,}{{Otor} et~al.}{2016}]{Otor+16}
{Otor} O.~J.,  et~al., 2016, \mn@doi [\aj] {10.3847/0004-6256/152/6/165}, \href
  {http://adsabs.harvard.edu/abs/2016AJ....152..165O} {152, 165}

\bibitem[\protect\citeauthoryear{{Petrovich}}{{Petrovich}}{2015}]{Petrovich15b}
{Petrovich} C.,  2015, \mn@doi [\apj] {10.1088/0004-637X/799/1/27}, \href
  {http://adsabs.harvard.edu/abs/2015ApJ...799...27P} {799, 27}

\bibitem[\protect\citeauthoryear{{Petrovich} \& {Tremaine}}{{Petrovich} \&
  {Tremaine}}{2016}]{PetrovichTremaine16}
{Petrovich} C.,  {Tremaine} S.,  2016, \mn@doi [\apj]
  {10.3847/0004-637X/829/2/132}, \href
  {http://adsabs.harvard.edu/abs/2016ApJ...829..132P} {829, 132}

\bibitem[\protect\citeauthoryear{{Petrovich}, {Tremaine}  \&
  {Rafikov}}{{Petrovich} et~al.}{2014}]{Petrovich+14}
{Petrovich} C.,  {Tremaine} S.,   {Rafikov} R.,  2014, \mn@doi [ApJ]
  {10.1088/0004-637X/786/2/101}, \href
  {http://adsabs.harvard.edu/abs/2014ApJ...786..101P} {786, 101}

\bibitem[\protect\citeauthoryear{{Pfalzner}, {Vogel}, {Scharw{\"a}chter}  \&
  {Olczak}}{{Pfalzner} et~al.}{2005}]{Pfalzner+05}
{Pfalzner} S.,  {Vogel} P.,  {Scharw{\"a}chter} J.,   {Olczak} C.,  2005,
  \mn@doi [\aap] {10.1051/0004-6361:20042467}, \href
  {http://adsabs.harvard.edu/abs/2005A%26A...437..967P} {437, 967}

\bibitem[\protect\citeauthoryear{{Picogna} \& {Marzari}}{{Picogna} \&
  {Marzari}}{2015}]{PicognaMarzari15}
{Picogna} G.,  {Marzari} F.,  2015, \mn@doi [\aap]
  {10.1051/0004-6361/201526162}, \href
  {http://adsabs.harvard.edu/abs/2015A%26A...583A.133P} {583, A133}

\bibitem[\protect\citeauthoryear{{Pu} \& {Wu}}{{Pu} \& {Wu}}{2015}]{PuWu15}
{Pu} B.,  {Wu} Y.,  2015, \mn@doi [\apj] {10.1088/0004-637X/807/1/44}, \href
  {http://adsabs.harvard.edu/abs/2015ApJ...807...44P} {807, 44}

\bibitem[\protect\citeauthoryear{{Quillen}, {Bodman}  \& {Moore}}{{Quillen}
  et~al.}{2013}]{Quillen+13}
{Quillen} A.~C.,  {Bodman} E.,   {Moore} A.,  2013, \mn@doi [\mnras]
  {10.1093/mnras/stt1442}, \href
  {http://adsabs.harvard.edu/abs/2013MNRAS.435.2256Q} {435, 2256}

\bibitem[\protect\citeauthoryear{{Raghavan} et~al.,}{{Raghavan}
  et~al.}{2010}]{Raghavan+10}
{Raghavan} D.,  et~al., 2010, \mn@doi [\apjs] {10.1088/0067-0049/190/1/1},
  \href {http://adsabs.harvard.edu/abs/2010ApJS..190....1R} {190, 1}

\bibitem[\protect\citeauthoryear{{Rasio} \& {Ford}}{{Rasio} \&
  {Ford}}{1996}]{RasioFord96}
{Rasio} F.~A.,  {Ford} E.~B.,  1996, \mn@doi [Science]
  {10.1126/science.274.5289.954}, \href
  {http://adsabs.harvard.edu/abs/1996Sci...274..954R} {274, 954}

\bibitem[\protect\citeauthoryear{{Rauer} et~al.,}{{Rauer}
  et~al.}{2014}]{Rauer+14}
{Rauer} H.,  et~al., 2014, \mn@doi [Experimental Astronomy]
  {10.1007/s10686-014-9383-4}, \href
  {http://adsabs.harvard.edu/abs/2014ExA....38..249R} {38, 249}

\bibitem[\protect\citeauthoryear{{Raymond}, {Armitage}  \&
  {Gorelick}}{{Raymond} et~al.}{2009}]{Raymond+09}
{Raymond} S.~N.,  {Armitage} P.~J.,   {Gorelick} N.,  2009, \mn@doi [\apjl]
  {10.1088/0004-637X/699/2/L88}, \href
  {http://adsabs.harvard.edu/abs/2009ApJ...699L..88R} {699, L88}

\bibitem[\protect\citeauthoryear{{Raymond}, {Armitage}  \&
  {Gorelick}}{{Raymond} et~al.}{2010}]{Raymond+10}
{Raymond} S.~N.,  {Armitage} P.~J.,   {Gorelick} N.,  2010, \mn@doi [\apj]
  {10.1088/0004-637X/711/2/772}, \href
  {http://adsabs.harvard.edu/abs/2010ApJ...711..772R} {711, 772}

\bibitem[\protect\citeauthoryear{{Raymond} et~al.,}{{Raymond}
  et~al.}{2011}]{Raymond+11}
{Raymond} S.~N.,  et~al., 2011, \mn@doi [A\&A] {10.1051/0004-6361/201116456},
  \href {http://adsabs.harvard.edu/abs/2011A%26A...530A..62R} {530, A62}

\bibitem[\protect\citeauthoryear{{Raymond} et~al.,}{{Raymond}
  et~al.}{2012}]{Raymond+12}
{Raymond} S.~N.,  et~al., 2012, \mn@doi [A\&A] {10.1051/0004-6361/201117049},
  \href {http://adsabs.harvard.edu/abs/2012A%26A...541A..11R} {541, A11}

\bibitem[\protect\citeauthoryear{{Ricker} et~al.,}{{Ricker}
  et~al.}{2015}]{Ricker+15}
{Ricker} G.~R.,  et~al., 2015, \mn@doi [Journal of Astronomical Telescopes,
  Instruments, and Systems] {10.1117/1.JATIS.1.1.014003}, \href
  {http://adsabs.harvard.edu/abs/2015JATIS...1a4003R} {1, 014003}

\bibitem[\protect\citeauthoryear{{Rieke} et~al.,}{{Rieke}
  et~al.}{2005}]{Rieke+05}
{Rieke} G.~H.,  et~al., 2005, \mn@doi [\apj] {10.1086/426937}, \href
  {http://adsabs.harvard.edu/abs/2005ApJ...620.1010R} {620, 1010}

\bibitem[\protect\citeauthoryear{{Rowan} et~al.,}{{Rowan}
  et~al.}{2016}]{Rowan+16}
{Rowan} D.,  et~al., 2016, \mn@doi [\apj] {10.3847/0004-637X/817/2/104}, \href
  {http://adsabs.harvard.edu/abs/2016ApJ...817..104R} {817, 104}

\bibitem[\protect\citeauthoryear{{Rowe} et~al.,}{{Rowe} et~al.}{2014}]{Rowe+14}
{Rowe} J.~F.,  et~al., 2014, \mn@doi [ApJ] {10.1088/0004-637X/784/1/45}, \href
  {http://adsabs.harvard.edu/abs/2014ApJ...784...45R} {784, 45}

\bibitem[\protect\citeauthoryear{{Santerne} et~al.,}{{Santerne}
  et~al.}{2016}]{Santerne+16}
{Santerne} A.,  et~al., 2016, \mn@doi [\aap] {10.1051/0004-6361/201527329},
  \href {http://adsabs.harvard.edu/abs/2016A%26A...587A..64S} {587, A64}

\bibitem[\protect\citeauthoryear{{Shabram}, {Demory}, {Cisewski}, {Ford}  \&
  {Rogers}}{{Shabram} et~al.}{2016}]{Shabram+16}
{Shabram} M.,  {Demory} B.-O.,  {Cisewski} J.,  {Ford} E.~B.,   {Rogers} L.,
  2016, \mn@doi [\apj] {10.3847/0004-637X/820/2/93}, \href
  {http://adsabs.harvard.edu/abs/2016ApJ...820...93S} {820, 93}

\bibitem[\protect\citeauthoryear{{Shvartzvald} et~al.,}{{Shvartzvald}
  et~al.}{2016}]{Shvartzvald+16}
{Shvartzvald} Y.,  et~al., 2016, \mn@doi [\mnras] {10.1093/mnras/stw191}, \href
  {http://adsabs.harvard.edu/abs/2016MNRAS.457.4089S} {457, 4089}

\bibitem[\protect\citeauthoryear{{Socrates}, {Katz}, {Dong}  \&
  {Tremaine}}{{Socrates} et~al.}{2012}]{Socrates+12}
{Socrates} A.,  {Katz} B.,  {Dong} S.,   {Tremaine} S.,  2012, \mn@doi [\apj]
  {10.1088/0004-637X/750/2/106}, \href
  {http://adsabs.harvard.edu/abs/2012ApJ...750..106S} {750, 106}

\bibitem[\protect\citeauthoryear{{Sotiriadis}, {Libert}, {Bitsch}  \&
  {Crida}}{{Sotiriadis} et~al.}{2017}]{Sotiriadis+17}
{Sotiriadis} S.,  {Libert} A.-S.,  {Bitsch} B.,   {Crida} A.,  2017, \mn@doi
  [\aap] {10.1051/0004-6361/201628470}, \href
  {http://adsabs.harvard.edu/abs/2017A%26A...598A..70S} {598, A70}

\bibitem[\protect\citeauthoryear{{Spalding} \& {Batygin}}{{Spalding} \&
  {Batygin}}{2016}]{SpaldingBatygin16}
{Spalding} C.,  {Batygin} K.,  2016, \mn@doi [\apj]
  {10.3847/0004-637X/830/1/5}, \href
  {http://adsabs.harvard.edu/abs/2016ApJ...830....5S} {830, 5}

\bibitem[\protect\citeauthoryear{{Su} et~al.,}{{Su} et~al.}{2006}]{Su+06}
{Su} K.~Y.~L.,  et~al., 2006, \mn@doi [\apj] {10.1086/508649}, \href
  {http://adsabs.harvard.edu/abs/2006ApJ...653..675S} {653, 675}

\bibitem[\protect\citeauthoryear{{Thibault} \& {Haghighipour}}{{Thibault} \&
  {Haghighipour}}{2015}]{ThibaultHaghighipour15}
{Thibault} P.,  {Haghighipour} N.,  2015, in {Planetary Exploration and
  Science: Recent Results and Advances}. Springer Geophysics.
p.~309 (\mn@eprint {arXiv} {1406.1357}), \mn@doi{10.1007/978-3-662-45052-9}

\bibitem[\protect\citeauthoryear{{Thureau} et~al.,}{{Thureau}
  et~al.}{2014}]{Thureau+14}
{Thureau} N.~D.,  et~al., 2014, \mn@doi [\mnras] {10.1093/mnras/stu1864}, \href
  {http://adsabs.harvard.edu/abs/2014MNRAS.445.2558T} {445, 2558}

\bibitem[\protect\citeauthoryear{{Trilling} et~al.,}{{Trilling}
  et~al.}{2008}]{Trilling+08}
{Trilling} D.~E.,  et~al., 2008, \mn@doi [\apj] {10.1086/525514}, \href
  {http://adsabs.harvard.edu/abs/2008ApJ...674.1086T} {674, 1086}

\bibitem[\protect\citeauthoryear{{Tsiganis}, {Gomes}, {Morbidelli}  \&
  {Levison}}{{Tsiganis} et~al.}{2005}]{Tsiganis+05}
{Tsiganis} K.,  {Gomes} R.,  {Morbidelli} A.,   {Levison} H.~F.,  2005, \mn@doi
  [Nature] {10.1038/nature03539}, \href
  {http://adsabs.harvard.edu/abs/2005Natur.435..459T} {435, 459}

\bibitem[\protect\citeauthoryear{{Uehara}, {Kawahara}, {Masuda}, {Yamada}  \&
  {Aizawa}}{{Uehara} et~al.}{2016}]{Uehara+16}
{Uehara} S.,  {Kawahara} H.,  {Masuda} K.,  {Yamada} S.,   {Aizawa} M.,  2016,
  \mn@doi [\apj] {10.3847/0004-637X/822/1/2}, \href
  {http://adsabs.harvard.edu/abs/2016ApJ...822....2U} {822, 2}

\bibitem[\protect\citeauthoryear{{Van Eylen} \& {Albrecht}}{{Van Eylen} \&
  {Albrecht}}{2015}]{VanEylenAlbrecht15}
{Van Eylen} V.,  {Albrecht} S.,  2015, \mn@doi [\apj]
  {10.1088/0004-637X/808/2/126}, \href
  {http://adsabs.harvard.edu/abs/2015ApJ...808..126V} {808, 126}

\bibitem[\protect\citeauthoryear{{Veras} \& {Armitage}}{{Veras} \&
  {Armitage}}{2005}]{VerasArmitage05}
{Veras} D.,  {Armitage} P.~J.,  2005, \mn@doi [\apjl] {10.1086/428831}, \href
  {http://adsabs.harvard.edu/abs/2005ApJ...620L.111V} {620, L111}

\bibitem[\protect\citeauthoryear{{Veras} \& {Armitage}}{{Veras} \&
  {Armitage}}{2006}]{VerasArmitage06}
{Veras} D.,  {Armitage} P.~J.,  2006, \mn@doi [\apj] {10.1086/504582}, \href
  {http://adsabs.harvard.edu/abs/2006ApJ...645.1509V} {645, 1509}

\bibitem[\protect\citeauthoryear{{Veras}, {Mustill}, {G{\"a}nsicke},
  {Redfield}, {Georgakarakos}, {Bowler}  \& {Lloyd}}{{Veras}
  et~al.}{2016}]{Veras+16}
{Veras} D.,  {Mustill} A.~J.,  {G{\"a}nsicke} B.~T.,  {Redfield} S.,
  {Georgakarakos} N.,  {Bowler} A.~B.,   {Lloyd} M.~J.~S.,  2016, \mn@doi
  [\mnras] {10.1093/mnras/stw476}, \href
  {http://adsabs.harvard.edu/abs/2016MNRAS.458.3942V} {458, 3942}

\bibitem[\protect\citeauthoryear{{Vigan} et~al.,}{{Vigan}
  et~al.}{2012}]{Vigan+12}
{Vigan} A.,  et~al., 2012, \mn@doi [\aap] {10.1051/0004-6361/201218991}, \href
  {http://adsabs.harvard.edu/abs/2012A%26A...544A...9V} {544, A9}

\bibitem[\protect\citeauthoryear{{Villaver}, {Livio}, {Mustill}  \&
  {Siess}}{{Villaver} et~al.}{2014}]{Villaver+14}
{Villaver} E.,  {Livio} M.,  {Mustill} A.~J.,   {Siess} L.,  2014, \mn@doi
  [ApJ] {10.1088/0004-637X/794/1/3}, \href
  {http://adsabs.harvard.edu/abs/2014ApJ...794....3V} {794, 3}

\bibitem[\protect\citeauthoryear{{Volk} \& {Gladman}}{{Volk} \&
  {Gladman}}{2015}]{VolkGladman15}
{Volk} K.,  {Gladman} B.,  2015, \mn@doi [\apjl] {10.1088/2041-8205/806/2/L26},
  \href {http://adsabs.harvard.edu/abs/2015ApJ...806L..26V} {806, L26}

\bibitem[\protect\citeauthoryear{{Wang}, {Fischer}, {Xie}  \& {Ciardi}}{{Wang}
  et~al.}{2014}]{Wang+14}
{Wang} J.,  {Fischer} D.~A.,  {Xie} J.-W.,   {Ciardi} D.~R.,  2014, \mn@doi
  [\apj] {10.1088/0004-637X/791/2/111}, \href
  {http://adsabs.harvard.edu/abs/2014ApJ...791..111W} {791, 111}

\bibitem[\protect\citeauthoryear{{Wang} et~al.,}{{Wang} et~al.}{2015}]{Wang+15}
{Wang} J.,  et~al., 2015, \mn@doi [\apj] {10.1088/0004-637X/815/2/127}, \href
  {http://adsabs.harvard.edu/abs/2015ApJ...815..127W} {815, 127}

\bibitem[\protect\citeauthoryear{{Weidenschilling} \&
  {Marzari}}{{Weidenschilling} \& {Marzari}}{1996}]{WeidenschillingMarzari96}
{Weidenschilling} S.~J.,  {Marzari} F.,  1996, \mn@doi [\nat]
  {10.1038/384619a0}, \href {http://adsabs.harvard.edu/abs/1996Natur.384..619W}
  {384, 619}

\bibitem[\protect\citeauthoryear{{Weiss} \& {Marcy}}{{Weiss} \&
  {Marcy}}{2014}]{WeissMarcy14}
{Weiss} L.~M.,  {Marcy} G.~W.,  2014, \mn@doi [ApJL]
  {10.1088/2041-8205/783/1/L6}, \href
  {http://adsabs.harvard.edu/abs/2014ApJ...783L...6W} {783, L6}

\bibitem[\protect\citeauthoryear{{Wittenmyer} et~al.,}{{Wittenmyer}
  et~al.}{2016}]{Wittenmyer+16}
{Wittenmyer} R.~A.,  et~al., 2016, \mn@doi [\apj] {10.3847/0004-637X/819/1/28},
  \href {http://adsabs.harvard.edu/abs/2016ApJ...819...28W} {819, 28}

\bibitem[\protect\citeauthoryear{{Wu} \& {Lithwick}}{{Wu} \&
  {Lithwick}}{2011}]{WuLithwick11}
{Wu} Y.,  {Lithwick} Y.,  2011, \mn@doi [ApJ] {10.1088/0004-637X/735/2/109},
  \href {http://adsabs.harvard.edu/abs/2011ApJ...735..109W} {735, 109}

\bibitem[\protect\citeauthoryear{{Wu} \& {Murray}}{{Wu} \&
  {Murray}}{2003}]{WuMurray03}
{Wu} Y.,  {Murray} N.,  2003, \mn@doi [ApJ] {10.1086/374598}, \href
  {http://adsabs.harvard.edu/abs/2003ApJ...589..605W} {589, 605}

\bibitem[\protect\citeauthoryear{{Wyatt}}{{Wyatt}}{2008}]{Wyatt08}
{Wyatt} M.~C.,  2008, \mn@doi [\araa] {10.1146/annurev.astro.45.051806.110525},
  \href {http://adsabs.harvard.edu/abs/2008ARA%26A..46..339W} {46, 339}

\bibitem[\protect\citeauthoryear{{Xie} et~al.,}{{Xie} et~al.}{2016}]{Xie+16}
{Xie} J.-W.,  et~al., 2016, \mn@doi [Proceedings of the National Academy of
  Science] {10.1073/pnas.1604692113}, \href
  {http://adsabs.harvard.edu/abs/2016PNAS..11311431X} {113, 11431}

\bibitem[\protect\citeauthoryear{{Zuckerman}}{{Zuckerman}}{2014}]{Zuckerman14}
{Zuckerman} B.,  2014, \mn@doi [\apjl] {10.1088/2041-8205/791/2/L27}, \href
  {http://adsabs.harvard.edu/abs/2014ApJ...791L..27Z} {791, L27}

\makeatother
\end{thebibliography}

\bsp    
\label{lastpage}

\end{document}